# Signs of Possible High-Temperature Superconductivity in Graphite Intercalated with Lithium-Based Alloys

*Vadim Ksenofontov*, Vasily S. Minkov*, Alexander P. Drozdov, Ulrich Pöschl, Mikhail I. Eremets*

Max Planck Institute for Chemistry; Hahn-Meitner-Weg 1, 55128 Mainz, Germany

E-mail: v.ksenofontov@mpic.de (V.K.); v.minkov@mpic.de (V.S.M)

"† Mikhail I. Eremets passed away on November 16, 2024."

**Abstract:** We report experimental results indicating possible high-temperature superconductivity in graphite intercalation compounds synthesized with lithium-based alloys. Temperature-dependent measurements of magnetic moment, trapped magnetic flux, and electrical resistance reveal transitions with critical temperatures ($T_c$) in the range of 240-350 K at ambient pressure, depending on the intercalation metals and synthesis conditions. The highest $T_c$ values and largest high-$T_c$ volume fractions (< 0.1 %) were observed in graphite samples intercalated with ternary Sr-Ca-Li alloy. Our results and analyses suggest that the observed transitions originate from local superconductivity rather than intrinsic magnetic properties of the intercalated graphite, which may be an alternative explanation. Accordingly, we propose and intend to pursue further investigations to test and confirm the nature of the observed high-$T_c$ transitions, and to obtain larger high-$T_c$ fractions by optimizing the intercalation materials and synthesis methods.

## 1. Introduction

Theoretical predictions and recent experimental studies suggest that superconductivity at room temperature may be achievable in substances consisting of light elements.[1-4] For example, high-temperature superconductivity at critical temperatures up to $T_c \approx 250$ K has been observed for hydrides at megabar pressures.[5-7] Carbon-based materials are also considered as promising candidates for high-temperature superconductivity, because delocalized $\pi$-electron systems, low-dimensional frameworks, and high vibrational frequencies can enhance electron pairing and electron-phonon interactions.[8]

The tuneable electronic properties of carbon allotropes such as graphite, graphene, fullerenes, nanotubes, etc. enable a variety of potentially superconducting phases. For example, superconductivity has been observed in boron-doped amorphous Q-carbon with $T_c \approx 55$ K [9] and in $Cs_2RbC_{60}$ with $T_c \approx 33$ K.[10] Superconducting transition temperatures above 100 K at ambient pressure have been predicted for $C_{24}$-network systems containing Na, Mg, Al, In, and Tl. [11]

Signs of high-temperature superconductivity have been observed even in non-intercalated/undoped graphite, although these results remain controversial,[12-15] and claims of superconductivity in graphite above room temperature remain to be confirmed.[16] Superconducting transitions below 0.5 K have been observed in rhombohedral tetra- and penta-layer graphene,[17] while higher transition temperatures can be achieved in graphite intercalated with alkali or alkaline-earth metals and other species.[18] For example, superconductivity with $T_c \approx 11.5$ K has been reported for $CaC_6$[19] and $Ca_2Li_3C_6$[20] and can be described by the BCS theory.[21,22]

In a companion study, we obtained experimental evidence for possible superconductivity in graphite intercalated with calcium-ammonia solutions. [23] In this study, we intercalated graphite with lithium-based alloys and observed high $T_c$ transitions at temperatures up to 350 K in measurements of magnetic moment and magnetoresistance. The following features suggest that the observed transitions originate from local superconductivity rather than intrinsic magnetic properties of the intercalated graphite: 1) temperature-dependent magnetic moment in zero-field cooling (ZFC) and field cooling (FC) mode indicating expulsion of applied magnetic field below $T_c$ (Meissner effect); 2) field-dependent magnetic moment characteristic for type II superconductors; 3) magnetic flux trapping and magnetic flux creep; 4) shift to lower temperatures and broadening of transitions with increasing magnetic field.

## 2. Results and Discussion

### 2.1 Samples

In the main text of this paper, we focus on three types of samples obtained by intercalation of expanded graphite sheets (EGS) with melts of pure Li metal, binary Ca-Li alloy, and ternary Sr-Ca-Li alloys at temperatures in the range of 423 K to 623 K over a period of one to ten minutes as specified in Table 1. Details on the intercalation methods and complementary results obtained with other samples are presented in the Supporting Information (SI), including highly oriented pyrolytic graphite (HOPG) and EGS intercalated with other metals and alloys. The results obtained with HOPG were qualitatively similar but less pronounced, which may be due to the high surface-to-volume ratio of EGS[24], which favors rapid diffusion and intercalation.

**Table 1.** Metals/alloys (atomic ratios), reaction temperatures, and reaction times applied in the synthesis of the EGS intercalation samples presented and discussed in the main text.

| Sample name | Metal/Alloy (atomic ratio) | Intercalation Temperature (K) | Intercalation Time (min) |
|---|---|---|---|
| Sample I | Li | 623 | 1 |
| Sample II | Ca-Li (1:3.5) | 623 | 1 |
| Sample IIIa | Sr-Ca-Li (1:1:20) | 523 | 1 |
| Sample IIIb | Sr-Ca-Li (1:2:20) | 623 | 1 |
| Sample IIIc | Sr-Ca-Li (1:2:20) | 473 | 10 |
| Sample IIId | Sr-Ca-Li (1:2:20) | 423 | 10 |

## 2.2 Temperature-dependent magnetic moment

Figure 1 shows the temperature dependence of magnetic moment, *m(T)*, for EGS intercalated with binary Ca-Li alloy (Sample II) measured in zero-field cooling (ZFC) and field-cooling (FC) mode. It reflects a well-established superconducting transition near 11.5 K, which can be attributed to a low-temperature superconducting phase of $CaC_6$[21] or $Ca_2Li_3C_6$[20] (low-$T_c$ transition, Fig. 1a). A substantially smaller but still pronounced step in the ZFC signal indicates another transition at 240 K (high-$T_c$ transition, Fig. 1b), which is consistent with similar high-$T_c$ transitions observed in graphite samples intercalated with Ca–$NH_3$ solutions in the companion study. [23] Decreasing the Ca content in Ca:Li alloy from 1:3.5 to a 1:20 resulted in a reduction of signal intensity for both transitions, while the high-$T_c$ transition remained at about 240 K. Lower intercalation temperature below 523 K reduced the low-$T_c$ transition and increased the high-$T_c$ transition signal at $T_c$ values up to 265 K (Fig. S13, SI). Overall, the samples obtained with binary Ca-Li alloys exhibited high-$T_c$ transitions in the range of 240 K to 265 K (Fig. S14, SI). The ratio between high-$T_c$ and low-$T_c$ transition signals varied significantly with intercalation temperature, time, and Ca content in the Ca–Li alloys, indicating that the high-$T_c$ transition is not related to the low-$T_c$ phase. The intercalation kinetics presented in the SI (Fig. S15) support this conclusion: longer intercalation times at temperatures above 623 K lead to a growth of low-$T_c$ transition signal, while the high-$T_c$ transition signal vanishes. Graphite samples intercalated with Li only (Sample I) did not show any high-$T_c$ anomalies in magnetic moment measurements (Fig. S16, SI), which confirms that the observed high-$T_c$ anomaly is due to intercalated Ca atoms in accordance with the result of our companion study[23].

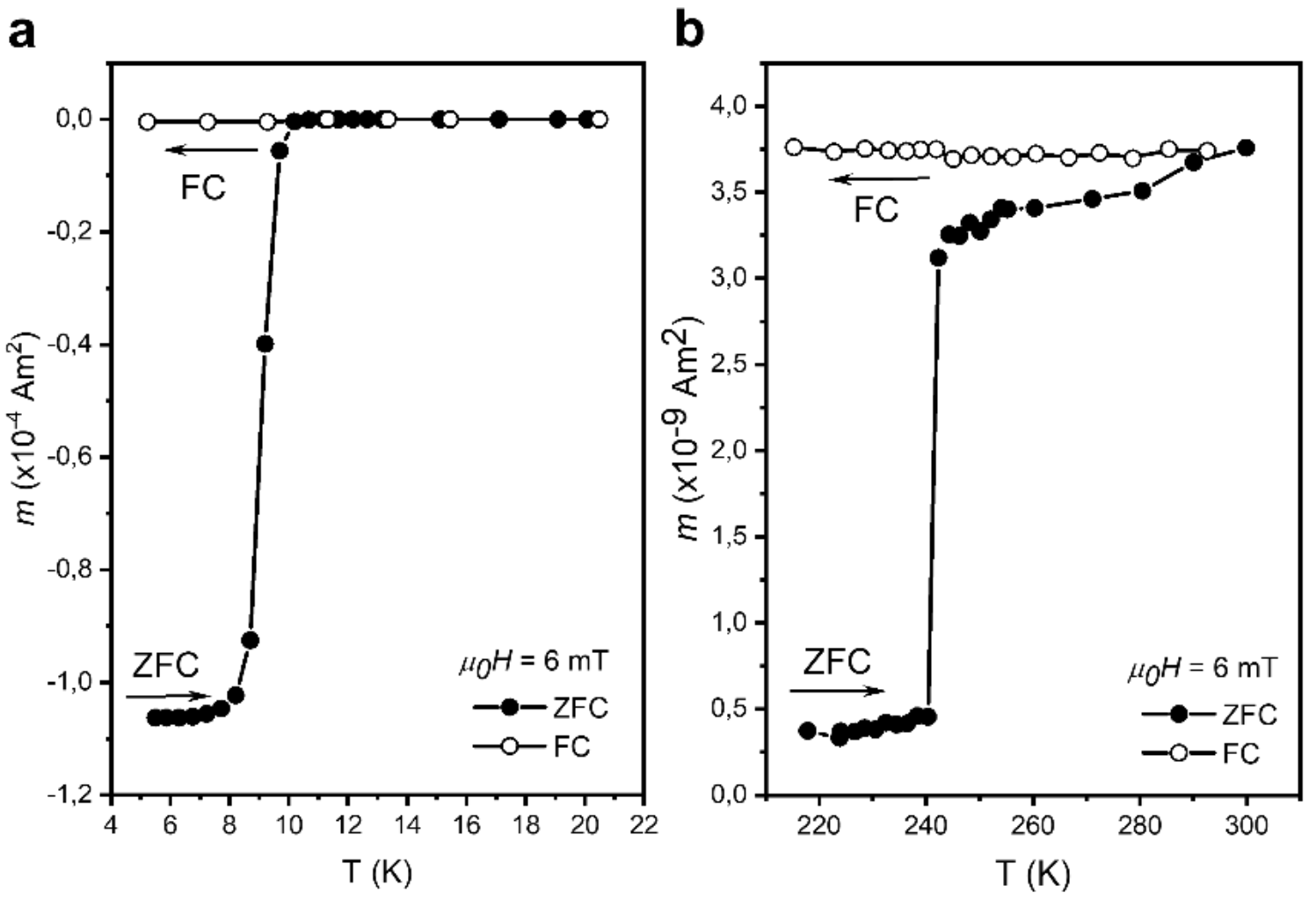

**Figure 1.** Temperature-dependent magnetic moment, *m(T)*, of a graphite sample intercalated with binary Ca-Li alloy (Sample II). Magnetic field is perpendicular to graphene layers. a) low-$T_c$ transition near 11.5 K attributable to known superconductivity in $CaC_6$[21] or $Ca_2Li_3C_6$[20], b) high-$T_c$ transition near 240 K. Measurement data points from zero-field cooling (ZFC) and field-cooling (FC) modes (open/filled circles); connecting lines to guide the eye.

Like calcium, strontium can form a $SrC_6$ phase with a significantly lower superconducting $T_c$ of 1.65 K.[25] In a graphite sample intercalated with ternary Sr-Ca-Li alloy (Sample IIIa), we did not observe a low-$T_c$ superconducting transition above the experimental lower limit of 5 K (Fig. 2a). On the other hand, the co-intercalation with Sr increased the high-$T_c$ transition temperature to 261 K. Apparently, co-intercalation with Sr suppresses the formation of the $CaC_6$ phase and may increase the amount of intercalated Ca in the graphite. Doubling the Ca content in the alloy and increasing the intercalation temperature raised the high-$T_c$ transition temperature in Sample IIIb further to ≈ 270 K (Fig. 2b). Compared to binary Ca–Li alloys, graphite samples intercalated with ternary Sr–Ca–Li alloys exhibited high-$T_c$ transitions at higher temperatures up to 350 K (Fig. S17, SI). For instance, Sample IIIc exhibits a high-$T_c$ transition near 350 K (Fig. 2c).

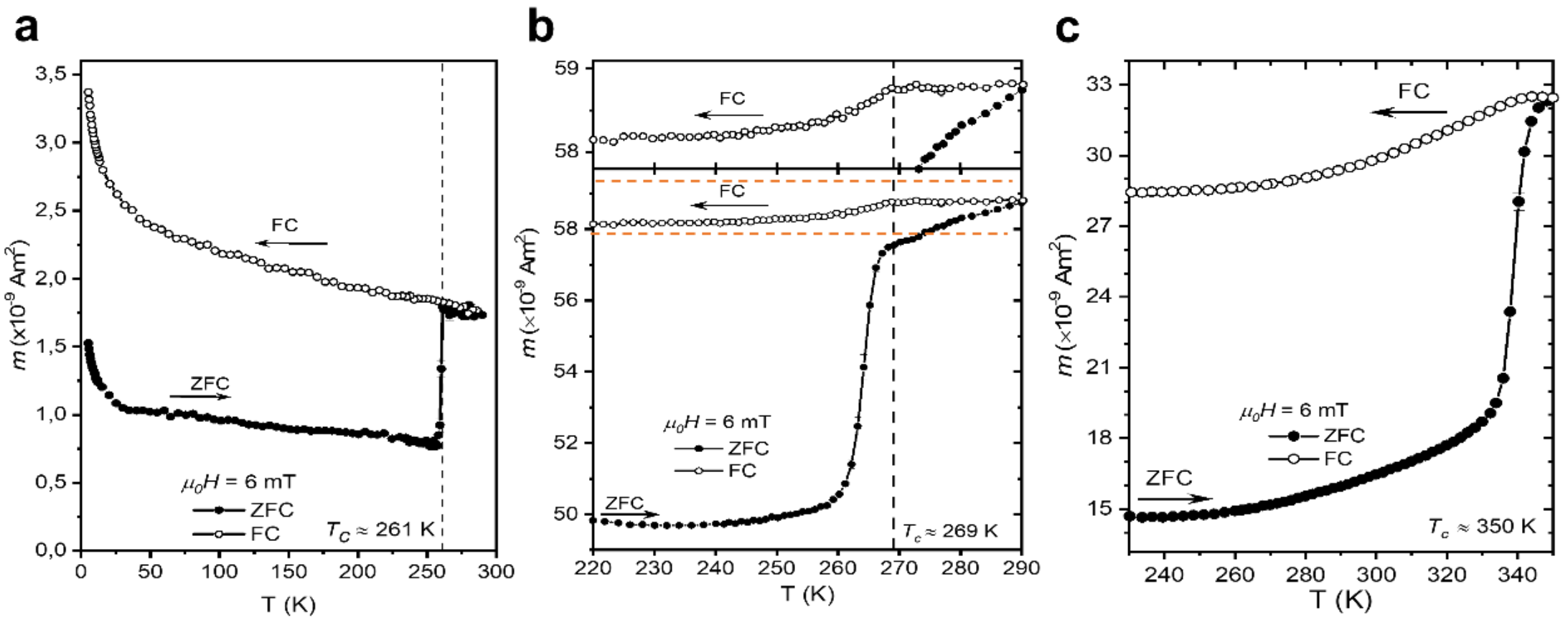

**Figure 2.** Temperature-dependent magnetic moment, *m(T)*, of graphite samples intercalated with ternary Sr-Ca-Li alloys (Samples IIIa, IIIb, IIIc): a) high-$T_c$ transition at ≈ 261 K in Sample IIIa. b) high-$T_c$ transition at ≈ 270 K in Sample IIIb (zoomed region displayed at top of panel). c) high-$T_c$ transition near 350 K in Sample IIIc. Measurement data points from zero-field cooling (ZFC) and field-cooling (FC) modes (open/filled circles); connecting lines to guide the eye.

### 2.3 Trapped magnetic flux

The high-$T_c$ transitions detected by temperature-dependent magnetic moment measurements may be attributable to magnetic or superconducting properties of the investigated samples. To elucidate the physical mechanism, we investigated the trapping of magnetic flux as an important hallmark of superconductivity[26,27], which was effective in the probing of high-temperature superconductivity in hydrogen-rich compounds under high pressure[28] and other superconductors at atmospheric pressure.[29,30] The trapped magnetic flux occurs due to persistent currents in a superconducting sample after exposing it to a magnetic field. The induced magnetic moment, $m_{trap}$, decreases upon warming of the sample and disappears at $T_c$. Figure 3b shows trapped flux measurements in graphite intercalated with a ternary Sr-Ca-Li alloy (Sample IIIb). The temperature dependence of the trapped magnetic moment, $m_{trap}(T)$, induced by application and subsequent switching of a magnetic field of 1 T at 5.2 K, shows a clear stepwise transition at approximately 270 K, which is in good agreement with the high-$T_c$ transition temperature obtained in conventional temperature-dependent magnetic moment measurements of the same sample (Fig. 3a). The magnetic moment measured at reverse cooling from ≈ 286 K to 5 K is almost temperature-independent (run III, Fig. 3b).

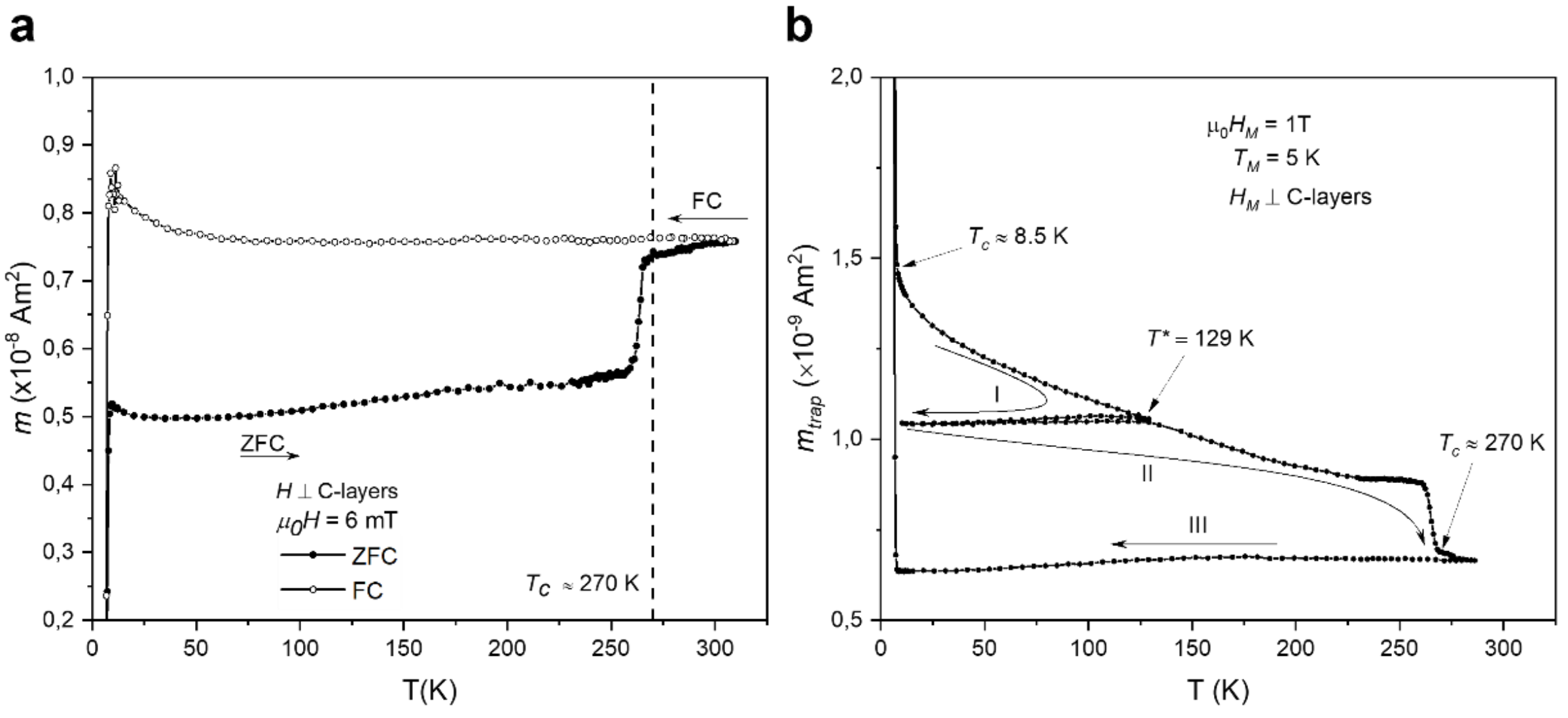

**Figure 3.** Trapping of magnetic flux in graphite intercalated with ternary Sr-Ca-Li alloy (Sample IIIb): a) temperature-dependent magnetic moment, *m(T)*, measured in magnetic field of 6 mT perpendicular to graphene layers. Diamagnetic signal below 8.5 K attributed to concomitant low-$T_c$ superconducting phase ($CaC_6/Ca_2Li_3C_6$). b) Temperature dependence of trapped magnetic moment, $m_{trap}(T)$, measured in zero field. Trapped magnetic moment was generated at a magnetic field of 1 T at 5.2 K (ZFC protocol, $H_M$ perpendicular to graphene layers). Strong increase of $m_{trap}(T)$ below 8.5 K attributed to trapped magnetic flux in concomitant low-$T_c$ superconducting phase ($CaC_6/Ca_2Li_3C_6$). Black arrows indicate the course of temperature change in the experiment (runs I, II, III).

To check if induced magnetic moment could be of magnetic rather than superconducting origin, we also investigated the behaviour of $m_{trap}(T)$ in the trapped magnetic flux sample, which lost part of its induced magnetic moment upon heating to an intermediate temperature $T^*$ = 129 K. When the temperature was again decreased below $T^*$ (run I, Fig. 3b), $m_{trap}(T)$ remained essentially stable during a subsequent cooling and heating cycle in the range of 10 K –129 K (intermediate plateau in Fig. 3b). When the temperature was further increased above $T^*$ = 129 K (run II, Fig. 3b), $m_{trap}$ follows the initial trajectory, ending with a step at $T_c \approx 270$ K. This observed behaviour suggests that the induced magnetic moment is due to superconductivity rather than intrinsic magnetism. If it were due to intrinsic magnetism, $m_{trap}(T)$ would not be expected to exhibit a stable plateau value during the cooling and heating cycle between 129 K and 10 K; instead, $m_{trap}(T)$ would be expected to follow the original heating curve or exhibit a similar temperature dependence with negative slope $\frac{\partial m_{trap}}{\partial T} < 0$. The characteristic plateau of $m_{trap}(T)$ was first observed in trapped flux investigations of superconducting hydrogen-rich compounds under high pressure[28] and was recently confirmed in superconducting Sn-Pb alloys at atmospheric pressure[30]. In contrast, ferromagnetic materials usually do not exhibit such characteristic plateaus with zero or positive slope as illustrated for weak ferromagnetic $FeF_3$ (Fig. S18, SI).

The slight, steady decrease in $m_{trap}(T)$ with $\frac{\partial m_{trap}}{\partial T} > 0$ observed upon cooling in the plateau region (Fig. 3b) can be attributed to the thermally activated creep of superconducting vortices, which according to the Anderson-Kim theory leads to a logarithmic decay over time of the trapped magnetic moment. Fig. 4 shows the decay of $m_{trap}(t)$ observed over a period of 20 hours for Sample IIIb in ZFC mode after application and switching off the magnetic field of $\mu_0 H_M$ = 1 T at $T_M$ = 12.2 K in two different orientations relative to the graphene layers. We fitted the measured $m_{trap}(t)$ data by an equation with logarithmic time dependence $m_{trap}(t) = m_{trap,0} - m_1 \times \ln(t)$ (green lines in Fig. 4), using $m_{trap,0}$(12.2 K) ≈ 7.51×10$^{-9}$ A m$^2$ for $H_M \perp$ C–layers and $m_{trap,0}$(12.2 K) ≈ 11.72×10$^{-9}$ A m$^2$ for $H_M$ || C–layers, both estimated from $m_{trap}(T)$ trapped flux measurements. The rate of logarithmic relaxation $S = -\frac{1}{m_{trap,0}}\frac{dm_{trap}}{d\ln t} = -\frac{k_B T}{U_{eff}}$ depends on the temperature and activation energy of motion of magnetic vortices $U_{eff}$. At $T$ = 12.2 K relaxation rate $S(H_M \perp$ C–layers) ≈ - 0.006 and $S(H_M$ || C–layers) = - 0.005 , whereas at higher temperatures its absolute value, as expected, increases: $S(H_M \perp$ C–layers) ≈ - 0.007 at $T$ = 130 K and $S(H_M \perp$ C–layers) ≈ - 0.013 at $T$ = 260 K.

Overall, the trapped flux experiments suggest that the observed high-$T_c$ transition in the investigated samples is related to superconductivity rather than intrinsic magnetic phenomena.

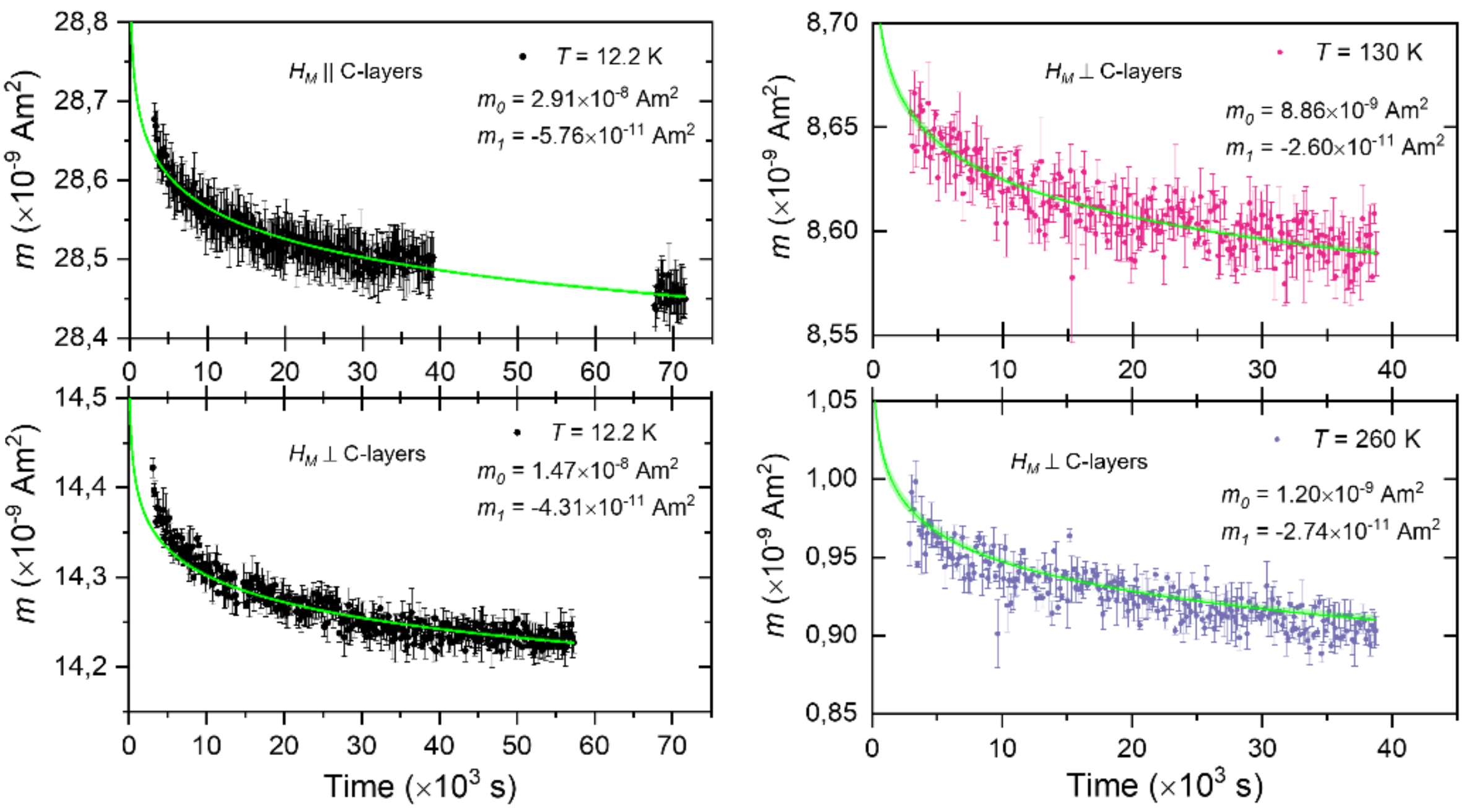

**Figure 4.** Time dependence of trapped magnetic moment at $T$ = 12.2 K, 130 K, and 260 K in graphite intercalated with ternary Sr-Ca-Li alloy (Sample IIIb). Trapped magnetic flux was created at $\mu_0 H_M$ = 1 T and $T$ = 12.2 K under ZFC protocol with two different orientations of the sample relative to the applied magnetic field (black circles): when $H_M$ is parallel (upper panel) and perpendicular to graphene layers (lower panel). $m_{trap}(t)$ datasets were also measured at higher temperatures, $T$ = 130 K (magenta circles) and $T$ = 260 K (blue circles). For these measurements trapped flux were generated at $\mu_0 H_M$ = 1 T, and $T$ = 130 K and $T$ = 260 K under ZFC protocol, respectively. Error bars represent the standard deviation of the mean value averaged over three independent measurements. Green curves are fits of experimental data by equation $m_{trap}(t) = m_0 - m_1 \times \ln(t)$.

**2.4 Magnetic field dependence of high-$T_c$ transition**

To investigate the effect of magnetic fields on the observed high-$T_c$ transition, we performed a series of $m(T)$ measurements in Sample IIIb at different magnetic field strengths and different orientations of the sample: parallel ($H$ || C-layers) and perpendicular ($H \perp$ C-layers) to the graphene layers (Figs. 5a and 5b; Fig. S19, SI). With increasing magnetic field, $T_c$ shifts to lower temperature (Fig. 5c) and a broadening of the transition occurs (Figs. 5a and 5b), which is characteristic for superconductivity. The broadening is more pronounced when the magnetic field is parallel to the graphene layers and the circulating currents propagate vertically along the $c$-axis across inhomogeneous interlayer space in contrast to the highly conductive graphene layers.

As illustrated in Fig. 5d, the change of magnetic moment during the high-$T_c$ transition, $|\Delta m|$, exhibits a sharp initial increase and reaches a maximum at $\mu_0 H \approx$ 70 mT ($H$ || C-layers) and $\mu_0 H \approx$ 100 mT ($H\perp$ C-layers), respectively. As the applied magnetic field is further increased, $|\Delta m|$ decreases continuously and finally disappears. The observed behaviour is characteristic for type-II superconductors and indicates a mixed state. The near-linear initial increase of $|\Delta m|$ resembles the virgin curve of magnetic moment due to the Meissner effect. Note that the behavior of $m(T)$ in the FC measurement of a Sample IIIb (Fig. 2b) and a Sample IIIc (Fig. 2c) also indicates a Meissner effect. The subsequent continuous decrease of $|\Delta m|$ is consistent with the penetration of vortices through the sample in a mixed superconducting state, thereby reducing the shielding of the applied magnetic fields by the Meissner state. The applied magnetic field at which $|\Delta m|$ reaches a maximum (Fig. 5d) can serve as an upper estimate for the penetration field $H_p$, above which the magnetic field begins to penetrate into a superconducting sample. For fields parallel to the conducting layers, the shielding currents are more suppressed than for fields perpendicular to the planes, where the shielding currents flow within the homogeneous graphene layers. This leads to a lower first critical field $H_{c1}$ in the parallel configuration and consequently to a lower penetration field value of 70 mT recorded in our experiment. For Sample IIIc with a high-$T_c$ transition near 350 K, similar results were obtained as illustrated in Fig. S21 (SI) analogous to Fig. 5.

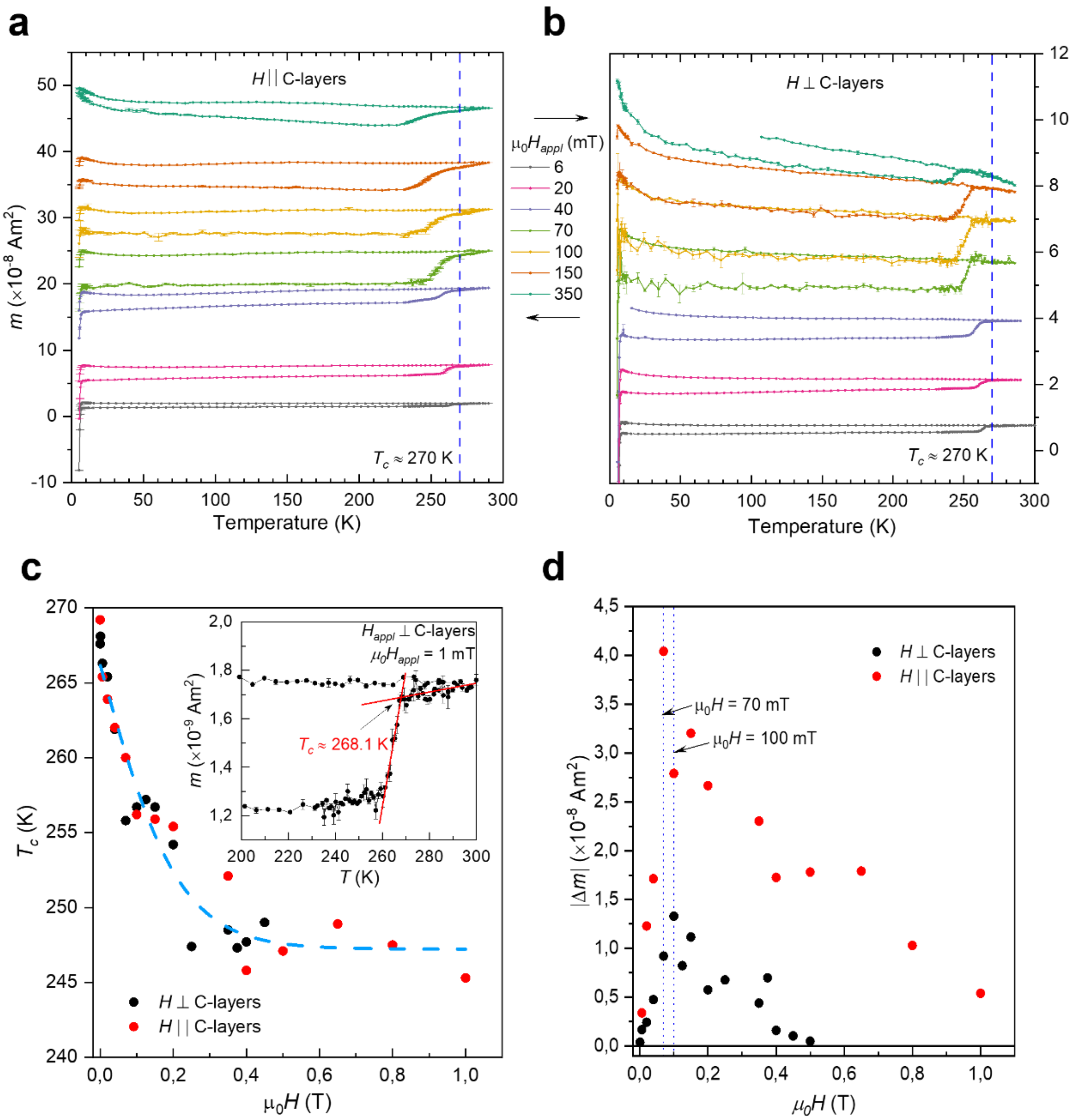

**Figure 5.** Magnetic field dependence of transitions in graphite intercalated with ternary Sr-Ca-Li alloy (Sample IIIb): temperature-dependent magnetic moment, *m(T)*, measured in ZFC and FC modes at different applied magnetic fields in two orientations: a) parallel to graphene layers; and b) perpendicular to graphene layers; dashed blue line marks high-$T_c$ transition at zero field near 270 K. c) decrease of $T_c$ with increasing magnetic field; $T_c$ defined as intersection of two linear fits to the data across and above the transition (inset); dashed blue line is a fit based on the hyperbolic tangent function as specified in the SI (Sect. S.4.7) d) change in magnetic moment $|\Delta m|$ during high-$T_c$ transition as a function of magnetic field.

As shown in Fig. 5c, the decrease of $T_c$ is exceptionally strong in relatively weak magnetic fields, whereas the field dependence of $T_c$ weakens at higher magnetic fields. A magnetic field of 0.2 T reduces $T_c$ by ≈ 15 K, corresponding to a slope of approx. 75 K/T, which is much more than expected according to known orbital decoupling mechanisms. For comparison, the high-temperature superconductor $H_3S$ with $T_c$ ≈ 203 K exhibits a field-induced decrease in $T_c$ of approximately 1.3 K/T.[31] The unusually strong field dependence of $T_c$ in the intercalated graphite may be related to Pauli paramagnetism of conducting electrons as reported for lithium-intercalated graphite,[32] and their polarization in external magnetic fields. Due to strong background signals of the intercalation samples prepared with lithium alloys, we were not able to observe the characteristic signature of superconductivity in magnetic hysteresis measurements, which was achieved for the graphite samples intercalated with Ca–$NH_3$ solutions in our companion study.[23]

### 2.5 Low-$T_c$ and High-$T_c$ volume fractions

Figure 6 shows the magnetic field dependence of the change in magnetic moment (*Δm*) observed during the low-$T_c$ transition in graphite intercalated with ternary Sr-Ca-Li alloy (Sample IIIb), which is characteristic for type-II superconductors. From the slope of the *Δm(H)* curve in the descending branch (0 to 6 mT), we obtain anisotropic values of volume magnetic susceptibility ($\chi_V = (d\Delta m/dH)/V$) when the magnetic field is oriented parallel (in-plane) or perpendicular (out-of-plane) to the graphitic graphene layers, respectively: $\chi_{V\parallel}$ = (-2.9±0.1) $\cdot 10^{-3}$ and $\chi_{V\perp}$ = (-1.7±0.1) $\cdot 10^{-2}$.

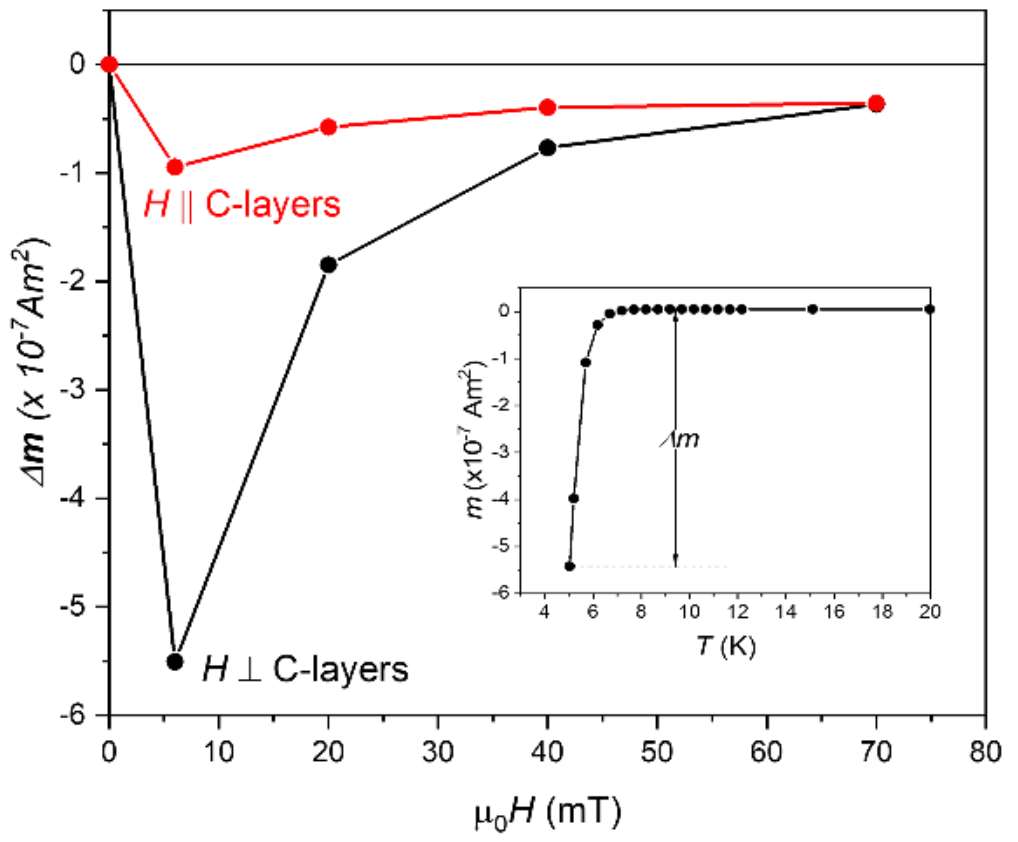

**Figure 6.** Change in magnetic moment *Δm* during low-$T_c$ transition in graphite intercalated with ternary Sr-Ca-Li alloy (Sample IIIb) plotted against the applied magnetic field. The red and black curves demonstrate the dependence of *Δm(H)* for the applied field oriented parallel and perpendicular to the graphene layers, respectively. The inset shows the determination of *Δm* as the difference in magnetic moment values at 5.2 K and above the transition temperature

measured in an external field of $\mu_0 H = 6$ mT applied perpendicular to the layer planes. All $\Delta m$ values are determined relative to 5.2 K and correspond to measurements performed in the ZFC mode.

This experimental result is consistent with the anisotropic properties of $CaC_6$ and/or $Li_3Ca_2C_6$ phases, to which the low-$T_c$ transition in Sample IIIb is attributable.[20] As detailed in the SI (Sect. S.4.8.1),[33][34][35] we obtain estimates of lower critical field at zero temperature ($\mu_0 H_{c1\perp}(0) = 14.4\pm6.4$ mT; $\mu_0 H_{c1||}(0) = 2.9\pm1.2$ mT) and penetration fields ($B_{p\perp} = 5.3\pm2.3$ mT out-of-plane, $B_{p||} = 2.1\pm1$ mT in-plane), respectively. For the perpendicular orientation (black curve in Fig. 6), the applied magnetic field in the descending part of the $\Delta m(H)$ curve is of similar magnitude as the estimated penetration field, which implies that the low-$T_c$ phase is on the verge of the Meissner state. In parallel orientation, however, (red curve in Fig. 6), the applied field exceeds the penetration field and the sample is expected to be in a mixed Shubnikov state. In the Meissner state, the volume magnetic susceptibility is related to the volume fraction of the superconducting phase ($f$): $\chi_V = -\frac{f}{1-N\cdot f}$ , where $N$ is the sample demagnetization factor.[35] To account for the demagnetization factor of Sample IIIb, we normalize the measurement result by the volume magnetic susceptibility of a superconducting lead sample of the same shape and dimensions ($\chi_V(Pb) = -1.53\pm0.01$, Sect. S.4.7.2) and obtain the following estimate for the volume fraction of the low-$T_c$ phase in sample IIIb: $f_L \approx \chi_{V\perp}/\chi_V(Pb) \approx 1.1\pm0.1$ %. Performing the same analysis for the change of magnetic moment observed at the high-$T_c$ transition (Fig. 5d), we obtain the following anisotropic values of volume magnetic susceptibility for the high-$T_c$ phase: $\chi_{V||} = (1.0\pm0.1)\cdot10^{-4}$ and $\chi_{V\perp} = (2.3\pm0.2)\cdot10^{-5}$. The anisotropy can be tentatively explained by superconducting flakes in a non-magnetic, non-superconducting matrix, where the flakes are oriented perpendicular to the graphene layers and the penetration field is higher for this orientation of the applied field. Assuming a Meissner state for the high-$T_c$ phase in this orientation to the magnetic field, we obtain the following estimate for the volume fraction of the high-$T_c$ phase in Sample IIIb: $f_H \approx \chi_{V||}/|\chi_V(Pb)| \approx (6.5\pm0.7)\cdot10^{-5} \approx 0.01$ %. The largest high-$T_c$ volume fractions are observed in ternary compounds intercalated at temperatures below 523 K, when the low-temperature phase is absent. The maximum value $f_H = (7.2\pm1.3)\cdot10^{-4}$ is obtained for Sample IIIc (Sect. 4.8.2, Table S4, SI). Overall, the intercalated samples are characterized by $f_H < 0.1$ %.

## 2.6 Resistivity

As shown in Fig. 7a, the measurements in zero magnetic field of the graphite intercalated with ternary Sr-Ca-Li alloy (Sample IIId) do not exhibit substantial changes of resistivity at the high-$T_c$ transition temperature about 350 K (Fig. 7b), which can be attributed to the low abundance of the high-$T_c$ fraction ($f_H < 0.1$ %). Zero resistivity of the high-$T_c$ fraction can be estimated to reduce the overall resistivity ρ (approx. $10^{-6}$ Ω m) of the intercalated graphite sample by $f_H \cdot \rho < 1 \cdot 10^{-9}$ Ω m, which is less than the average noise level of approx. $2 \cdot 10^{-10}$ Ω m of our four-probe resistance measurements. Nevertheless, the conductivity of the Sample IIId is several times higher than for non-superconducting graphite intercalated with Li metal only (Sample I, Fig. 7a), indicating higher density of conducting electrons as discussed below. Both resistivities exhibit a power-law temperature dependence up to approx. 100 K followed by a linear increase up to approx. 200 K, which is characteristic for metallic conductors. At higher temperatures, however, the resistivities exhibit a steeper increase deviating from the linear trend, which may be explained by increased charge carrier scattering due to the thermally activated motion of $Li^+$ ions as detected by $\mu^+$SR in $LiC_6$ and $LiC_{12}$.[36]

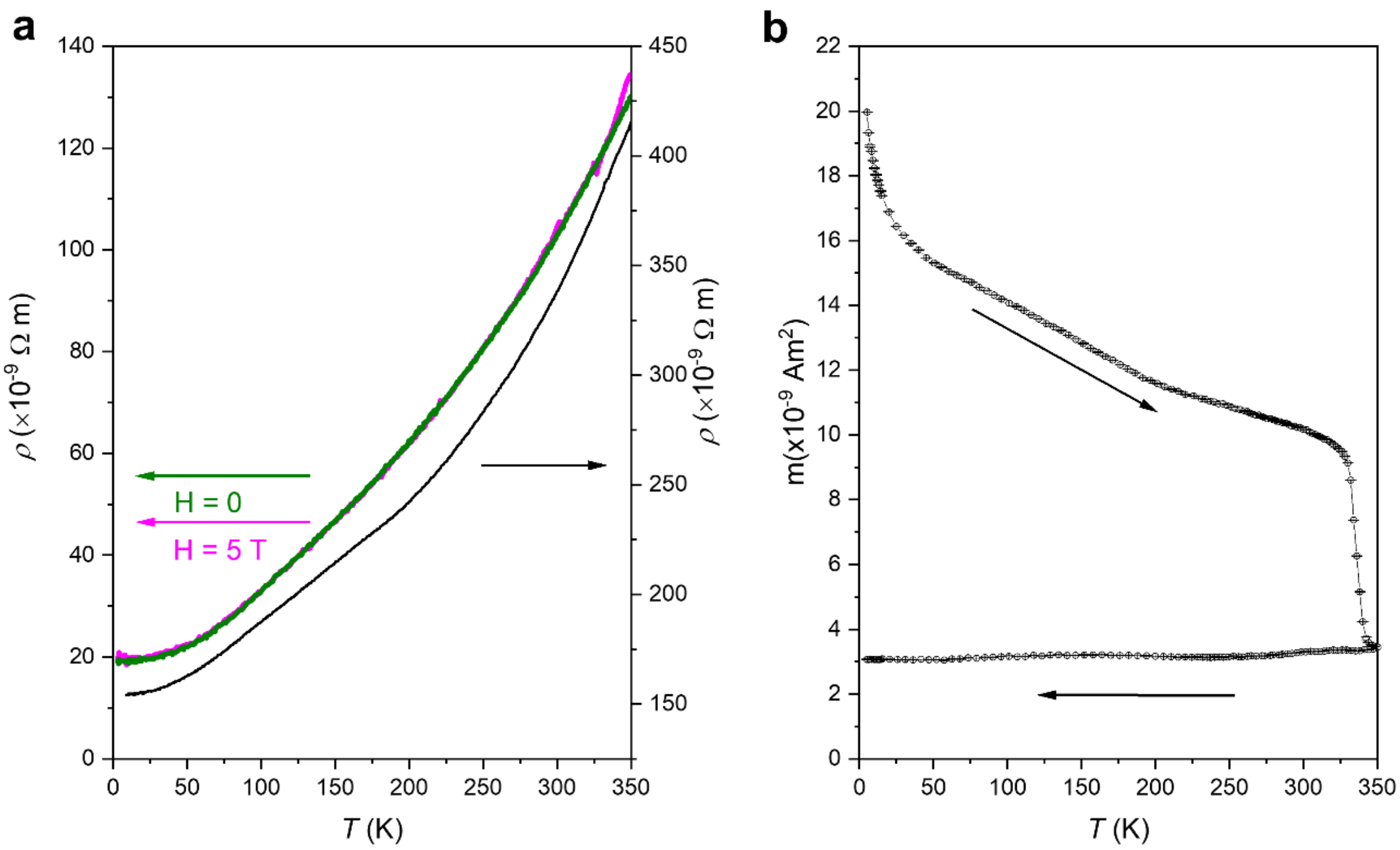

**Figure 7.** Temperature-dependent resistivity and trapping of magnetic flux in graphite sample intercalated with ternary Sr-Ca-Li alloy (Sample IIId): a) Four-probe resistance measurements of graphite intercalated with pure Li metal (Sample I, black curve, right axis) and with ternary Sr-Ca-Li alloy (Sample IIId), green curve at zero magnetic field; magenta curve at 5 T parallel to graphene layers, heating mode; left axis. b) Trapped magnetic flux in Sample IIId.

Temperature dependence of trapped flux generated at 1 T and 5.2 K indicates high-$T_c$ transition near 350 K.

Up to 300 K, the resistivities of Sample IIId measured in a magnetic field of 5 T and zero magnetic field coincide within the noise level of our resistivity measurements of approx. $2 \cdot 10^{-10}$ Ω m. We plan to extend our resistance measurements of intercalated graphite samples in variable magnetic fields across and above the high-$T_c$ transition in forthcoming follow-up studies (Sect. S.4.5.3, Fig. S23, SI).

The absence of pronounced resistivity changes at high $T_c$ transitions in zero magnetic field does not contradict possible superconductivity if the high-$T_c$ fraction is distributed in the form of isolated flakes rather than in the form of interconnected filaments. Due to the low volume fraction ($f_H < 0.1\%$), the average distance between compact flakes would be over 10 times larger than their size.

## 3. Conclusions

At present, we cannot yet fully clarify the origin of the observed high-$T_c$ transitions, but the available experimental evidence suggests that it may be granular or local superconductivity as previously reported for undoped Grafoil (microcrystalline graphite similar to EGS) with a transition temperature of 14 K.[37] Our results and interpretation are consistent with other studies reporting signs of high-$T_c$ superconductivity in different types of graphitic materials. Precker *et al*. reported signs of granular superconductivity in natural graphite at temperatures as high as 350 K (flux trapping and partial expulsion) that were attributed to 2D interfaces between rhombohedral and Bernal phases.[38] Stiller *et al.* mapped the remanent magnetic flux trapped in natural graphite samples by magnetic force microscopy to support the hypothesis of localized superconductivity with $T_c \approx 370$ K.[39] The importance of structural disorder for superconductivity in graphite was also discussed by Layek *et al.* [40], who reported room-temperature granular superconductivity in potassium intercalated and subsequently deintercalated HOPG samples, a process that results in highly disordered graphene stacking and a non-uniform distribution of potassium atoms within the residual high-stage compound, which predominantly exhibits a $KC_{72}$ structure. The authors[40] estimated volume fractions $< 3 \cdot 10^{-5}$ for the superconducting phase and concluded that the superconducting regions are lamellar grains that become superconducting only with appropriate doping, which is consistent with our results for intercalated EGS samples. In a subsequent study from the same research group, Rousset-Zenou *et al.* [41] reported that granular superconducting signatures (up to 500 K) are present in entirely undoped, off-the-shelf graphite materials with a high density of defects or structural disorder and "ill-stacked" interfaces. It is worth noting that the small magnitude of the effects observed in these studies requires careful and precise processing of the magnetic measurement data – including the subtraction of the intrinsic diamagnetism of the graphite and the subsequent deconvolution of the ferromagnetic and superconducting-type hysteresis loops.

The chemical and physical steps used to prepare our EGS samples can explain why the superconducting flakes appear to be oriented perpendicular to the graphite basal planes. When graphite is rapidly heated to form expanded graphite, the intercalated acids are volatilized and create massive internal pressure that causes the graphene layers to buckle, twist, and deform into a highly porous, “worm-like” structures folding up by 90° relative to their original orientation.[42] Superconducting Cooper pairs in doped graphite form at step-edged interfaces or rhombohedral stacking faults, where the electronic density of states becomes exceptionally high (flat bands superconductivity).[43-47] Because these shear planes and tearing defects run vertically through the exfoliating graphene layers, the resulting superconducting channels from localized “flakes” running perpendicular to the remaining flat basal planes.

X-ray powder diffraction confirmed the presence of the hexagonal and rhombohedral phases in our EGS starting material (Fig. S4 and S5, SI) and also in the intercalated sample (Sect. S.4.6, SI). Theoretical studies forecast that interfaces between rhombohedral graphite and hexagonal graphite (Bernal graphite) are especially promising for high-$T_c$ superconductivity due to flat bands at the Fermi energy.[43-47] Moreover, an enhancement of the conducting electron density is also expected to favour high-$T_c$ superconductivity.[43] Our resistance measurements indicate that the conducting electron densities in the high-$T_c$ samples intercalated with ternary Sr-Ca-Li alloy are higher than in non-superconducting graphite intercalated with Li metal only. Summarizing the essence of this hypothesis, possible high-$T_c$ superconductivity is driven by a synergistic mechanism: the interfaces between coexisting hexagonal and rhombohedral phases (flakes) provide the necessary flat-band structural template, while binary or ternary alloy intercalation enhance the electronic carrier density to activate the state.

Alternative hypothetical explanation is based on the observation of electronic stripes - charge density waves (CDWs) - in the graphitic superconductor $CaC_6$ via high-resolution scanning tunneling microscopy.[48] This observation demonstrated that graphene sheets can host a pristine, purely electronic CDW driven by a topological reconstruction of the Fermi surface and coupled to the softest phonon modes of the intercalant lattice around 250 K. These electronic stripes are highly likely to exist within EGS intercalated with a ternary Li-Ca-Sr alloy. A theoretical study based on the results and data of our experimental investigations suggests that localized nanoscale Sr-Ca-Li metal flakes or particles adopt the phonon spectrum of the surrounding graphite matrix, thereby generating conventional electron-phonon superconductivity in these particles at room temperature.[49] The soft phonons associated with the extracted Einstein temperature ($\Theta_E$) right around 250 K provide the pairing mechanism required to push isolated grains into a superconducting state. The CDW domain walls can provide a natural mosaic network that physically partitions the material, stabilizing these high-temperature superconducting flakes within a normal metallic matrix. To definitively clarify these interwoven states, magnetoresistive measurements of intercalated EGS are of paramount importance. To address these delicate interlayer dynamics and further investigate the coupling mechanisms, magneto-transport measurements are envisaged (Fig. S23, SI).

Critically reviewing our results, we have also considered other alternative explanations, such as ferromagnetism,[50] magnetic impurities, superparamagnetism, or other intrinsic magnetic properties of the intercalated graphite samples. However, the following features speak against an intrinsic magnetic nature and for a superconducting origin of the observed high-$T_c$ transitions:

i) magnetic moment *m(T)* in ZFC and FC modes, consistent with type-II superconductors. Below the high-$T_c$ transition, *m(T)* is almost temperature-independent in a broad temperature range that is not specific for spin-glass or superparamagnetic properties (Fig. 5 a,b);

ii) field-dependent magnetic moment consistent with type-II superconductors (Fig. 5d);

iii) magnetic flux trapping and magnetic flux creep; characteristic plateau in the temperature dependence of trapped magnetic moment (Fig. 3b);

iv) shift of the high-$T_c$ transition to lower temperatures and its broadening with increasing magnetic field.

Ongoing experiments show that the observed high-$T_c$ transitions and signs of possible superconductivity are not limited to Ca-doped systems. Graphite samples intercalated by K–Li, Sr–Li, Mg–Li, Zn–Li, and Mg–Zn–Li alloys exhibit similar effects and are summarized in the SI (Fig. S28 – S30, SI).

In summary, we present reproducible observations of high-$T_c$ transitions suggesting possible superconductivity in EGS intercalated with Li-based binary and ternary metal alloys at ambient pressure, exhibiting $T_c$ values near and above room temperature. Further experimental and theoretical studies are required to independently confirm and further elucidate the observed phenomena.

**Acknowledgements**

The authors acknowledge funding and support by the Max Planck Society. We thank Jan Leitner for the SIMS characterization of the original EGS and HOPG materials. V.K. thanks Angela Möller, Anna Katharina Weber and Frank Ludwig for X-ray powder diffraction measurements and helpful discussions. The paper is dedicated to the memory of Mikhail I. Eremets, who passed away before its completion. An earlier version of this manuscript was posted in arxiv.org on 26 September 2025.

**Data Availability Statements**

Data will be published in suitable online repository.

**Supporting Information**

The authors have cited additional references within the Supporting Information.[51-62]

**Author Contributions**

M.I.E. initiated the study and developed it together with V.K. and V.S.M., who contributed equally to the overall research project. The samples presented in this paper were prepared by V.K. The measurements and analyses presented in this paper were performed by V.K. with contributions from V.S.M.  V.K. and U.P. wrote the manuscript with contributions from V.S.M. and A.P.D. All authors discussed the results and conclusions.

**References**

1. V. L. Ginzburg, "Once again about high-temperature superconductivity," *Contemporary Physics* 33 (1992): 15.
2. V. L. Ginzburg, "High-temperature superconductivity (history and general review)," *Soviet Physics Uspekhi* 34 (1991): 283.
3. N. W. Ashcroft, "Metallic Hydrogen: A High-Temperature Superconductor?," *Physical Review Letters* 21 (1968): 1748.
4. N. W. Ashcroft, "Hydrogen Dominant Metallic Alloys: High Temperature Superconductors?," *Physical Review Letters* 92 (2004): 187002.

5. A. P. Drozdov, M. I. Eremets, I. A. Troyan, V. Ksenofontov, S. I. Shylin, "Conventional superconductivity at 203 K at high pressures," *Nature* 525 (2015): 73.

6. A. P. Drozdov, P. P. Kong, V. S. Minkov, S. P. Besedin, M. A. Kuzovnikov, S. Mozaffari, L. Balicas, F. Balakirev, D. Graf, V. B. Prakapenka, E. Greenberg, D. A. Knyazev, M. Tkacz, M. I. Eremets, "Superconductivity at 250 K in lanthanum hydride under high pressures," *Nature* 569 (2019): 528.

7. M. Somayazulu, M. Ahart, A.K. Mishra, Z.M. Geballe, M. Baldini, Y. Meng, V.V. Struzhkin, and R.J. Hemley, "Evidence for superconductivity above 260 K in lanthanum superhydride at megabar pressures," *Physical Review Letters* 122 (2019): 027001.

8. M. L. Cohen, "Superconductivity in modified semiconductors and the path to higher transition temperatures." *Superconductor Science and Technology* 28 (2015): 043001.

9. A. Bhaumik, R. Sachan, S. Gupta, J. Narayan, "Discovery of High-Temperature Superconductivity ($T_c$ = 55 K) in B-Doped Q-Carbon," *ACS Nano* 11 (2017): 11915.

10. K. Tanigaki, T.W. Ebbesen, S. Saito, J. Mizuki, J.S. Tsai, Y. Kubo and S. Kuroshima, "Superconductivity at 33 K in $Cs_xRb_yC_{60}$," *Nature* 352 (1991): 222.

11. Y.-L. Hai, M.-J. Jiang, H.-L. Tian, G.-H. Zhong, W.-J. Li, C.-L. Yang, X.-J. Chen, and H.-Q. Lin, "Superconductivity Above 100 K Predicted in Carbon-Cage Network," *Advanced Science* 10 (2023): 2303639.

12. T. Scheike, P. Esquinazi, A. Setzer, W. Böhlmann, "Granular superconductivity at room temperature in bulk highly oriented pyrolytic graphite samples," *Carbon* 59 (2013): 140.

13. P. Esquinazi, "Invited review: Graphite and its hidden superconductivity," *Papers in Physics* 5 (2013): 050007.

14. A. Champi, C.E. Precker and P.D. Esquinazi, "Hints of granular superconductivity in natural graphite verified by trapped flux transport measurements," *New Journal of Physics* 25 (2023): 093029.

15. Y. V. Kopelevich, J. H. S. Torres, R. Ricardo da Silva, M. C. Diamantini, C. A. Trugenbergerand, V. M. Vinokur, "Global room-temperature superconductivity in graphite," *Advanced Quantum Technologies* (2023): 2300230.

16. M. Núñez-Regueiro, T. Devillers, E. Beaugnon, A. de Marles, T. Crozes, S. Pairis, C. Swale, H. Klein, O. Leynaud, A. Hadj-Azzem, F. Gay, D. Dufeu, "Magnetic field sorting of superconducting graphite particles with T > 400 K." *arXiv* (2024): arXiv:2410.18020.

17. T. Han, Z. Lu, Z. Hadjri, L. Shi, Z. Wu, W. Xu, Y. Yao, A.A. Cotten, O.S. Sedeh, H. Weldeyesus, J. Yang, J. Seo, S. Ye, M. Zhou, H. Liu, G. Shi, Z. Hua, K. Watanabe, T.

Taniguchi, P. Xiong, D.M. Zumbühl, L. Fu & L. Ju, "Signatures of chiral superconductivity in rhombohedral graphene," *Nature* 643 (2025): 654.

18. T. Enoki, M. Suzuki, M. Endo, *Graphite Intercalation Compounds and Applications*, Oxford University Press, New York, 2003.

19. N. Emery, C. Hérold, M. d'Astuto, V. Garcia, Ch. Bellin, J. F. Marêché, P. Lagrange, and G. Loupias, "Superconductivity of Bulk $CaC_6$," *Physical Review Letters* 95 (2005): 087003.

20. N. Emery, C. Herold, J.-F. Mareche, C. Bellouard, G. Loupias, P. Lagrange, "Superconductivity in $Li_3Ca_2C_6$ intercalated graphite," *Journal of Solid State Chemistry* 179 (2006): 1289.

21. I.I. Mazin, L. Boeri, O.V. Dolgov, A.A. Golubov, G.B. Bachelet, M. Giantomassi, O.K. Andersen, "Unresolved problems in superconductivity of $CaC_6$." *Physica C: Superconductivity and its Applications* 460–462 (2007): 116.

22. I. I. Mazin, "Intercalant-Driven Superconductivity in $YbC_6$ and $CaC_6$," *Physical Review Letters* 95 (2005): 227001.

23. V. S. Minkov, V. Ksenofontov, M. I. Eremets, "Possible High-Temperature Superconductivity in Graphite Intercalated with Calcium-Ammonia Solutions," *Carbon Trends* (2026), doi: https://doi.org/10.1016/j.cartre.2026.100672.

24. M. Wissler, "Graphite and carbon powders for electrochemical applications." *Journal of Power Sources* 156 (2006): 142.

25. J. S. Kim, L. Boeri, J. R. O'Brien, F. S. Razavi, and R. K. Kremer, "Superconductivity in heavy alkaline-earth intercalated graphites," *Physical Review Letters* 99 (2007): 027001.

26. A. B. Pippard, "Trapped flux in superconductors," *Philosophical Transactions of the Royal Society of London. Series A* 248 (1955): 97.

27. K A. Müller, M. Takashige & J.G. Bednorz, "Flux trapping and superconductive glass state in $La_2CuO_{4-y}$:Ba," *Physical Review Letters* 58 (1987): 1143.

28. V.S. Minkov, V. Ksenofontov, S.L. Budko, E.F. Talantsev & M.I. Eremets, "Magnetic flux trapping in hydrogen-rich high-temperature superconductors," *Nature Physics* 19 (2023): 1293.

29. S.L Bud'ko, M. Xu & P.C. Canfield, "Trapped flux in pure and Mn-substituted $CaKFe_4As_4$ and $MgB_2$ superconducting single crystals," *Superconductor Science and Technology* 36 (2023): 115001.

30. Y. Mizuguchi, T. Murakami, M.R. Kasem & H. Arima, "Large self-heating by trapped-flux reduction in Sn–Pb solders," *Europhysics Letters* 147 (2024): 36003.

31. S. Mozaffari, D. Sun, V.S. Minkov, A.P. Drozdov, D.A. Knyazev, J.B. Betts, M. Einaga, K. Shimizu, M. I. Eremets, L. Balicas & F. F. Balakirev, "Superconducting phase diagram of $H_3S$ under high magnetic fields," *Nature Communications* 10 (2019): 2522.
32. K. Mukai, T. Inoue, "Magnetic susceptibility measurements on Li-intercalated graphite," *Carbon* 123 (2017): 645.
33. T. Shiroka, G. Lamura, R. De Renzi, M. Belli, M. Emery, H. Rida, S. Cahen, J.-F. Marêché, P. Lagrange & C. Hérold, "μSR investigation of the intercalated graphite superconductor $CaC_6$," *New Journal of Physics* 13 (2011): 013038.
34. R. Cubitt, J. S. White, M. Laver, M. R. Eskildsen, C. D. Dewhurst, D. McK. Paul, A. J. Crichton, M. Ellerby, C. Howard, Z. Kurban, and F. Norris, "Small-angle neutron scattering measurements of the vortex lattice in $CaC_6$," *Physical Review B* 75 (2007):140516.
35. R. Prozorov and V. G. Kogan, "Effective demagnetizing factors of diamagnetic samples of various shapes," *Physical Review Applied* 10 (2018): 014030
36. I. Umegaki, S. Kawauchi, H. Sawada, H. Nozaki, Y. Higuchi, K. Miwa, Y. Kondo, M. Månsson, M. Telling, F.C. Coomer, S.P. Cottrell, T. Sasaki, T. Kobayashi and J. Sugiyama, "Li-ion diffusion in Li-intercalated graphite probed by $\mu^+$SR," *Physical Chemistry Chemical Physics* 19 (2017): 19058.
37. F. Arnold, J. Nyéki, and J. Saunders, "Superconducting sweet-spot in microcrystalline graphite revealed by point-contact spectroscopy," *JETP Letters* 107 (2018): 577.
38. C. E. Precker, P. D. Esquinazi, A. Champi, J. Barzola-Quiquia, M. Zoraghi, S. Muiños, N. Gradwohl, "Identification of a possible superconducting transition above room temperature in natural graphite crystals," *New Journal of Physics* 18(11) (2016): 113041.
39. M. Stiller, P. D. Esquinazi, J. Barzola-Quiquia, C. E. Precker, "Local Magnetic Measurements of Trapped Flux Through a Permanent Current Path in Graphite," *Journal of Low Temperature Physics* 191 (2018): 105.
40. S. Layek, M. Monteverde, G. Garbarino, M.-A. Méasson, A. Sulpice, N. Bendiab, P. Rodière, R. Cazali, A. Hadj-Azzem, V. Nassif, D. Bourgault, F. Gay, D. Dufeu, S. Pairis, J.-L. Hodeau, M. Núñez-Regueiro, "Possible high temperature superconducting transitions in disordered graphite obtained from room temperature deintercalated $KC_8$," *Carbon* 201 (2023): 667.
41. R. Rousset-Zenou, S. Layek, M. Monteverde, F. Gay, D. Dufeu, M. Núñez-Regueiro, "Hidden Granular Superconductivity Above 500K in off-the-shelf graphite materials," *arXiv* (2022): arXiv: 2207.09149.

42. D.D.L. Chung, "Exfoliation of graphite," *Journal of Materials Science* 22 (1987): 4190.
43. N.B. Kopnin, M. Ijäs, A. Harju & T.T. Heikkilä, "High-temperature surface superconductivity in topological flat-band systems," *Physical Review B* 83 (2011): 220503.
44. N.B. Kopnin, M. Ijäs, A. Harju & T.T. Heikkilä, "High-temperature surface superconductivity in rhombohedral graphite," *Physical Review B* 87 (2013): 140503.
45. P. Esquinazi, T.T. Heikkilä, Y.V. Lysogorskiy, D.A. Tayurskii, G.E. Volovik, "On the superconductivity of graphite interfaces," *JETP Letters* 100 (2014): 336.
46. T. T. Heikkilä, G. E. Volovik, "Flat bands as a route to high-temperature superconductivity in graphite," in *Basic Physics of Functionalized Graphite*, Springer, 2016, p. 123.
47. G. E. Volovik, "Graphite, graphene and flat-band superconductivity," *JETP Letters* 107 (2018): 516.
48. K. R. Rahnejat, C. A. Howard, N. E. Shuttleworth, S. R. Schofield, K. Iwaya, C. F. Hirjibehedin, C. Renner, G. Aeppli, M. Ellerby, "Charge density waves in the graphene sheets of the superconductor $CaC_6$," *Nature Communications* 2 (2011): 558.
49. E. F. Talantsev, "Einstein and Debye temperatures, electron-phonon coupling constant and a probable mechanism for ambient-pressure room-temperature superconductivity in intercalated graphite," *Annalen der Physik* 538 (2026): e00546.
50. V. Cuartero, J. Blasco, J. García, J. A. Rodríguez-Velamazán, C. Ritter, "Metamagnetic transition in $Tb_2MnCoO_6$," *EPJ Web of Conferences* 40 (2013): 15002.
51. M. I. Eremets, V. S. Minkov, P. P. Kong, A. P. Drozdov, S. Chariton & V. B. Prakapenka, "Universal diamond edge Raman scale to 0.5 terapascal," *Nature Communications* 14 (2023): 907.
52. M. Sepioni, R. R. Nair, I. Tsai, A. K. Geim, I. V. Grigorieva, "Limits on intrinsic ferromagnetism in graphene and graphite," *Europhysics Letters* 97 (2012): 47001.
53. A. Sadezky, H. Muckenhuber, H. Grothe, R. Niessner, U. Pöschl, "Raman microspectroscopy of soot and related carbonaceous materials," *Carbon* 43 (2025): 1731.
54. T. Latychevskaia, S.-K. Son, Y. Yang, D. Chancellor, M. Brown, S. Ozdemir, I. Madan, G. Berruto, F. Carbone, A. Mishchenko, K. Novoselov, "Stacking transition in rhombohedral graphite," *Frontiers of Physics* 14 (2019): 13608.
55. C. W. Bale, A. D. Pelton, "The Ca–Li system," *Journal of Phase Equilibria* 8 (1987): 125.

56. C. W. Bale, A. D. Pelton, “The Li–Sr system,” *Bulletin of Alloy Phase Diagrams* 10 (1989): 278.
57. A.A. Nayeb-Hashemi, J.B. Clark, A.D. Pelton, “The Li–Mg system,” *Bulletin of Alloy Phase Diagrams* 5 (1984): 365.
58. H. Okamoto, “Li–Zn (Lithium–Zinc),” *Journal of Phase Equilibria and Diffusion* 33 (2012): 345.
59. M. S. Dresselhaus, G. Dresselhaus, "Intercalation compounds of graphite," *Advances in Physics* 51 (2002): 1.
60. N. Galy, S. Cahen, P. Lagrange, C. Hérold, "Crystal structure of first stage strontium-graphite intercalation compound," *Carbon* 167 (2020): 864.
61. D. Guerard, A. Hérold, "Intercalation of heavy alkaline earth metals in graphite", *Carbon* 13 (1975): 337.
62. D. Billaud, E. McRae, A. Hérold, "Synthesis and structural properties of the first-stage lithium-graphite compound $LiC_6$," *Materials Research Bulletin* 12 (1977): 1101.

# Supporting Information

## Signs of Possible High-Temperature Superconductivity in Graphite Intercalated with Lithium-Based Alloys

*Vadim Ksenofontov*, Vasily S. Minkov*, Alexander P. Drozdov, Ulrich Pöschl, Mikhail I. Eremets*

### S.1 Measurement methods

#### S.1.1 Magnetic susceptibility measurements

Direct current (D.C.) magnetic susceptibility measurements were performed in a S700X SQUID magnetometer (Cryogenic Limited) and PPMS DynaCool 14 Tesla Cryogen-free base system (Quantum Design). Several small pieces of sample (length and width ≈2-3 mm, thickness ≈0.1-2 mm) with a total mass in the range of 5-30 mg were covered by anhydrous kerosene or silicone oil and put into a gelatine capsule (inner diameter ≈4 mm). Samples were predominantly oriented with graphene layers perpendicular to the applied magnetic field. For some measurements, samples were oriented with graphene layers parallel to the magnetic field.

The trapped magnetic flux was generated under zero field cooling (ZFC) and field colling (FC) conditions. The typical ZFC protocol included cooling of the sample at zero field from $T > T_c$ to the target temperature below $T_c$. At the target temperature the magnetic field was applied for approximately 1 hour, then the applied magnetic field was gradually decreased to 0 T. In FC mode the sample was cooled from its normal state at the applied magnetic field to the target temperature below $T_c$, waiting for approximately 1 hour, and then the applied magnetic field was gradually decreased to 0 T. After removal of the applied magnetic field in both ZFC and FC modes the SQUID magnetometer was allowed to stand at 0 T for approximately 4 hours to let the system relax and minimize noise associated with jumps of the magnetic flux in the superconducting magnet of the SQUID magnetometer. The fields applied for magnetization ranged from 16 mT to 2 T. Successive $m_{trap}(T)$ measurements were performed upon warming of the sample with temperature steps of 1-3 K. Before each cycle of magnetization, the sample was converted to its normal metal state by warming the sample to $T > T_c$ at zero field, and the superconducting magnet of the SQUID magnetometer was degaussed to eliminate remnant fields.

### S.1.2 Electrical resistance measurements

Four-probe resistance measurements were performed on bulk samples of rectangular shape. The samples were prepared and electrically connected using a special holder with gold-plated contact pins in a glove box under Argon atmosphere ($O_2$/$H_2O$ < 0.1 ppm). Excitation current from an AC resistance bridge (Model 372, Lake Shore) with amplitudes in the range of 33 μA – 1 mA at a frequency of 13 Hz were applied to ensure minimal heating of the sample with low heating power (approx. 0.1 nW). Measurements in zero field and magnetic fields were performed in the S700X SQUID magnetometer using the same AC resistance bridge.

### S.1.3 Scanning electron microcopy and energy dispersive X-ray analysis.

Scanning electron microcopy of pristine and intercalated graphite samples was done using a Ferra III Focus ion beam device (Tescan). The energy dispersive X-ray analysis was performed using the Oxford Instruments device integrated in Ferra III FIB.

### S.1.4 X-ray powder diffraction.

X-ray diffraction data were collected at room temperature on a STOE Stadi P powder diffractometer (STOE&Cie GmbH Darmstadt), using MoK$\alpha_1$ radiation, Ge-(111) monochromator and a Dectris MYTHEN 1 K detector.

### S.1.5 Raman spectroscopic characterization.

Raman spectroscopic measurements were performed at room temperature on bulk samples of highly oriented pyrolytic graphite (HOPG) and expanded graphite (EGS) using a self-built Raman spectrometer equipped with a Newport N-LHP-928 HeNe laser with a wavelength of 632.817 nm the HORIBA / JOBIN YVON HR 460 spectrometer.[51]

### S.1.6 Nanoscale Secondary Ion Mass Spectrometry (NanoSIMS)

Nanoscale Secondary Ion Mass Spectrometry (NanoSIMS) characterization of the compositions of highly oriented pyrolytic graphite (HOPG) and expanded graphite (EGS) was performed on bulk samples using a CAMECA NanoSIMS 50 ion probe. To characterize transition elements (Fe, Co, Ni) within a graphite matrix, an oxygen primary ion beam ($O^-$; d ~500 nm, 39–50 pA) was rastered over 15×15 µm²-sized regions of interest.

**S.1.7 Measurement uncertainties/error estimates.**

The S700X SQUID magnetometer allows to measure magnetic signals (total moment) with experimental errors of the order of $10^{-11}$ A m². In external magnetic fields up to 300 Oe ("low field" option), the experimental noise of our installation is ≈ $1\cdot10^{-10}$ A m². At higher magnetic fields ("field" option), especially in field range between 500 Oe – 1000 Oe the noise level was ≈ $2\cdot10^{-9}$ A m².

The experimental standard deviation of our four-probe resistivity measurements (Sample IIId) using the AC resistance bridge with an excitation current of 1 mA was ≈ 0.2 $10^{-9}$ Ω m (determined at $T < 20$ K).

**S.2 Characterization of starting graphite materials**

In the present study, we used two types of graphite for intercalation. Most experiments were performed with expanded graphite sheets (purity ≥ 99,85 %, SGL CARBON GmbH) with high specific volume[24] and with the following chemical composition specified by the supplier:

**Table S1.** Chemical composition of HOPG and EGS samples

| Element | Content |
|---|---|
| Carbon | ≥ 99,85% |
| Ash content | ≤ 0,15% |
| Sulphur | < 300 ppm |
| Chlorides | ≤ 10 ppm |
| Fluoride | ≤ 10 ppm |
| Halogens | ≤ 40 ppm |

Additional experiments were performed with HOPG (purity 99,99%, Magnametals). Energy dispersive X-ray analysis confirmed the high purity of both graphite starting materials (Figures S1 and S2).

Prior to intercalation, the graphite samples were annealed at 723 K for 2 hours in an Ar flow with $H_2$ (5%) to remove adsorbed gases and water. Untreated graphite samples were also tested but did not yield well reproducible results.

For intercalation, the following reactants were used: Ca (99.5%, redistilled, Alfa Aesar), Li (99.9%, Alfa Aesar), Sr (99.9%, Alfa Aesar), K (99.9%, Alfa Aesar), Mg (99.8%, Alfa

Aesar), Zn (99.999%, Alfa Aesar), anhydrous silicone oil and kerosene both dried with metallic sodium in a Ar glove box under Argon ($O_2$/$H_2O$ < 0.1 ppm ).

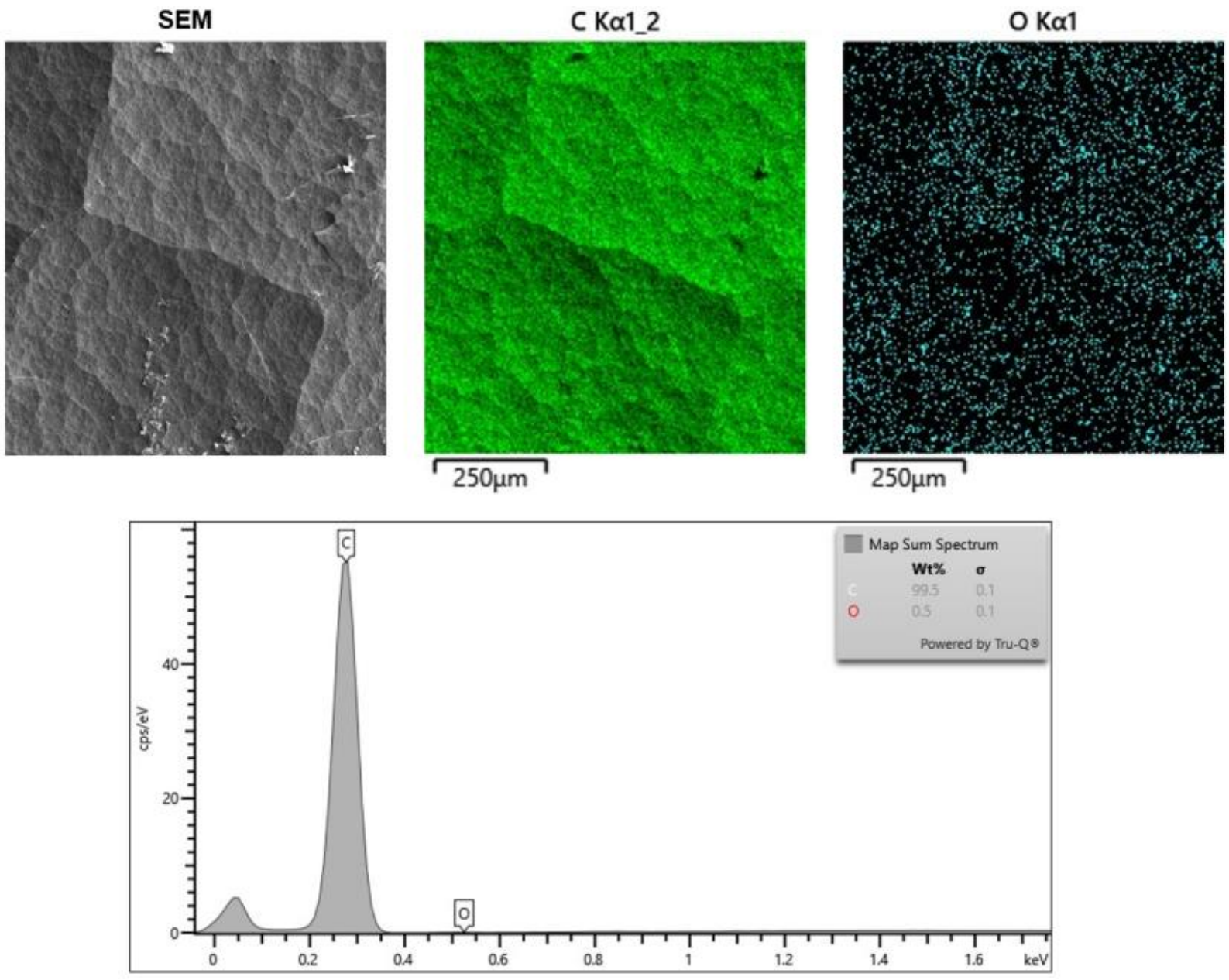

**Figure S1.** Energy dispersive X-ray analysis of HOPG. Top: scanning electron microscope image of the analyzed surface and maps showing the distribution of the determined elements. Bottom: EDX spectrum. The surface was cut parallel to graphene layers at a half of the thickness of the used material. The sample underwent the standard treatment prior the analysis: annealing at 723 K in Ar–$H_2$ flow (5% $H_2$) for 2 hours. The analysis confirmed high purity of the pristine graphite material.

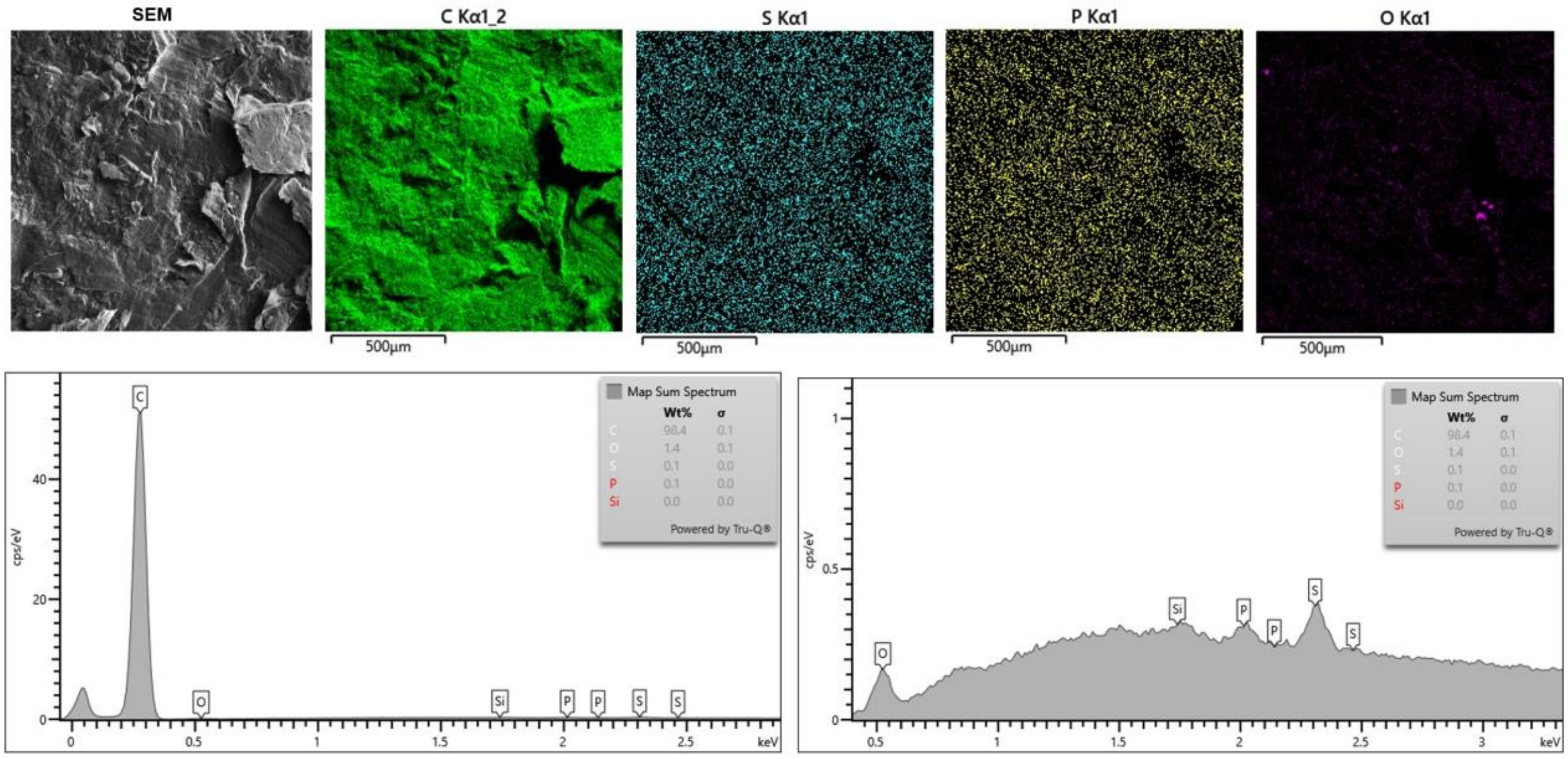

**Figure S2.** Energy dispersive X-ray analysis of EGS. Top: scanning electron microscope image of the analyzed surface and maps showing the distribution of the determined elements. Bottom: EDX spectrum with the zoomed area of interest, showing the peaks from impurity elements. The surface was cut parallel to graphene layers at a half of the thickness of the used material. The sample underwent the standard treatment prior the analysis: annealing at 723 K in Ar–$H_2$ flow (5% of $H_2$) for 2 hours. The analysis revealed impurities of phosphorus and sulfur (approx. 0.04 atomic percent each), likely from the technological process of compacting the expanded graphite.

In fact, traces of ferromagnetic elements in graphite, especially iron, can lie below the EDX detection limit. As shown by M. Sepioni *et al.*[52], EDX focusing is necessary to detect and measure potential ferromagnetic inclusions in HOPG. A clear indication of ferromagnetic inclusions also follows from magnetization measurements *M(H)* after subtraction of the diamagnetic background due to the graphite matrix. Following this method, we have proceeded magnetization measurements of the EGS sample in the temperature range between 2 K and 350 K. Figure S3a shows the temperature dependence of magnetization for the EGS sample in ZFC and FC modes. It demonstrates diamagnetic behavior and absence of both low-$T_c$ or high-$T_c$ anomalies observed in intercalation compounds synthesized with lithium-based alloys.

In studied temperature range EGS sample demonstrates a hysteretic behavior which becomes evident after subtraction of prevailing diamagnetic background. Measured magnetic moment normalized on the sample mass *M/m* vs applied field *H* measured at *T* = 300 K is shown in Figure S3b and Figure S3c, before and after subtraction of the linear diamagnetic background, respectively. According to M. Sepioni *et al.*, observed in our measurements saturation magnetization ≈ (1-2)·$10^{-3}$ emu/g stems from micron-size magnetic inclusions (presumably magnetite) in graphite and corresponds to the total number of Fe atoms in EGS between 1 and 6 ppm.

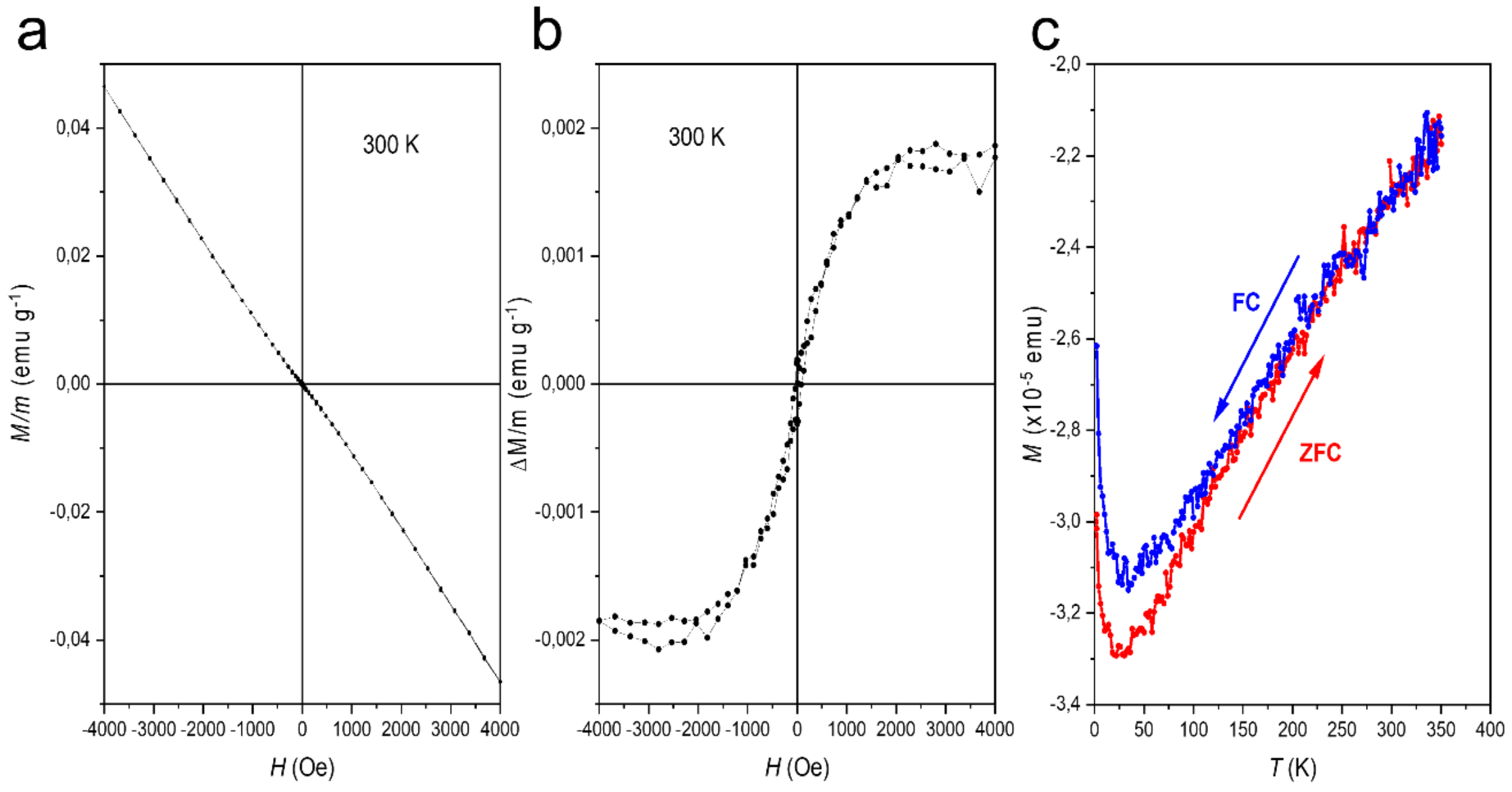

**Figure S3.** Specific magnetic moment *M/m* of the pristine EGS as a function of the applied field *H*, measured at T = 300 K: a) raw data and b) *ΔM/m* after subtracting the linear diamagnetic background (m is the sample mass). c) Temperature dependence of magnetization measured in an external field of 300 Oe in ZFC and FC modes. In all experiments, the magnetic field is perpendicular to the graphene layers.

Analysis of the composition of HOPG and EGS by NanoSIMS revealed the presence of iron and nickel impurities distributed extremely non-uniformly within the samples – predominantly in the form of "spots" ranging from 0.6 to 0.8 μm in size. No cobalt atoms were detected. An estimation of the iron and nickel contents relative to carbon is presented in Table S2. To convert the measured ion ratios into element ratios, relative sensitivity factors (RSF) were determined from measurements of a graphite standard with known Fe-, Co-, and Ni-concentrations. Given element ratios are geometric mean values of N=14 (HOPG) and N=16 (EGS) individual measurements, respectively.

Measurement errors are dominated by current uncertainties of the RSF and are about a factor of 2 for the given mean values.

**Table S2.** Estimation of iron and nickel content relative carbon in HOPG and EGS samples according to NanoSIMS characterization.

| Sample | Fe/C, (at%/at%) | Ni/C, (at%/at%) |
|---|---|---|
| HOPG (N=14) | $\sim 3\times10^{-5}$ | $\sim 9\times10^{-7}$ |
| EGS (N=16) | $\sim 3\times10^{-4}$ | $\sim 7\times10^{-6}$ |

The crystalline structure of HOPG is hexagonal, whereas EGS shows coexistence of both hexagonal and rhombohedral phases in relative amounts of ~ 60% and ~ 40%, respectively, as determined by XRD measurements (Figure S4).

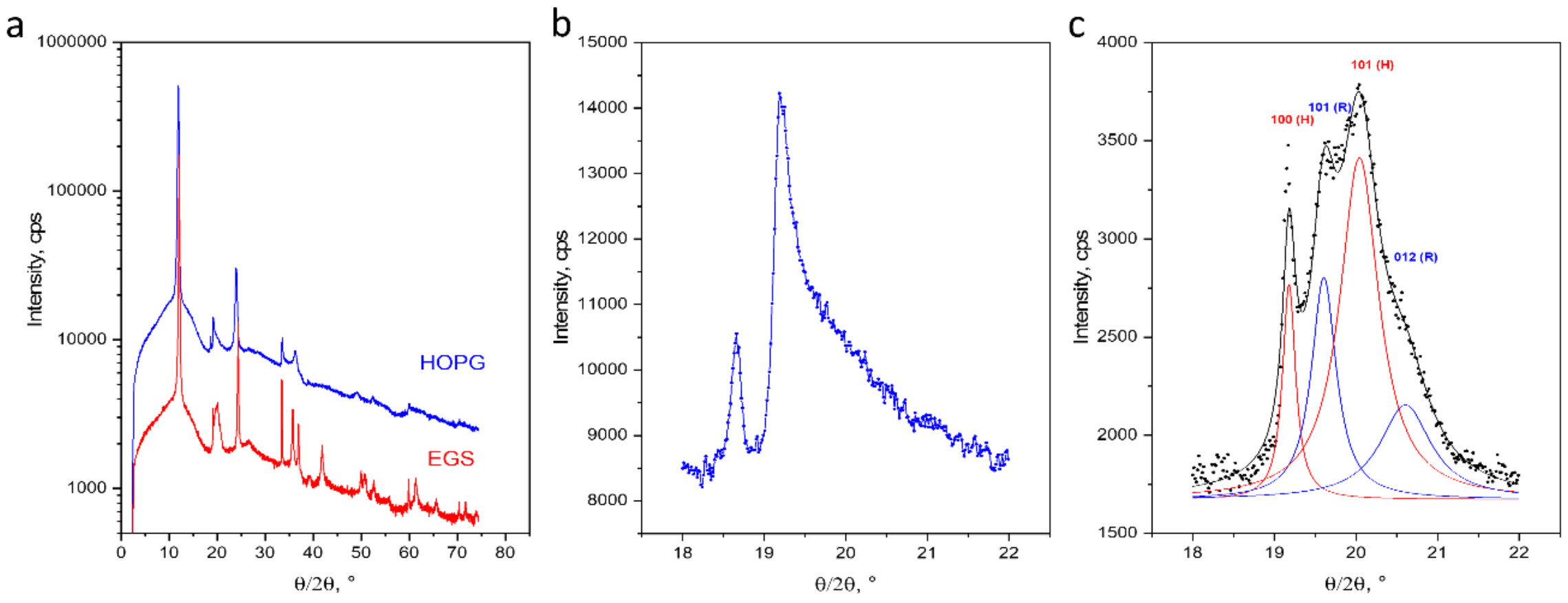

**Figure S4. X-ray powder diffraction (XRD) analysis. a)** HOPG and EGS. **b**) HOPG diffraction patterns indicating hexagonal phase. **c**) EGS diffraction patterns indicating hexagonal (H) and rhombohedral phases (R) in amounts of ~ 60% and ~ 40%, respectively.

Further structural characterization of pristine graphite samples was performed by Raman spectroscopy.

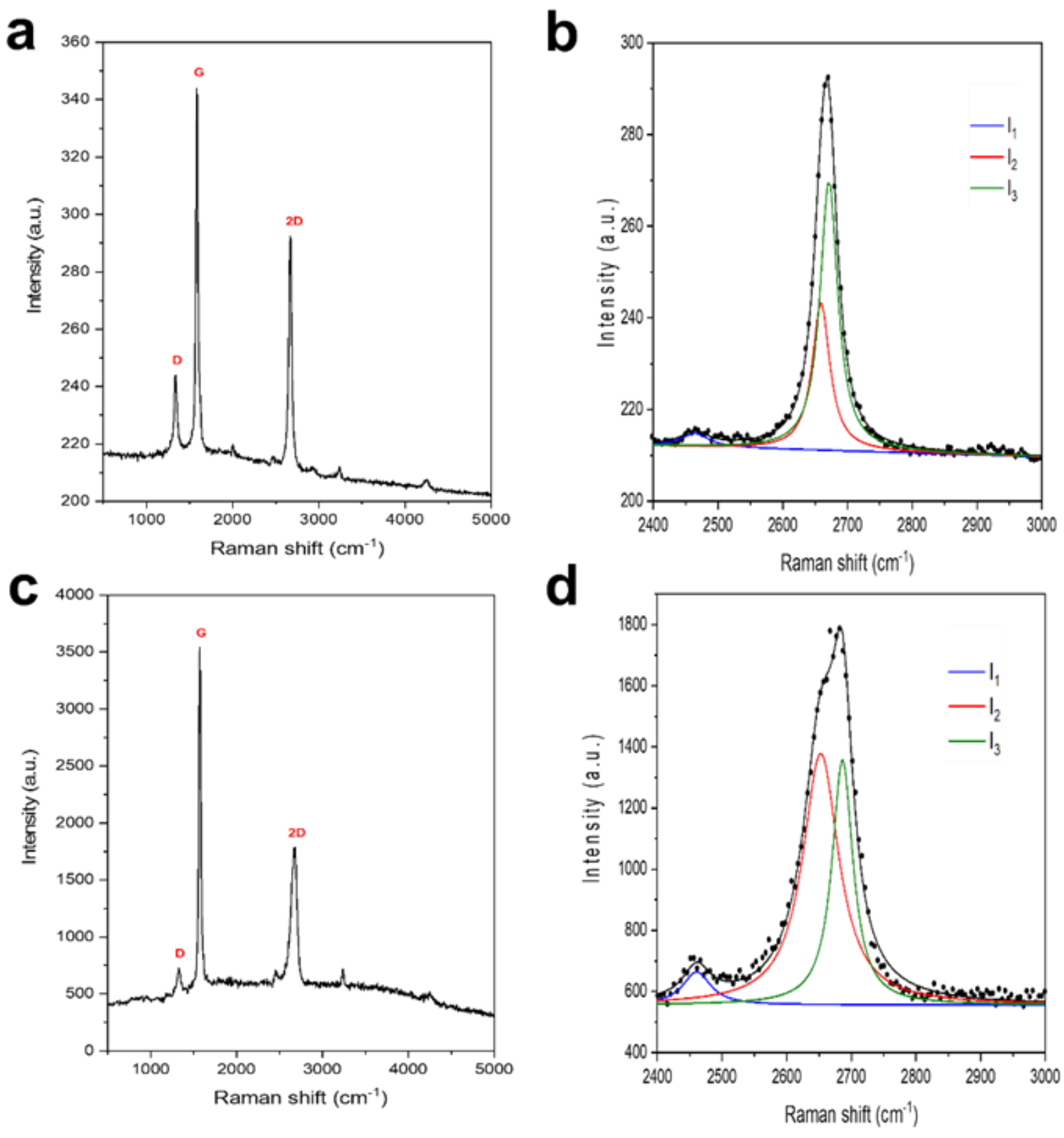

**Figure S5.** Raman spectra of HOPG (a,b) and EGS (c,d). b),d) 2D peaks of HOPG and EGS with Lorentzian line fit ($I_1$, $I_2$, $I_3$).

Figures S5 shows the Raman spectra of HOPG and EGS with D and G bands at ~1332.0 cm$^{-1}$ and ~1585.0 cm$^{-1}$ (HOPG) and ~1327.3 cm$^{-1}$ and ~1575.1 cm$^{-1}$ (EGS). The 2D band located at ~2670.0 cm$^{-1}$ for HOPG and ~2679.0 cm$^{-1}$ for EGS is the second-order overtone of the D band, alternatively also designated as D1 band.[53] According to Latychevskaia et al.[54], the linewidth of the 2D peak and its form are fingerprints of hexagonal and rhombohedral phases of graphite, and the stacking configuration can be determined from the ratio of Lorentzian subpeaks as shown in Figures S5b and S5d. The intensity ratio $R = I_3/I_2$ can be used to characterize the different stacking configurations in hexagonal and rhombohedral phases of graphite with $R > 1.7$ for ABA-stacking (hexagonal) and $R < 1.3$ for ABC stacking (rhombohedral) [54]. We obtained R = 1.8 for HOPG and R = 0.55 for EGS, confirming the hexagonal structure of HOPG and the simultaneous presence of hexagonal and rhombohedral phases in EGS, which is consistent with the XRD results.

Four-probe resistance measurements of the EGS show a temperature dependence similar to semiconductors (Figure S6).

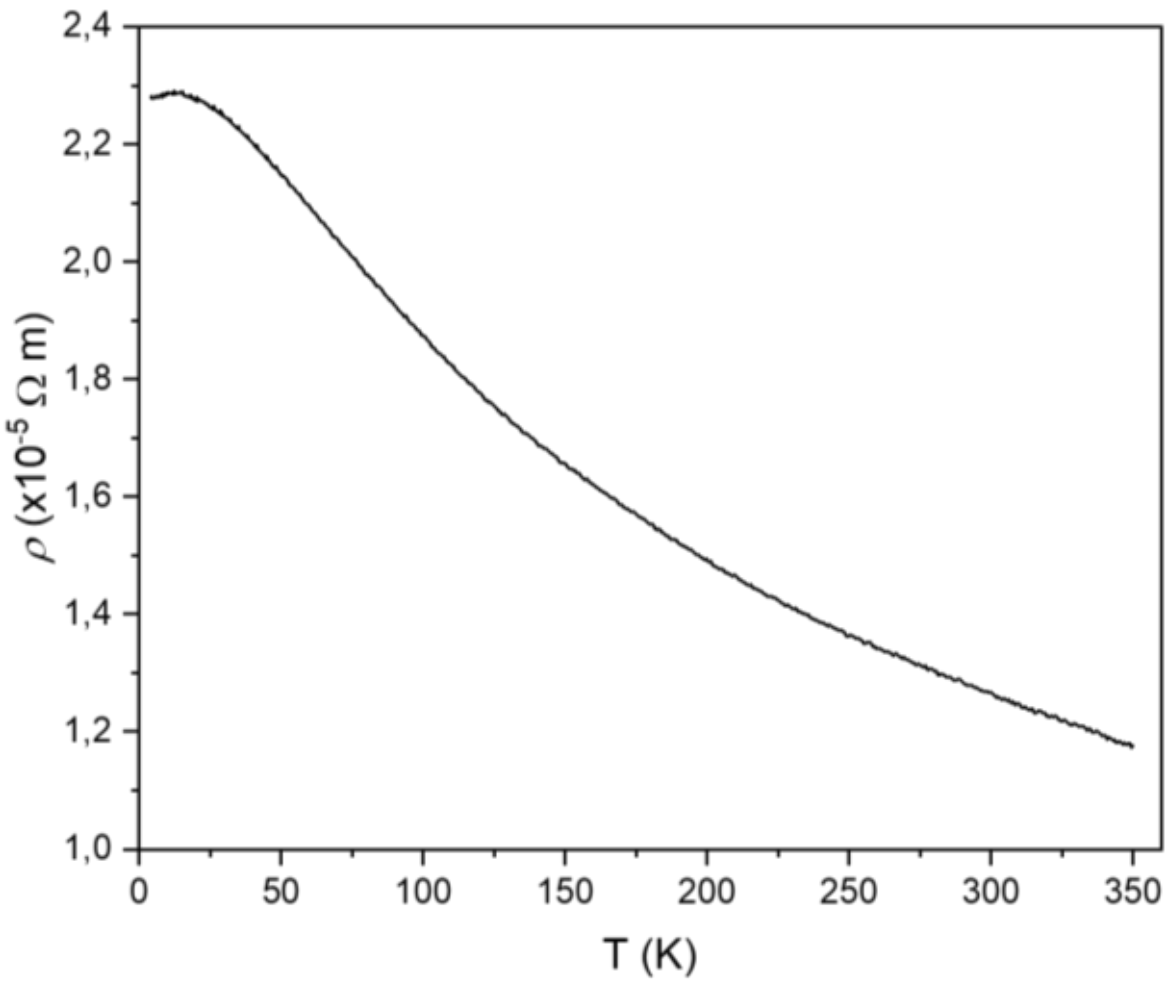

**Figure S6.** Temperature dependence of EGS resistivity.

### S.3 Synthesis and characterization of intercalated samples

Calcium, strontium, magnesium and zinc readily form alloys with lithium, which remain liquid at temperatures well below the melting points of the pure alkaline-earth metals[55-58] and can be used to intercalate graphite under well-controlled conditions. Intercalations were carried out in a stainless-steel crucible heated by a custom-designed miniature furnace under an inert argon atmosphere inside a glove box. Freshly cut pieces of the doping metal and lithium were weighed and placed in the crucible, then melted at temperatures ranging from 423 K to 723 K upon stirring to ensure a homogeneous alloy. The graphite samples were then immersed in the liquid alloy. The intercalation time was varied from 5 seconds to several days, after which the samples were removed from the melt and allowed to cool to room temperature. Prior to removal from the glove box, the prepared samples were covered by dried silicon oil or paraffin. At room temperature, the intercalated samples were stable as long protected from oxidants.

Importantly, the intercalation proceeded differently with different types of graphite. While the intercalation of HOPG samples required hours to days, EGS can be efficiently intercalated in seconds to minutes, which may be due to the high surface-to-volume ratio of EGS[24], which favors rapid diffusion and intercalation. Thus, we mainly used EGS in this study unless mentioned otherwise.

Figure S7 shows photographs of graphite samples intercalated with Ca–Li alloys. Electron micrographs and energy dispersive X-ray analyses of graphite intercalated with binary Li-Ca, Li-Sr and ternary Li-Ca-Sr allows are presented in Figures S8 – S11.

We also explored intercalation of graphite with other binary and ternary alloys, including Sr–Li, K–Li, Mg–Li, Zn–Li, and Mg–Zn–Li alloys. The preparation of these alloys and the subsequent intercalation of graphite was done the same way as for Sr-Ca-Li alloys. For each sample, the intercalation temperature and time are specified in the figure caption. The energy dispersive X-ray analysis of the EGS intercalated by binary Mg–Li alloy is presented in Figure S12.

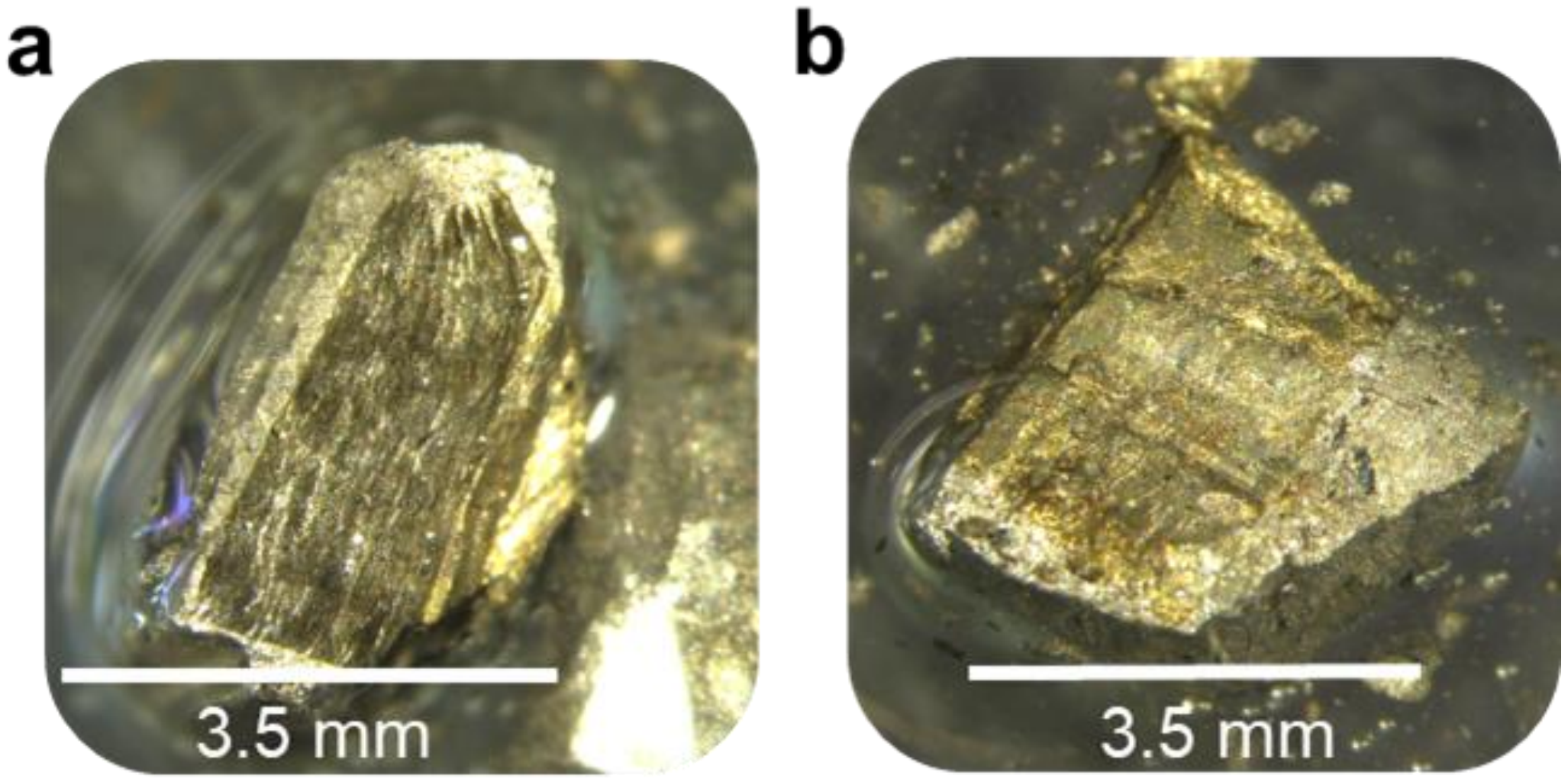

**Figure S7.** Photos of graphite sample intercalated with Ca–Li (1:3.5) alloy at 623 K for four days. Intercalated EGS sample cut (a) perpendicular and (b) parallel to the graphene layers.

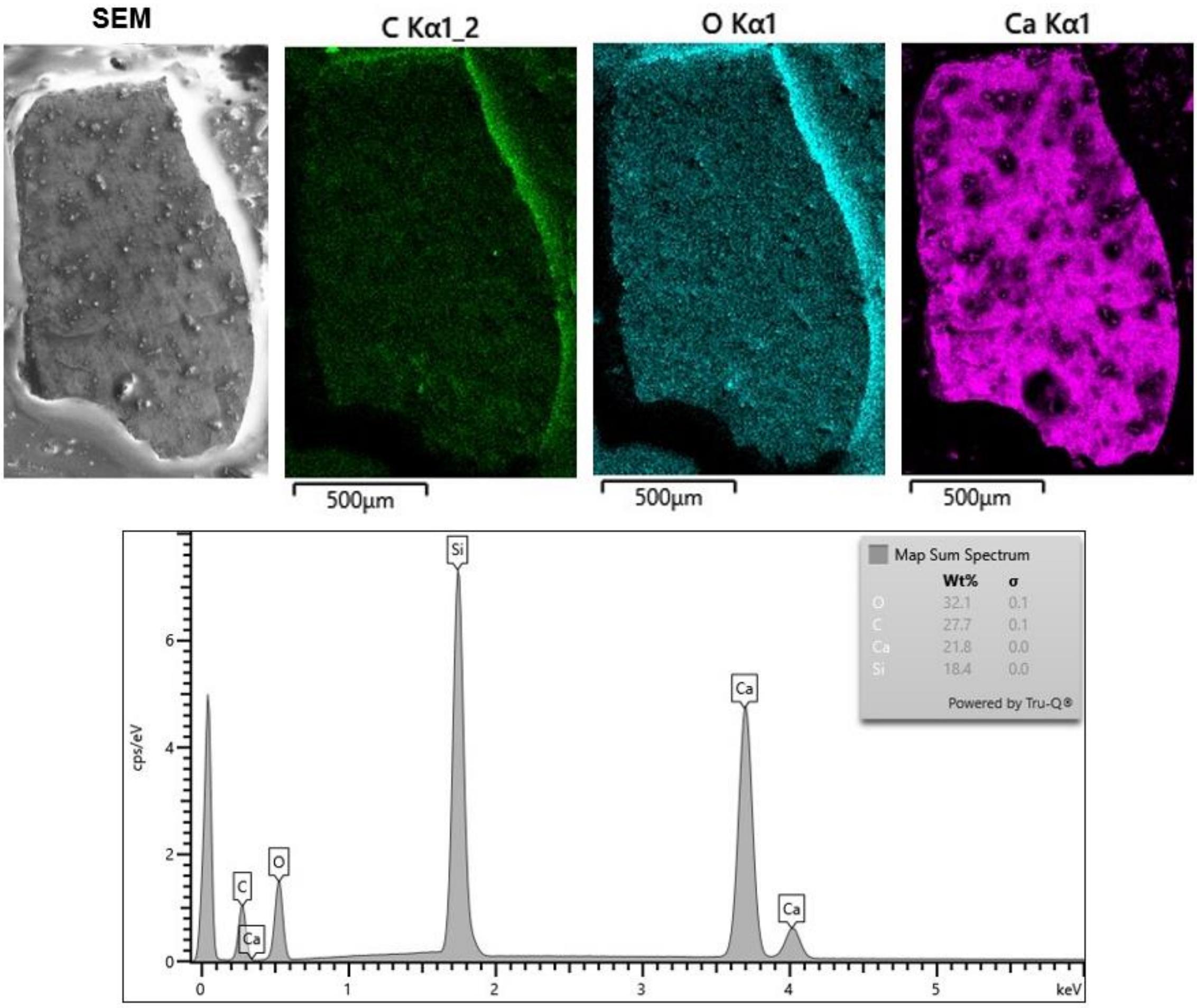

**Figure S8.** Energy dispersive X-ray analysis of the graphite intercalated with binary Ca–Li alloy. Top: scanning electron microscope image of the analyzed surface and colored maps showing the distribution of the determined elements. Bottom: EDX spectrum. The sample was synthesized by intercalation of graphite (HOPG) with Ca–Li (1:3.5) alloy at 623 K for 91 hours. The surface was cut parallel to graphene layers at a half of the thickness of the synthesized sample and covered by dry silicon oil. The analysis revealed the presence of doping Ca atoms intercalated into graphite. Ca content amounts approximately 9.9 atomic percent. The C:Ca:O ratio suggests the composition of $CaC_{4.2}$ and the derived Li:Ca atomic ratio of 3.0:1, assuming that Li and Ca were completely oxidized to $Li_2O$ and CaO and the Si:O ratio in the used silicon oil is 1:1.

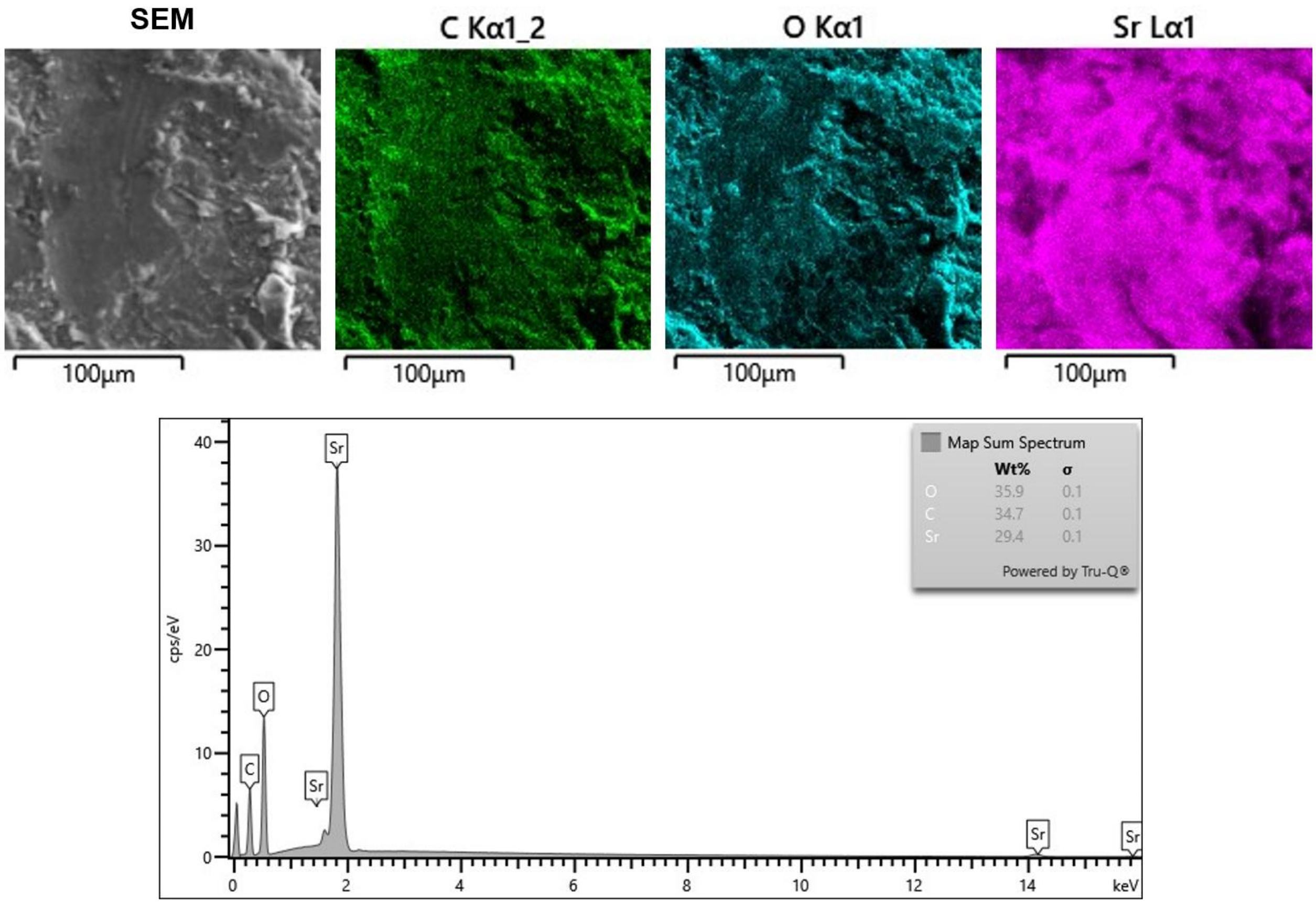

**Figure S9.** Energy dispersive X-ray analysis of the graphite intercalated with binary Sr–Li alloy. Top: scanning electron microscope image of the analyzed surface and colored maps showing the distribution of the determined elements. Bottom: EDX spectrum. The sample was synthesized by intercalation of EGS with Sr–Li (1:3) alloy at 623 K for 1 minute. The surface was cut parallel to graphene layers at a half of the thickness of the synthesized sample. The analysis revealed the presence of doping Sr atoms intercalated into graphite. Sr content amounts approximately 6.1 atomic percent. The C:Sr:O ratio suggests the composition of $SrC_{8.6}$ and the derived Li:Sr atomic ratio of 11.4:1, assuming that Li and Sr were completely oxidized to $Li_2O$ and SrO.

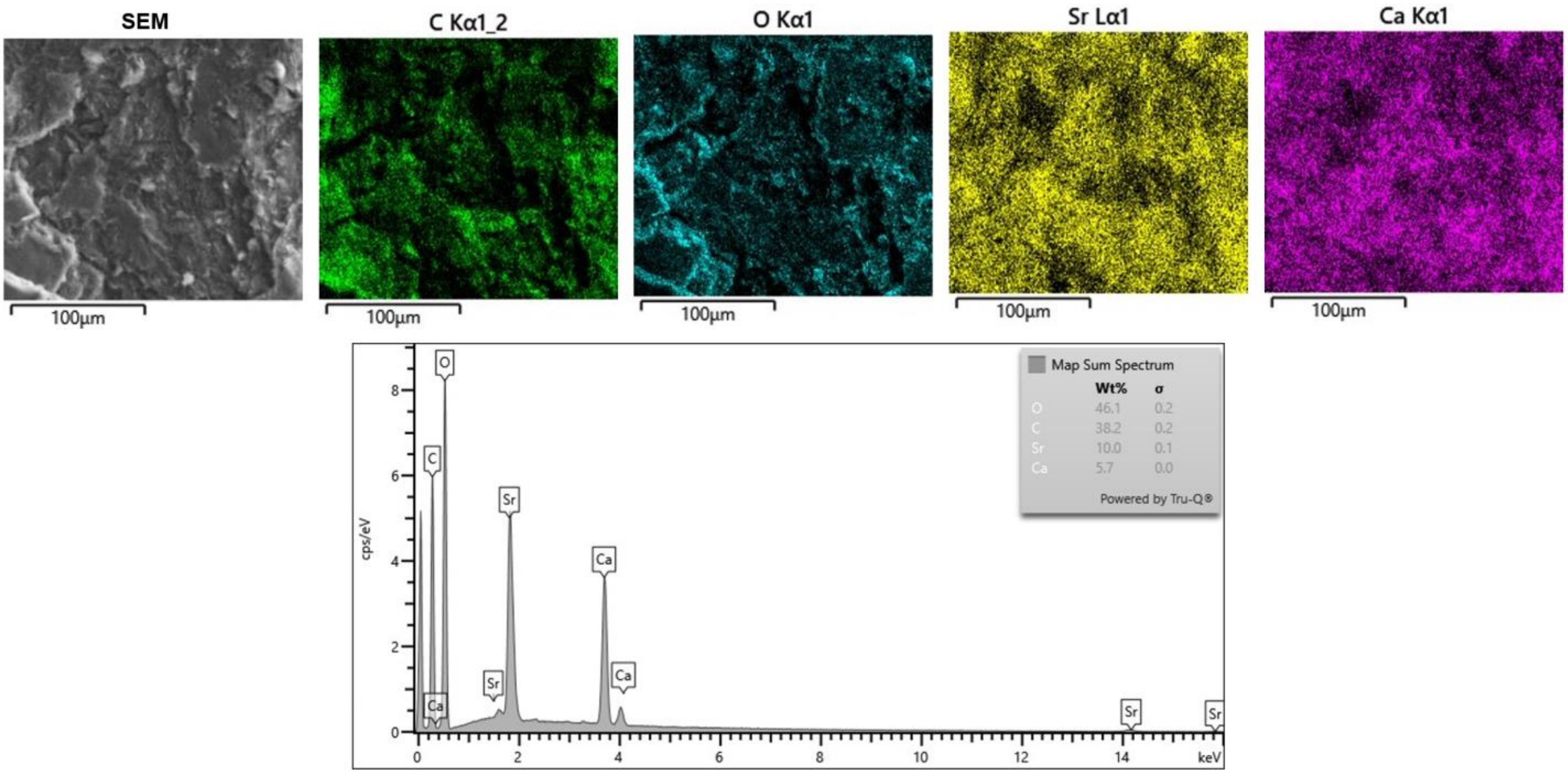

**Figure S10.** Energy dispersive X-ray analysis of the graphite intercalated with ternary Sr–Ca–Li alloy. Top: scanning electron microscope image of the analyzed surface and colored maps showing the distribution of the determined elements. Bottom: EDX spectrum. The sample was synthesized by intercalation of EGS with Sr–Ca–Li (1:1:20) alloy at 523 K for 1 minute. The surface was cut parallel to graphene layers at a half of the thickness of the synthesized sample. The analysis revealed the presence of both doping Ca and Sr atoms intercalated into graphite. Ca and Sr content amounts approximately 2.7 and 2.1 atomic percent, respectively. The C:Ca:Sr:O ratio suggests the composition of $Ca_1Sr_{0.8}C_{22.4}$ and the derived Sr:Ca:Li atomic ratio of 0.8:1: 22.8, assuming that Li, Ca, and Sr were completely oxidized to $Li_2O$, CaO, and SrO.

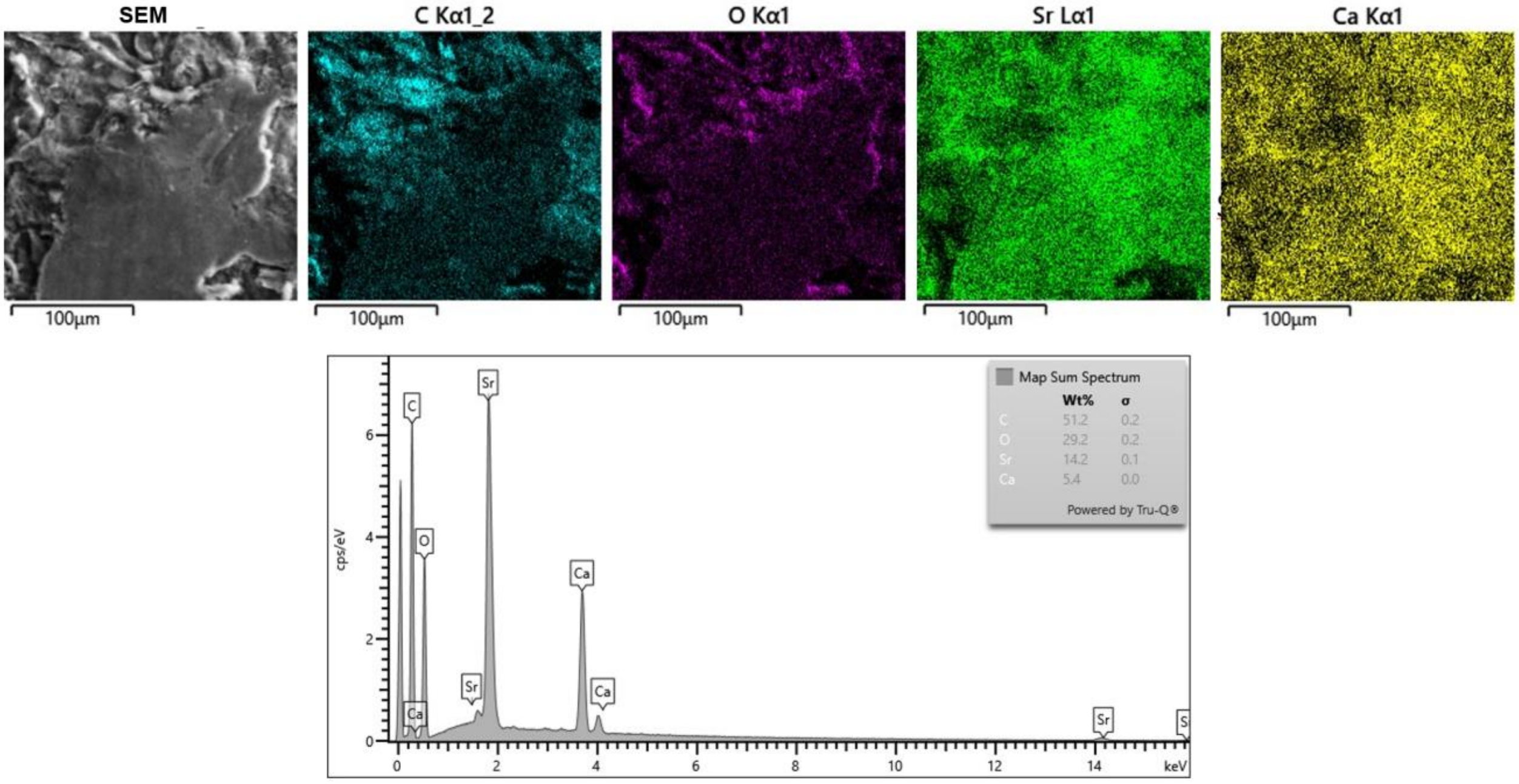

**Figure S11.** Energy dispersive X-ray analysis of the graphite intercalated with ternary Sr–Ca–Li alloy. Top: scanning electron microscope image of the analyzed surface and colored maps showing the distribution of the determined elements. Bottom: EDX spectrum. The sample was synthesized by intercalation of EGS with Sr-Ca–Li (1:1:20) alloy at 423 K for 1 minute. The surface was cut parallel to graphene layers at a half of the thickness of the synthesized sample. The analysis revealed the presence of both doping Ca and Sr atoms intercalated into graphite. Ca and Sr content amounts approximately 2.1 and 2.5 atomic percent, respectively. The C:Ca:Sr:O ratio suggests the composition of $Ca_1Sr_{1.2}C_{31.6}$ and the derived Sr:Ca:Li atomic ratio of 1.2:1: 22.6, assuming that Li, Ca, and Sr were completely oxidized to $Li_2O$, CaO, and SrO.

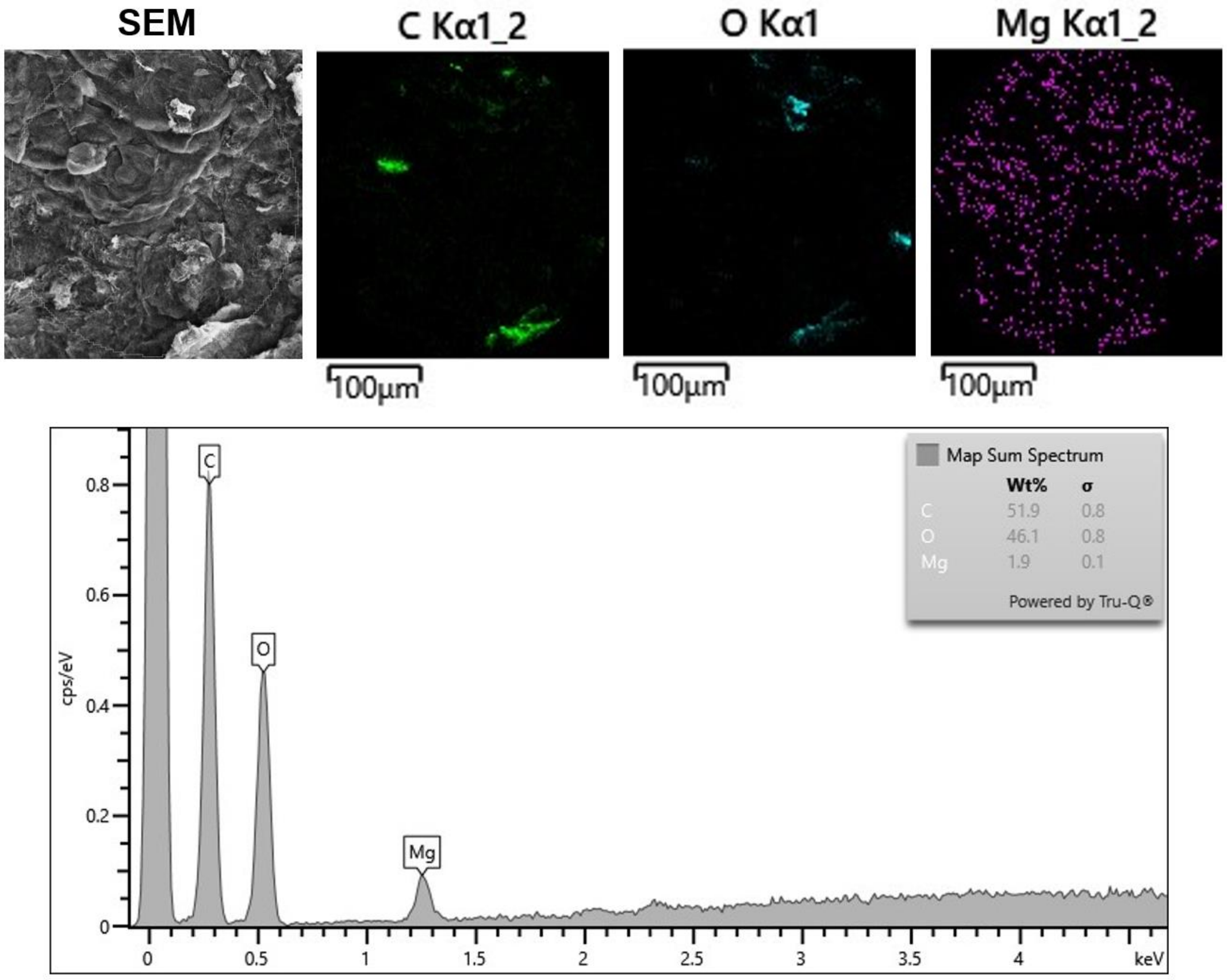

**Figure S12.** Energy dispersive X-ray analysis of the graphite intercalated by binary Mg–Li alloy. Top: scanning electron microscope image of the analyzed surface and colored maps showing the distribution of the determined elements. Bottom: EDX spectrum. The sample was synthesized by intercalation of EGS with Mg-Li (1:4) alloy at 623 K for 5 minutes. The surface was cut parallel to graphene layers at a half of the thickness of the synthesized sample. The analysis revealed the presence of doping Mg atoms intercalated into graphite. Mg content amounts approximately 1.1 atomic percent. The C:Mg:O ratio suggests the composition of $Mg_1C_{55.5}$ and the derived Mg:Li atomic ratio of 1:72, assuming that Li and Mg were completely oxidized to $Li_2O$ and MgO.

**Table S3.** Metals/alloys (molar ratios), reaction temperatures, reaction times and masses of the EGS intercalation samples used in magnetic measurements, discussed in the main text.

| Sample name | Metal/Alloy (molar ratio) | Intercalation Temperature (K) | Intercalation Time (min) | Sample masses (mg). Accuracy ± 1 mg. |
|---|---|---|---|---|
| Sample I | Li | 623 | 1 | 5 |
| Sample II | Ca-Li (1:3.5) | 623 | 1 | 16 |
| Sample IIIa | Sr-Ca-Li (1:1:20) | 523 | 1 | 6 |
| Sample IIIb | Sr-Ca-Li (1:2:20) | 623 | 1 | 7 |
| Sample IIIc | Sr-Ca-Li (1:2:20) | 473 | 10 | 8 |
| Sample IIId | Sr-Ca-Li (1:2:20) | 423 | 10 | 6 |

## S.4 Results of magnetic measurements

### S.4.1 Samples intercalated with binary Ca–Li alloys under different conditions

Figure S13(a,b) shows magnetization *m(T)* measurements of the sample synthesized at 623 K within 1 minute using EGS and a melt of binary Ca-Li alloy (atomic ratio 1:3.5, Sample II). It demonstrates well-established superconducting transition below 11.5 K (Fig. S13a) which could be attributed to $CaC_6$[19] or $Ca_2Li_3C_6$[20] low-temperature superconducting phases (hereinafter low-$T_c$). However, inspection of high temperature range indicates a small but pronounced step in $m(T)$ zero-field cooling (ZFC) run at ca. 240 K (hereinafter high-$T_c$ transition) (Fig. S13b), which is not present in the field-cooling mode (FC). Note that Fig. S13(a,b) is also shown and discussed in the main text (Fig. 1a,b).

Varying the synthesis conditions, we found that graphite samples intercalated by Ca–Li alloys at temperatures of 623 K and 723 K also exhibited low-$T_c$ and high-$T_c$ transitions. Decreasing the Ca content in Ca:Li alloy from 1:3.5 to a 1:20 resulted in a reduction of signal intensity for both transitions, while the high-$T_c$ transition remained at about 240 K (Fig. S13 (c,d). Lowering the intercalation temperature to 523 K reduced the magnitude of the low-$T_c$ transition and shifted the high-$T_c$ transition to 265 K (Fig. S13 e,f).

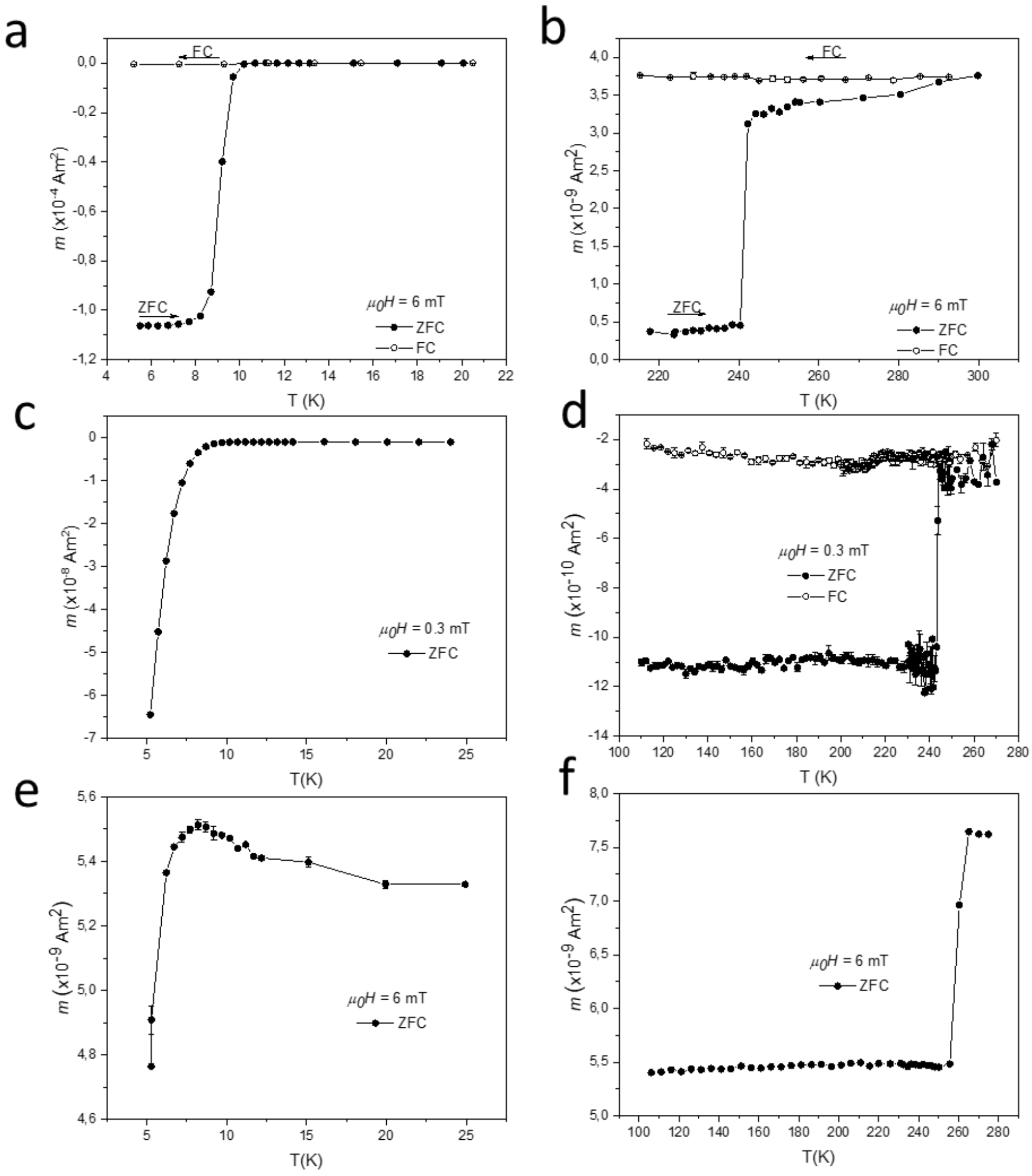

**Figure S13.** Temperature-dependent magnetization, *m(T)*, of graphite samples intercalated with binary Ca-Li alloys under different conditions. a,b) sample synthesized with Ca-Li (1:3.5) alloy at 623 K (1 min); a) low-$T_c$ transition near 11.5 K attributable to known superconductivity in $CaC_6$[19] or $Ca_2Li_3C_6$[20], b) high-$T_c$ transition near 240 K. c,d) sample synthesized with Ca-Li (1:20) alloy at 623 K (1 min). e,f) sample synthesized with Ca-Li (1:20) alloy at 523 K (1 min); magnitude of low-$T_c$ transition reduced and high- $T_c$ transition shifted to 265 K. Measurement data points from zero-field cooling (ZFC) and field-cooling (FC) modes (filled/open circles); connecting lines to guide the eye.

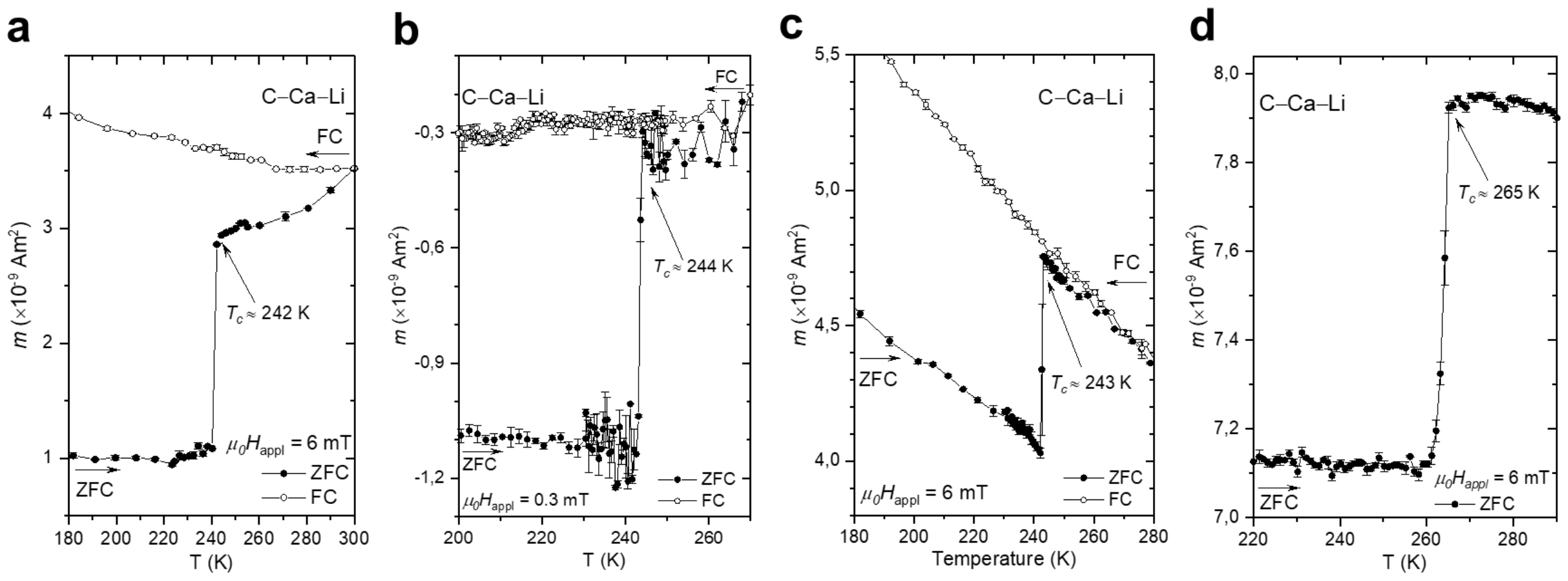

**Figure S14.** Temperature-dependent magnetization, *m(T)*, of graphite samples intercalated with binary Ca-Li alloys under different conditions. a) sample synthesized with Ca-Li (1:3.5) alloy at 623 K for 10 minutes; high-$T_c$ transition at ≈242 K. b) sample synthesized with Ca-Li (1:20) alloy at 623 K for 30 seconds; high-$T_c$ transition at ≈244 K. c) sample synthesized with Ca-Li (1:50) alloy at 723 K for 15 seconds; high-$T_c$ transition at ≈243 K. d) sample synthesized with Ca-Li (1:20) alloy at 523 K for 60 seconds; high-$T_c$ transition at ≈265 K. Applied magnetic fields were perpendicular to graphene layers in the samples. Measurement data points from zero-field cooling (ZFC) and field-cooling (FC) modes (filled/open circles); connecting lines to guide the eye.

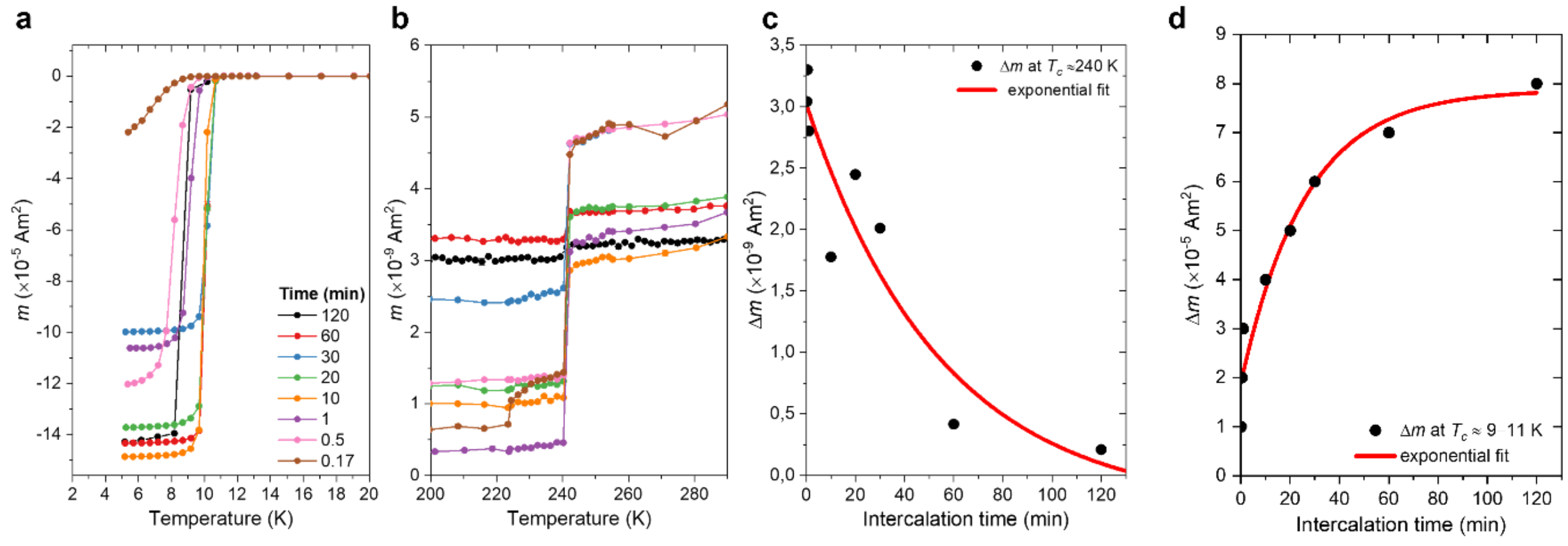

**Figure S15.** Low-$T_c$ and high-$T_c$ transitions of graphite samples intercalated with binary Ca-Li alloys for different reaction times. Panels a) and b) show ZFC *m(T)* data for sample synthesized with Ca-Li (1:3.5) alloy at 623 K with intercalation times from 10 seconds to 120 minutes. Panel a) shows low-$T_c$ superconducting transition at $T_c \approx$ 9–11 K; Panel b) shows high-$T_c$ transition at $T_c \approx$ 240 K. Panels c) and d) show the dependence of the observed magnetization change $\Delta m$ at the superconducting transitions in the low-$T_c$ (Panel a) and high-$T_c$ (Panel b) transition versus intercalation time. Solid line represents the exponential fit of experimental data. Applied magnetic fields were perpendicular to graphene layers in samples.

### S.4.2 Sample intercalated with lithium – Sample I

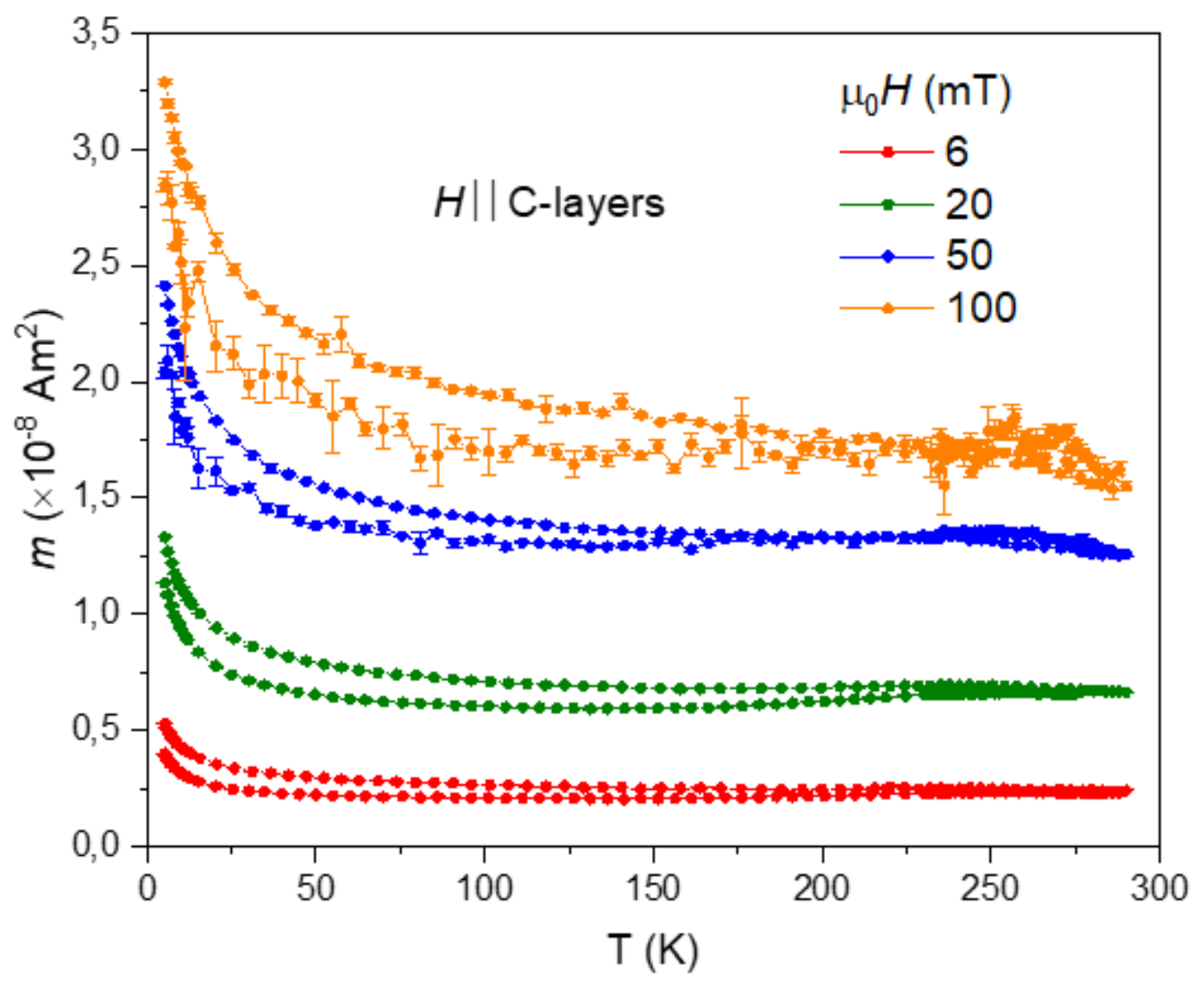

**Figure S16.** Temperature dependent magnetization, *m(T)*, of graphite sample intercalated with lithium only (Sample I, 623 K, 1 min). ZFC-FC measurements are done in different magnetic fields as indicated in the figure labels.

### S.4.3 Samples intercalated with ternary Sr-Ca-Li alloys

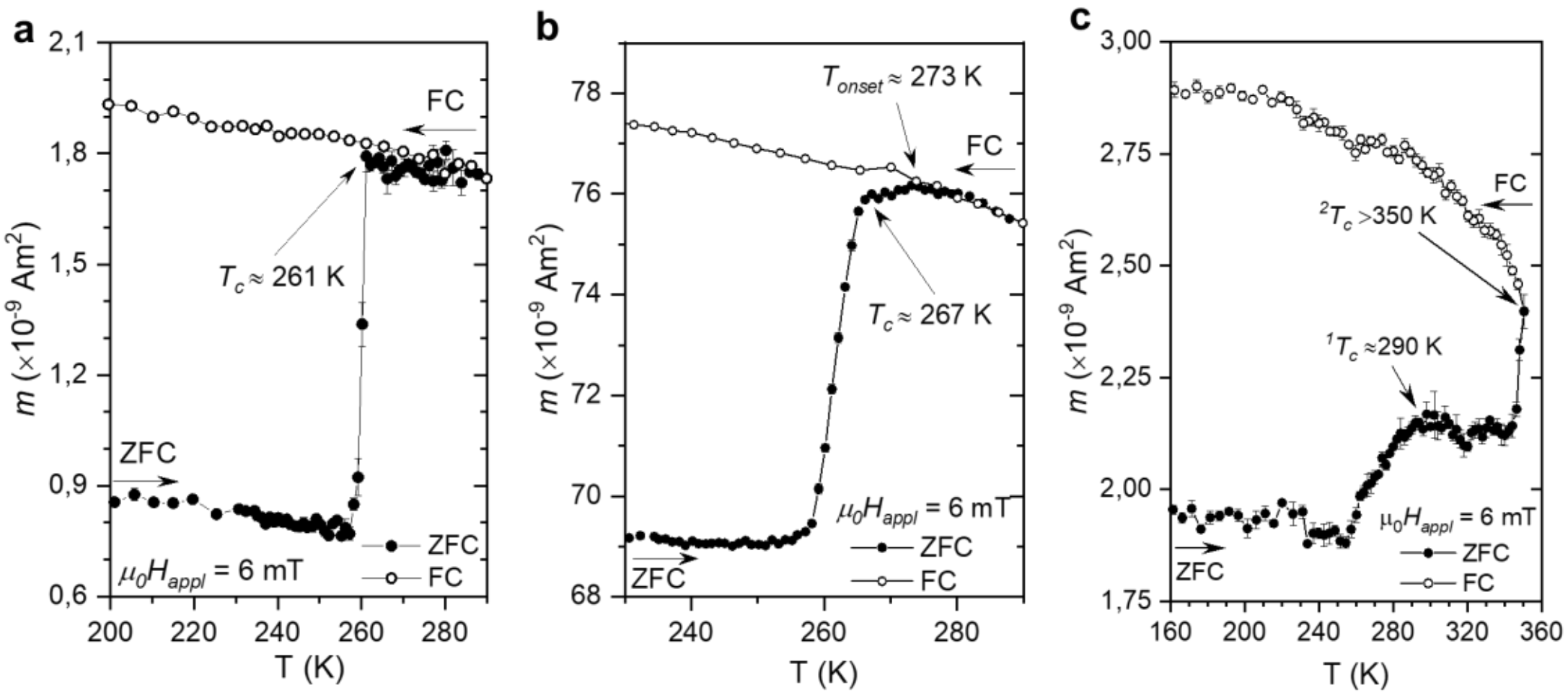

**Figure S17.** Magnetization *m(T)* measurements of graphite samples intercalated with ternary Sr-Ca-Li alloys. a) sample synthesized with Sr- Ca-Li (1:1:20) alloy at 523 K for 1 minute (Sample IIIa). Superconducting transition temperature $T_c \approx 261$ K. b) sample synthesized with Sr- Ca-Li (1:2:20) alloy at $T$ = 623 K for 1 minute. Begin of high-$T_c$ transition is smeared between $T_{onset} \approx 273$ K and $T_c \approx 267$ K. c) sample synthesized with Sr- Ca-Li (1:2:20) alloy at 473 K for 10 minutes. Magnetization data demonstrate the superconducting transition at ${}^2T_c$ above 350 K with a preceding step at ${}^1T_c \approx 290$ K, which may indicate the inhomogeneity of the superconducting phase(s). Applied magnetic fields were perpendicular to graphene layers in samples. Measurement data points from zero-field cooling (ZFC) and field-cooling (FC) modes (filled/open circles); connecting lines to guide the eye.

### S.4.4 Ferromagnetic reference sample of $FeF_3$

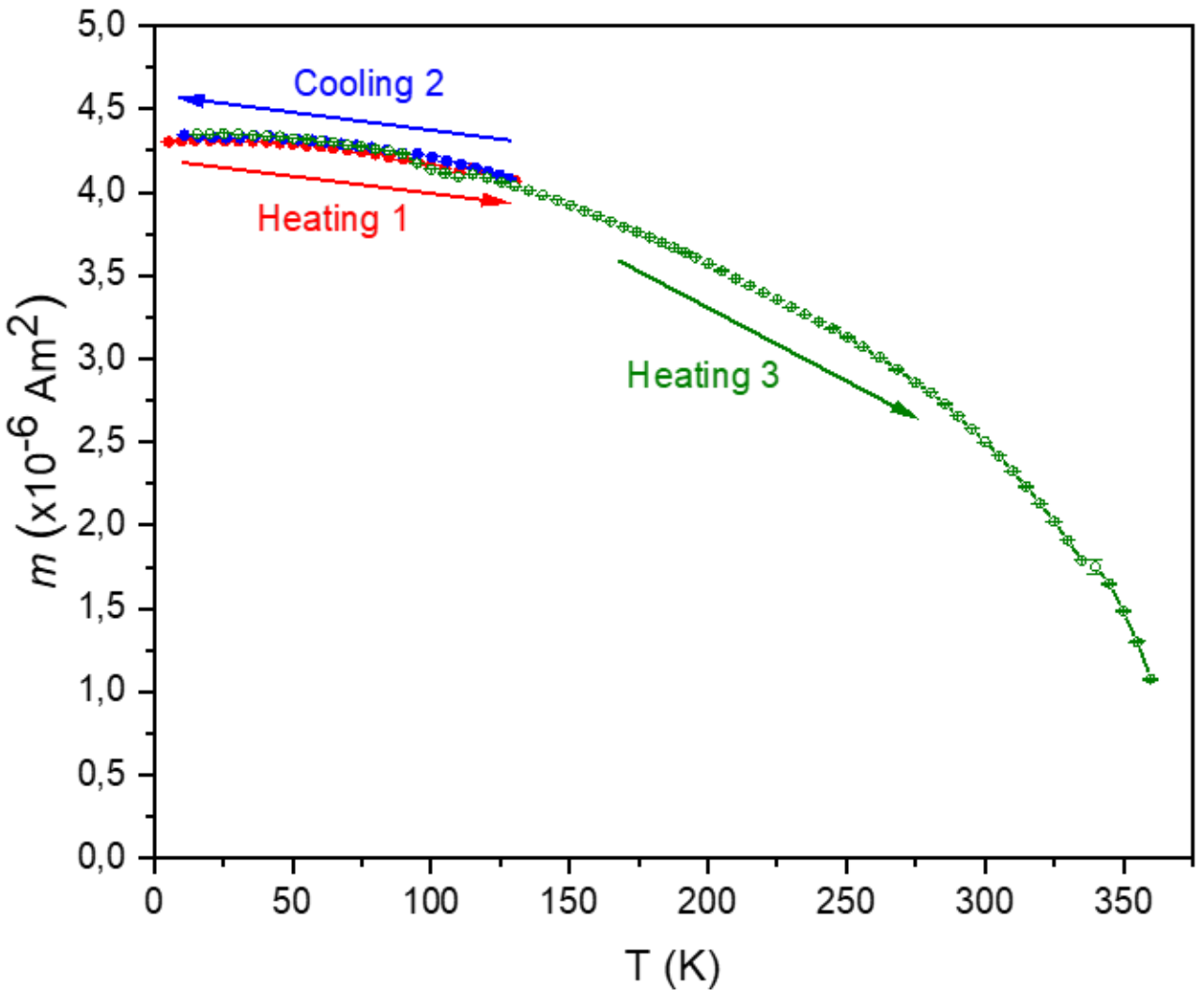

**Figure S18.** Temperature dependence of the remanent magnetization of weak ferromagnetic $FeF_3$ sample. After ZFC cooling a remanent magnetization was created by application and subsequent switching of magnetic field of 1 T at $T$ = 5.2 K. At Heating 1 run the sample was heated to $T$ = 130 K. At reverse Cooling 2 run the sample was cooled down to ca. 11 K and again heated to $T$ = 360 K (Heating 3 run).

### S.4.5 Magnetic field dependence of transitions in graphite intercalated with ternary Sr-Ca-Li alloy

#### S.4.5.1 Sample IIIb

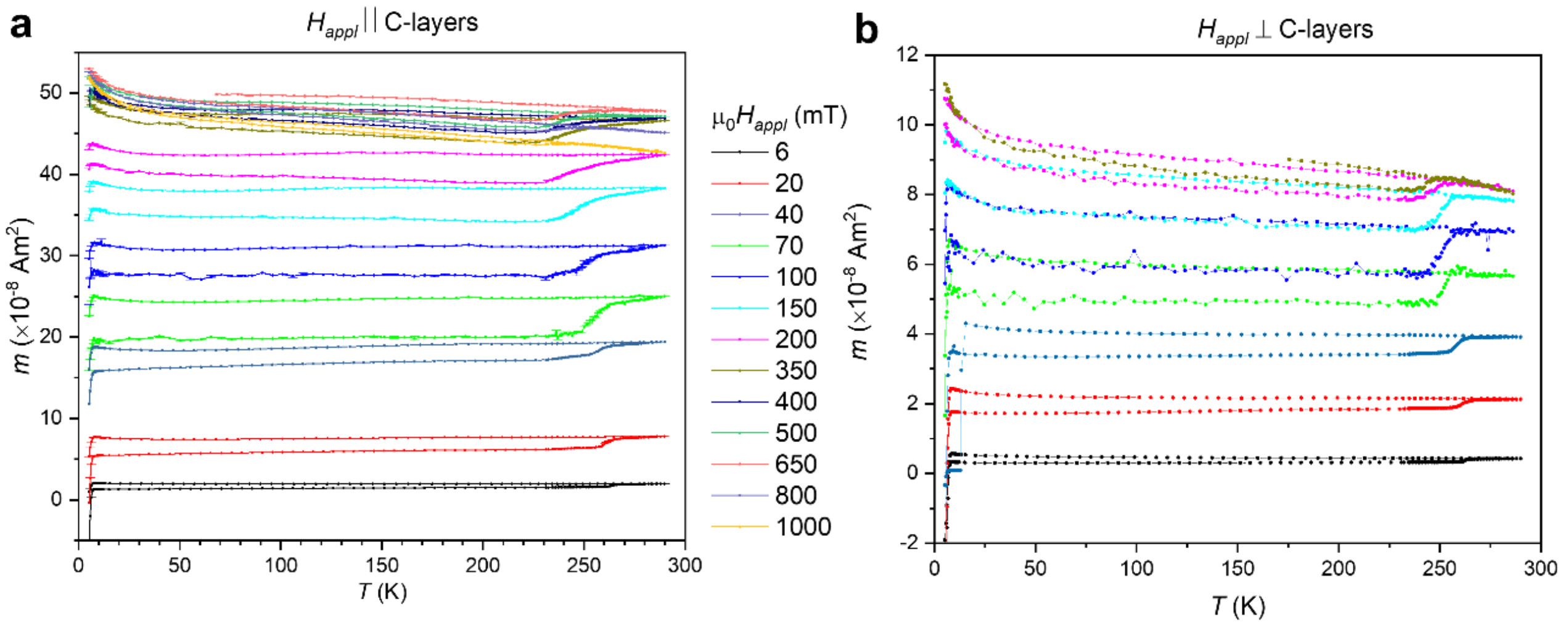

**Figure S19.** Magnetic field dependence of transitions in graphite sample intercalated with ternary Sr-Ca-Li alloy (Sample IIIb): temperature-dependent magnetization, *m(T)*, measured in ZFC and FC modes at different applied magnetic fields in two orientations: a) parallel to graphene layers; and b) perpendicular to graphene layers.

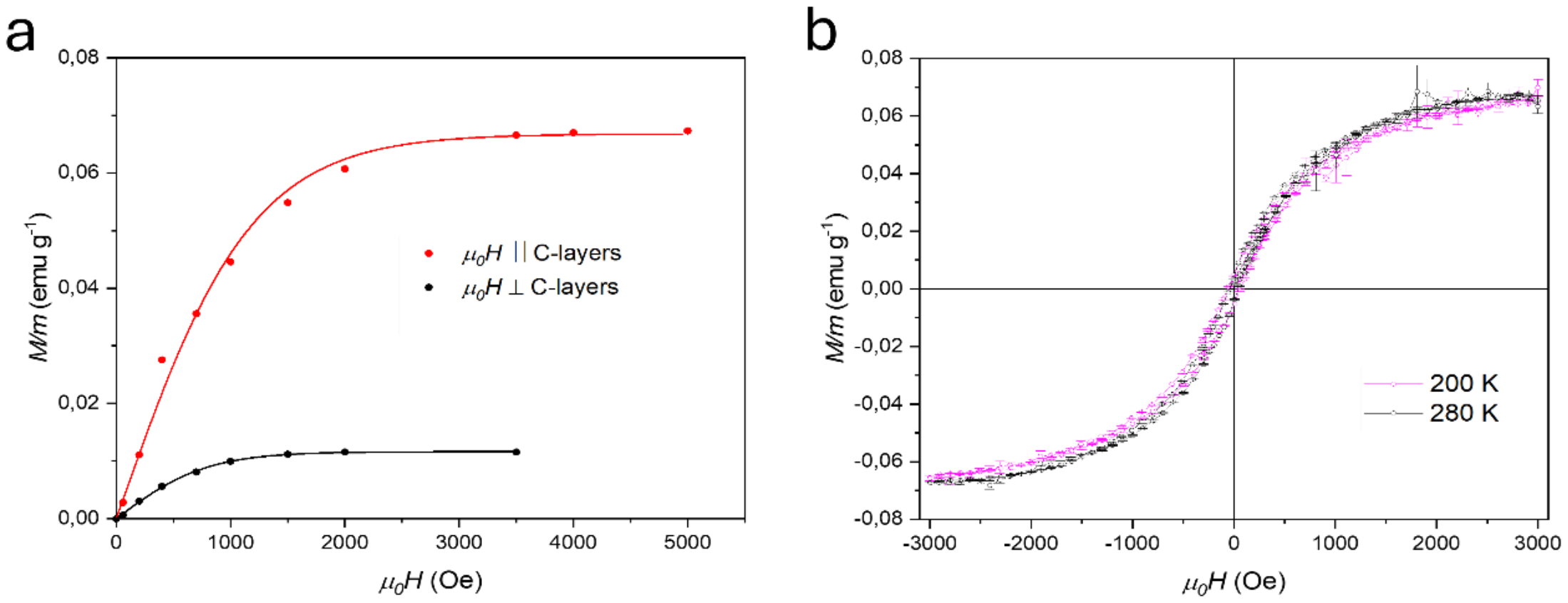

**Figure S20.** a) Magnetic field dependence of the specific magnetic moment, *M/m,* in Sample IIIb, measured parallel (red dots) and perpendicular (black dots) to the graphene layers at T = 285 K, with fit lines according to Eq. 1 (S.4.7). b) Specific magnetic moment, *M/m*, of Sample IIIb as a function of the magnetic field applied parallel to the graphene layers, measured below (T = 200 K, magenta circles) and above (T = 280 K, black circles) the high-$T_c$ transition.

Magnetization measurements of Sample IIIb demonstrate that *M(H)* is anisotropic and dominated by the Pauli paramagnetism of the intercalated graphene matrix. The signal of ferromagnetic impurities in the original EGS (Fig. S3b) is approximately 30 times lower than the saturation magnetization of the intercalated sample. In contrast to graphite intercalated with Ca–$NH_3$ solutions in our companion study,[23] the characteristic signature of superconductivity in the magnetic hysteresis loops is obscured in samples prepared with lithium alloys due to the strong paramagnetic background signal.

### S.4.5.2 Sample IIIc

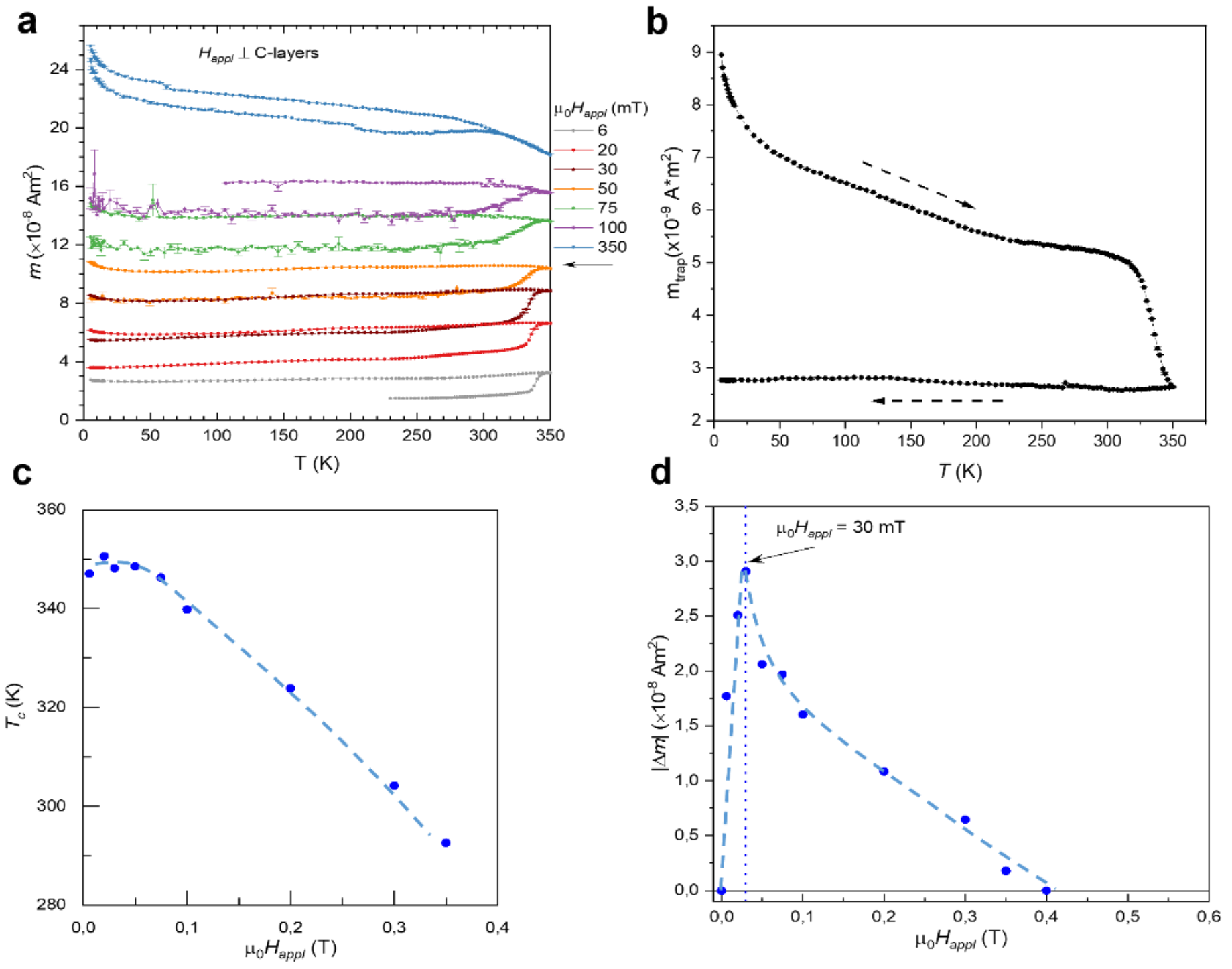

**Figure S21.** Magnetic field dependence of transitions in graphite sample intercalated with ternary Sr-Ca-Li alloy (Sample IIIc): a) *m(T)* data measured in ZFC and FC modes at different magnetic fields applied perpendicular to graphene layers. In contrast to the Sample IIIb, this sample does not show the superconducting transition at low temperatures due to $CaC_6/Ca_2Li_3C_6$ impurity phase. b) Temperature dependence of trapped flux generated at 1 T and 5.2 K indicates high-$T_c$ transition near 350 K. Panel c) shows the magnetic field induced shift of the transition onset temperature $T_c$. Blue dotted lines are guides for eye. d) Magnetic field dependence of the magnetization change $|\Delta m|$ during the high-$T_c$ transition. Magnetic fields are applied perpendicular to graphene layers.

### S.4.5.3 Sample IIId

Sample IIId was synthesized with ternary Sr-Ca-Li (1:2:20) alloy at 423 K for 10 minutes. Temperature behavior of trapped flux generated at $\mu_0 H_M$ = 1 T and $T_M$ = 5.2 K indicates high-$T_c$ transition near 350 K (Fig. 6b of the main text). Temperature-dependent magnetization measurements of this sample are presented in Fig. S21.

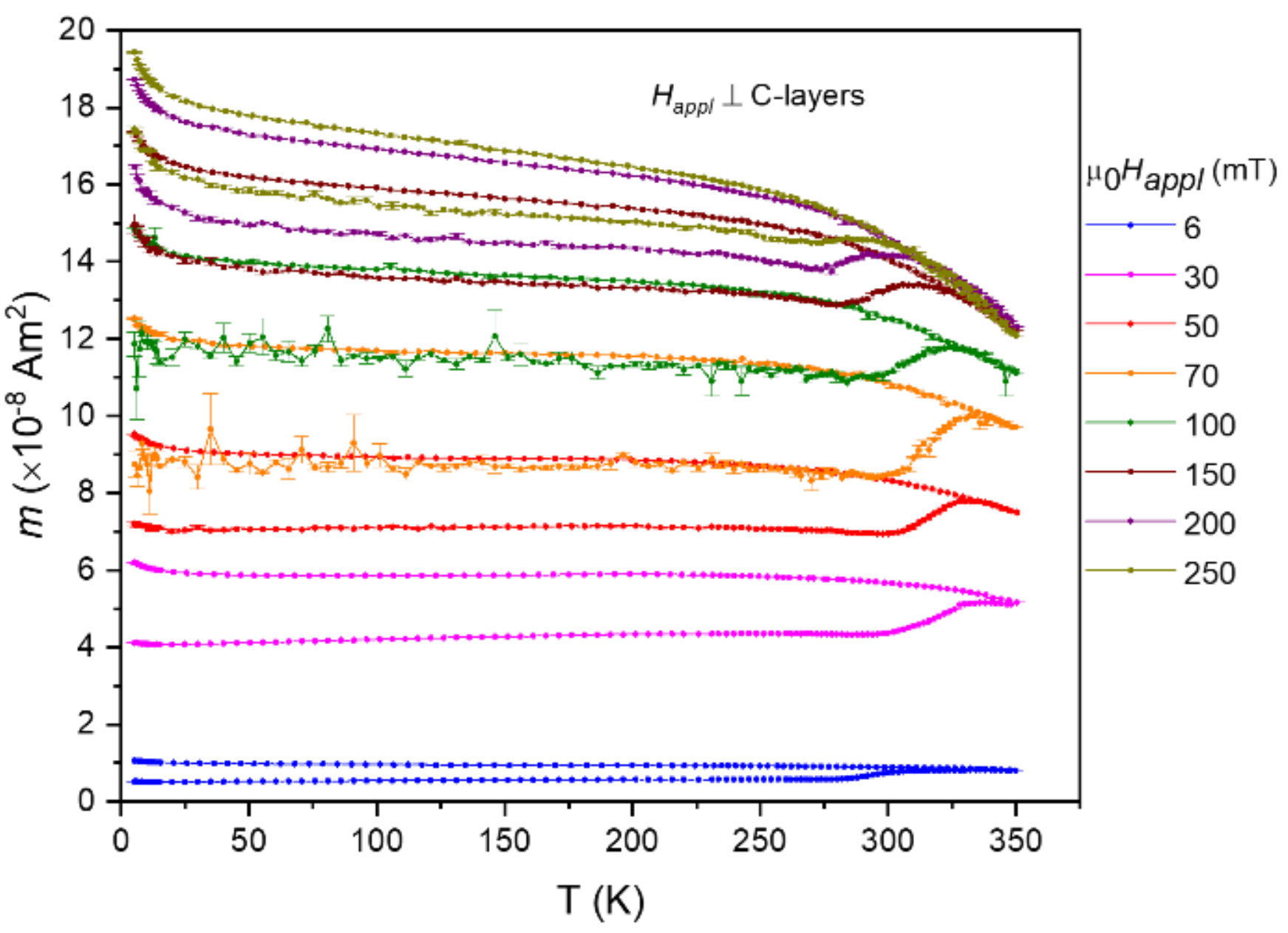

**Figure S22**. Temperature-dependent magnetization, *m(T)*, of Sample IIId used in electrical resistance measurements. ZFC-FC measurements are done in different magnetic fields perpendicular to graphene layers.

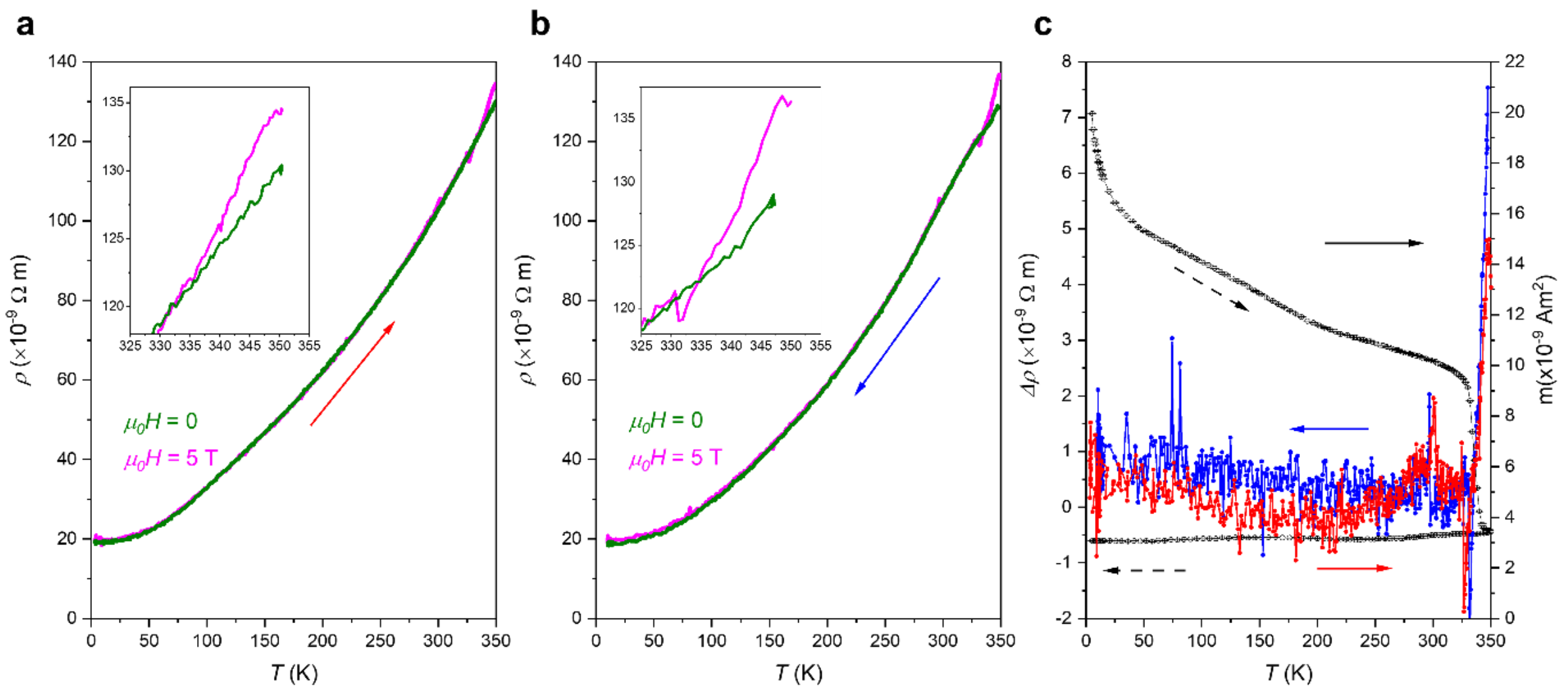

**Figure S23**. Temperature-dependent resistivity and trapping of magnetic flux in graphite sample intercalated with ternary Sr-Ca-Li alloy (Sample IIId): Four-probe resistance measurements of graphite intercalated with ternary Sr-Ca-Li alloy (Sample IIId), green curve at zero magnetic field; magenta curve at 5 T parallel to graphene layers in a) heating and b) cooling mode. Inset shows correspondent resistivities in the upper range of temperatures. c) Difference between resistivities measured in magnetic field of 5 T and zero field (red curve – heating, blue curve - cooling) and trapped magnetic flux (black curve) in Sample IIId. Temperature dependence of trapped flux generated at 1 T and 5.2 K indicates high-$T_c$ transition near 350 K.

Up to 300 K, the differences between the resistivities of Sample IIId measured in a magnetic field of 5 T and zero magnetic field were within the noise level of our resistivity measurements of approx. $2 \cdot 10^{-10}$ Ω m (Fig. S23c). A couple of spikes observed near 300 K and 330 K were slightly higher but still comparable to the noise level. At 350 K, however, the resistivity difference measured at 5 T was by a factor of 5 (heating mode) to 7 (cooling mode) higher than the noise. Its behaviour correlates with the high-$T_c$ transition as indicated by the trapped flux behavior of the Sample IIId at about 350 K (Fig. S23c). This observation can be plausibly explained if we assume that the magnetoresistance of the high- $T_c$ fraction in the normal (non-superconducting) state is substantially higher than the magnetoresistance of the non-superconducting surrounding in the intercalated sample. We suggest and intend to further explore and elucidate this effect in follow-up studies with extended resistance measurements of intercalated graphite samples in variable magnetic fields.

### S.4.6 Characterization of the Crystal Structure and Magnetism in a Selected Ternary Metal-Intercalated EGS

To elucidate the phase composition and formation kinetics within the expanded graphite host matrix during ternary intercalation, X-ray powder diffraction (XRD) measurements were performed on a Sample IIIe. This sample is a ”clone” of Sample IIId and was obtained by intercalating EGS with a ternary Sr-Ca-Li (1:2:20) alloy at 423 K for 10 minutes. This type of sample was chosen because it exhibits the best magnetic properties, featuring the high-$T_c$ anomaly and absence of low-$T_c$ transition (Section **S.4.5.3**).

#### S.4.6.1 Magnetization and Trapped Magnetic Flux in Sample IIIe

Magnetic characterization of the Sample IIIe was carried out via temperature-dependent magnetic susceptibility measurements. ZFC and FC measurements conducted under an applied magnetic field of $\mu_0 H$ = 60 Oe demonstrated a pronounced high-$T_c$ transition occurring at approximately 350 K. This observation is corroborated by trapped magnetic flux measurements. The remanent trapped flux temperature dependence precisely reproduced the same onset transition temperature, confirming that the phenomenon is an intrinsic bulk property of the sample. The low-temperature magnetization part of Sample IIIe showed a complete absence of any diamagnetic shielding or superconducting transitions down to liquid helium temperature, indicating the absence of a separate, macroscopic $CaC_6$ phase.[19] Apparently, magnetic properties of newly synthesized Sample IIIe completely coincide with those of the previously investigated Sample IIId.

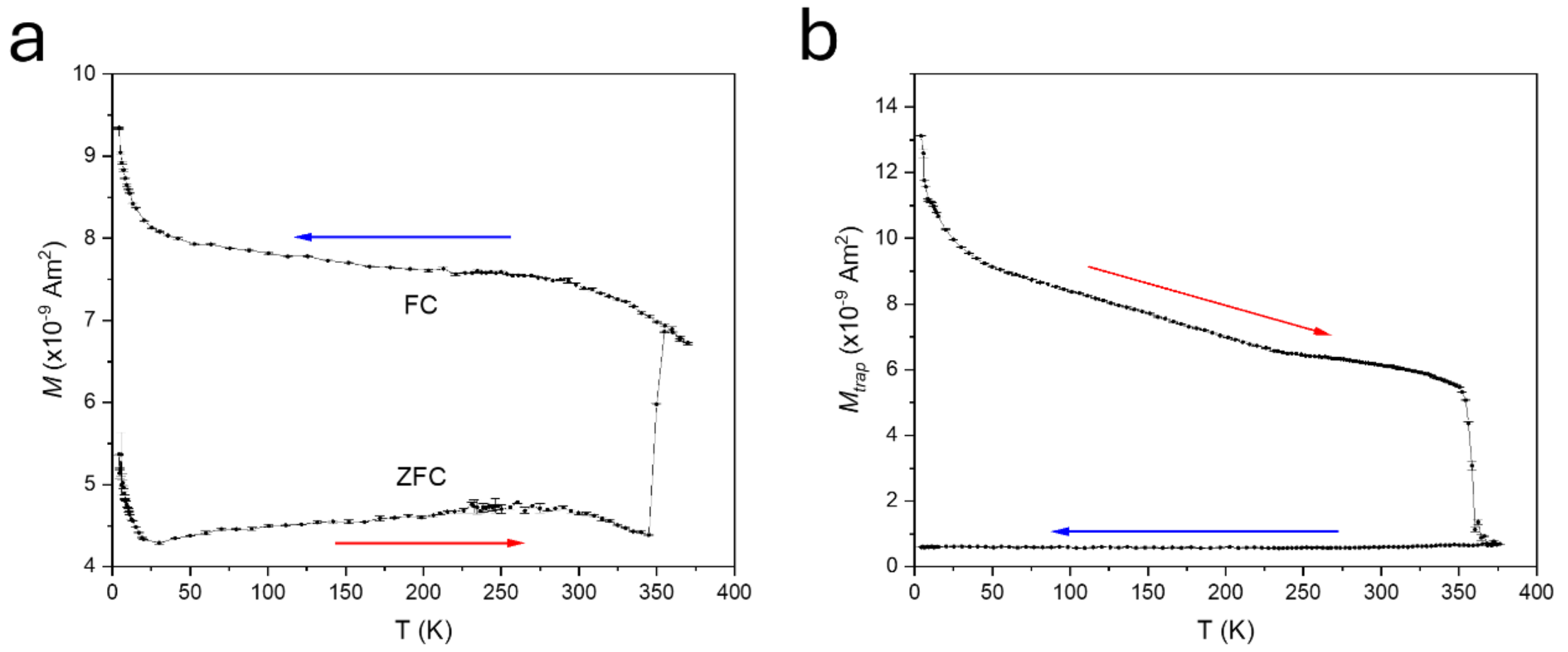

**Figure S24.** a) Temperature-dependent magnetization of Sample IIIe, used in X-ray characterization. ZFC-FC measurements were performed in a magnetic field of $\mu_0 H$ = 60 Oe applied parallel to the graphene layers. b) Trapping of magnetic flux in Sample IIIe. The temperature dependence of the trapped flux - generated at a magnetic field of 1 T applied parallel to the graphene layers at 5.2 K - indicates a high-$T_c$ transition around 350 K.

### S.4.6.2 X-ray Diffraction Analysis of Ternary Metal-Intercalated EGS

XRD measurements of the Sample IIIe were performed using Mo-$K_{\alpha}$ radiation ($\lambda$ = 0.71073 Å). The bulk sample was placed in a sample holder equipped with Kapton foil windows preventing decomposition due to atmospheric exposure. The XRD data obtained for the intercalated Sample IIIe and the EGS reference sample are presented in Figure SXX3. The XRD profile of the intercalated sample reveals a complex multi-phase composition, reflecting the competitive diffusion rates of the alkali and alkaline-earth alloy components within the graphite host.

The low-angle diffractogram range 7° < 2$\theta$ < 10° is dominated by a distinctive series of sharp, low-stage intercalation reflections located at 2$\theta$ = 7.48°, 7.84°, and 8.37°. These correspond to interlayer distances *d* of 5.443 Å, 5.196 Å, and 4.865 Å, respectively. These reflections directly match the (001) long-period spacing of graphite intercalation compounds (GICs). [59] Crucially, a secondary group of low-intensity reflections emerged at higher angles 2$\theta$ = 12.84°, 14.00°, and 15.12°. These peaks originate from secondary (002) structural harmonics of interplanar superlattices. They exhibit near-identical relative intensity ratios and possess *d*-spacings that are exactly halved, with a deviation < 0.05 %. Due to the constrained 10-minute intercalation duration, a significant fraction of the pristine EGS precursor core remained. This is evidenced by a prominent residual pristine (002) core reflection profile centered at 2$\theta$ = 12.1°. However, deconvolution of the line profiles of reflections within the range 18° < 2θ < 22° indicates that this core does not fully preserve the original configuration of the EGS precursor, which initially consisted of approximately 60% hexagonal (H) and 40% rhombohedral (R) phases (Figure S4). Instead, the fraction of the rhombohedral phase is reduced, and broadening has occurred, which also manifests as a left shoulder on the (002) peak.

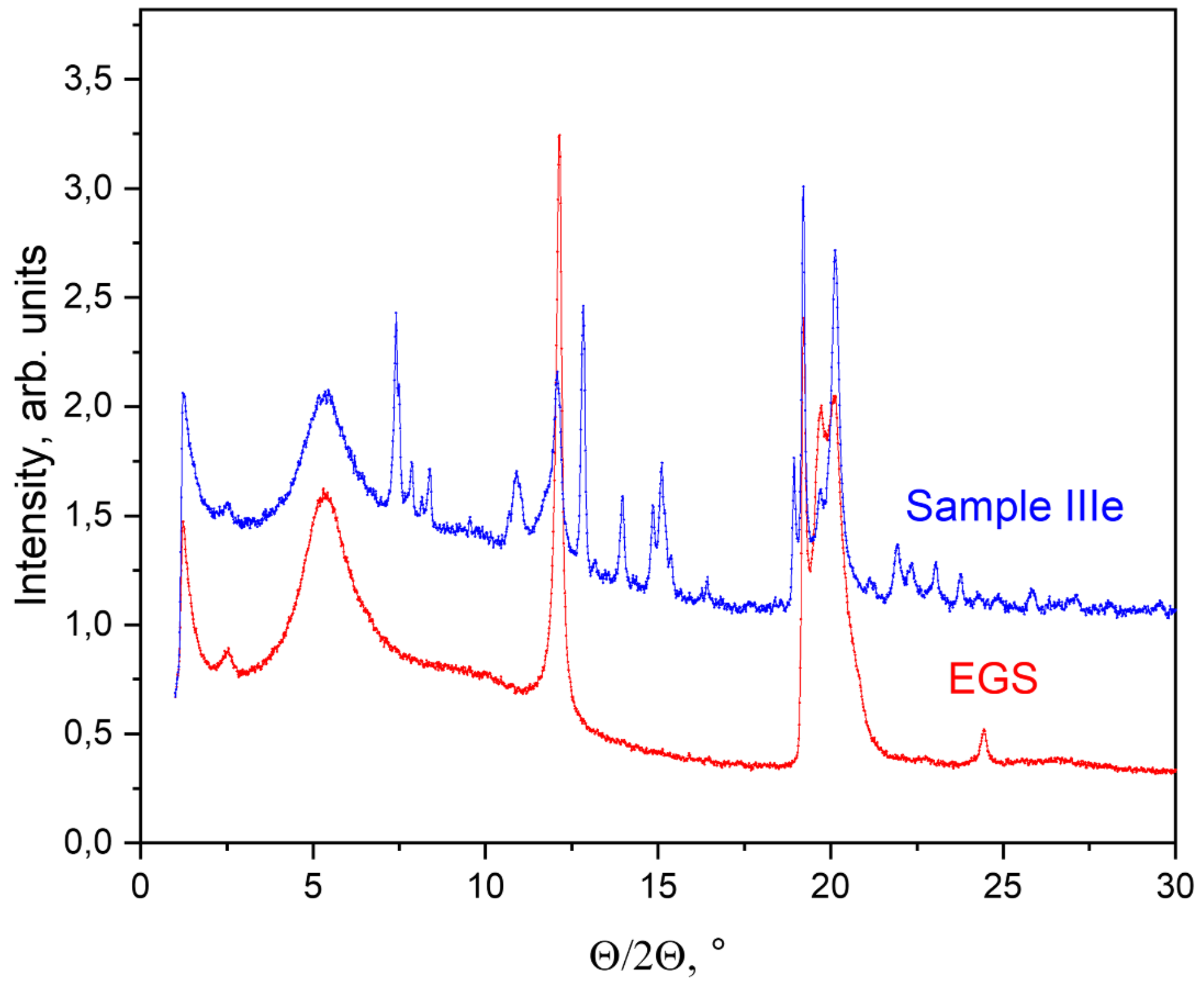

**Figure S25**. X-ray powder diffraction pattern (Mo-Kα, λ = 0.71073 Å) of EGS before and after intercalation with a ternary Sr-Ca-Li (1:2:20) alloy at 423 K for 10 minutes (Sample IIIe). The broad diffraction peak at $2\theta$ = 5.31° originates from the Kapton windows of the sample holder.

An analysis of the specific binary carbide contributions yields following results:

- **Absence of $CaC_6$ and $SrC_6$**. No diffraction signatures matching the characteristic $CaC_6$ Stage-1 GIC compound were observed.[19] The absence of $CaC_6$ in X-ray characterization is confirmed by the absence of a superconducting transition at ca. 11.5 K typically associated with this phase (Section S.4.5.1). This complete absence indicates that the $Ca^{2+}$ ions face a significant kinetic barrier at 150°C, preventing the nucleation of ordered calcium-carbide within the short time window. The absence of $SrC_6$ phase agrees with previous research establishing that it not possible to synthesize pure, bulk binary $SrC_6$ using a liquid lithium-strontium (Li–Sr) alloy.[60, 61]

- **Traces of $LiC_6$**. Distinct diffraction peaks corresponding to minor fraction of Stage-1 $LiC_6$ were indexed at $2\theta$ = 10.90°, 21.91°, and 22.36°. They correspond to reflections (001), (100) and (101), respectively.[62] The accelerated appearance of $LiC_6$ is a direct consequence of the small ionic radius and high mobility of $Li^+$ ions, which permits rapid diffusion between graphene layers and along the flake boundaries.

### S.4.7 Magnetic field dependence of transition temperature

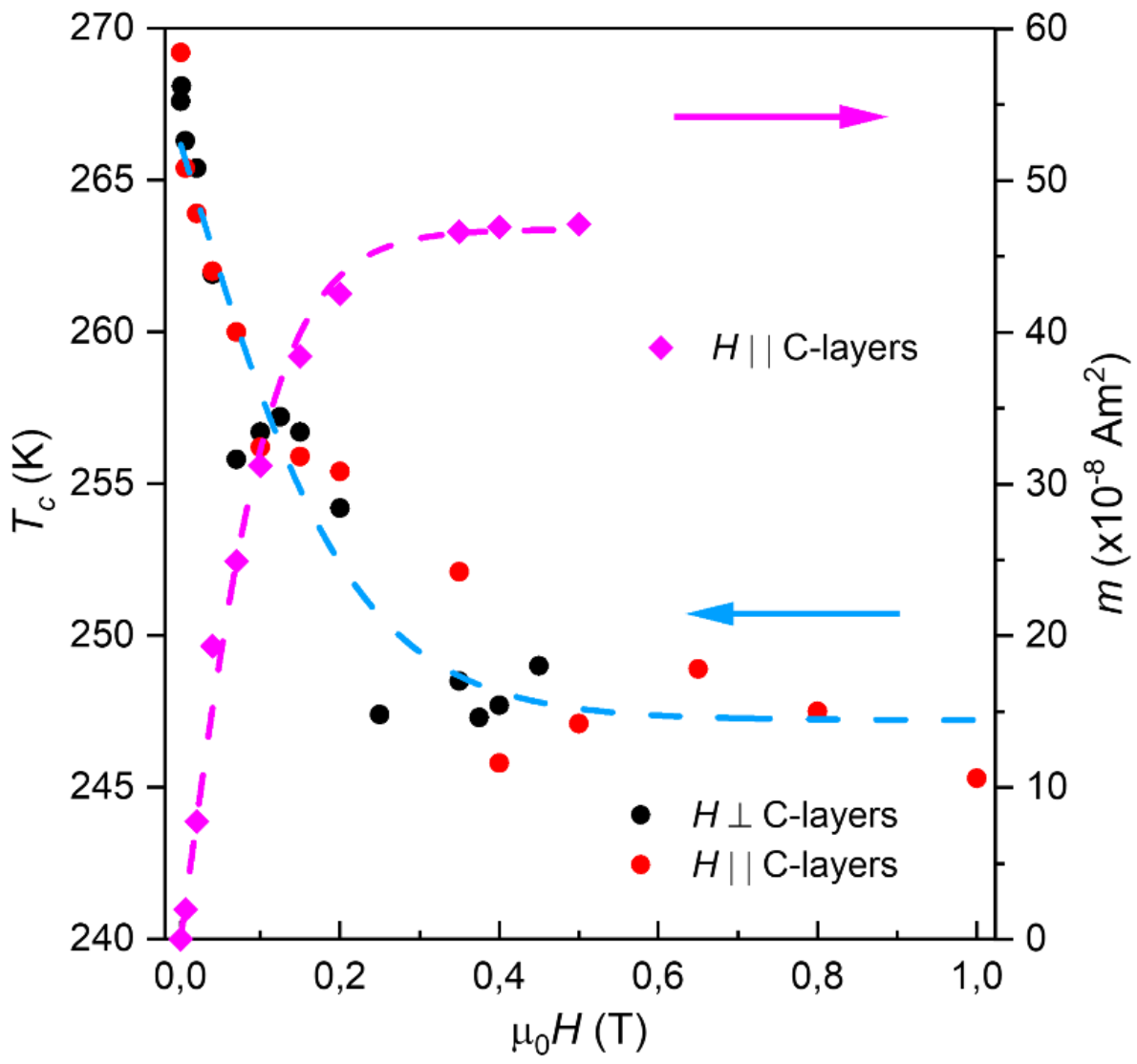

**Figure S26.** Magnetic field dependence of the transition temperature $T_c$ and magnetization in a graphite sample EGS intercalated with ternary Sr-Ca-Li alloy (Sample IIIb). Magnetization measured in magnetic field parallel to carbon layers at T = 285 K (magenta diamonds) with fit curve (dashed magenta line) according to the Eq. 1. Magnetic field dependence of the transition temperature $T_c$ (red and black points) at magnetic fields parallel (red points) and perpendicular (black points) to carbon layers. Dashed blue line is a fit based on the Equation 2.

Figure S26 shows magnetic field dependence of the transition temperature $T_c$ and magnetization *m* in a graphite sample intercalated with ternary Sr-Ca-Li alloy (Sample IIIb). The magnetization is measured at $T$ = 290 K as a function of magnetic field strength in parallel to the graphene layers. After an initial strong increase with the external magnetic field, the measured magnetization of the sample saturates above ~0.3 T, which coincides with the flattening of the field dependence of $T_c$ above ~0.3 T.

The magnetization *m* due to the polarization of the spin ½ conducting electrons in the external field *H* is described by the expression

$$m = N\mu_B \tanh\left(\frac{\mu_B H}{k_B T}\right) \quad (1)$$

where *N* is the amount of conducting electrons, $\mu_B$ and $k_B$ are the Bohr magneton and the Boltzmann constant. A fit of expression (1) to the magnetization data for Sample IIIb reflects the experimental features described above.

As shown in Fig. S22, the magnetic field dependence of $T_c$ demonstrates a similar trend: it demonstrates strong initial decrease with the external magnetic field below and saturates above ~0.3 T. A fit of experimental data $T_c(H)$ based on similar to expression 1 functional dependence

$T_c(H) = P1*\tanh(H*P2)+P3$....(2)

supports the relevance of Pauli paramagnetism for the observed magnetic field dependence of $T_c$. The fit parameters $P1$ = -19(1), $P2$ = 4.6(6), $P3$ = 266.2(9).

### S.4.8 Low-$T_c$ and High-$T_c$ fraction in samples intercalated with Sr–Ca–Li ternary alloys

### S.4.8.1 Estimation of the penetration fields for the low-$T_c$ phase of Sample IIIb

Using the London approximation for superconductors of second order, the out-of-plane lower critical field at zero temperature is

$$\mu_0 H_{c1\perp}(0) = \frac{\Phi_0}{4\pi\lambda_{ab}^2}\ln(\kappa)$$

With magnetic flux quantum $\Phi_0 = 2.0678 \cdot 10^{-15}$ T m$^2$, in-plane penetration depth $\lambda_{ab}$ = 62(4) nm, Ginzburg-Landau parameter κ = 1.4(2)[33] the out-plane lower critical field at zero temperature is $\mu_0 H_{c1\perp}(0) = 14.4 \pm 6.4$ mT.

The in-plane lower critical field at zero temperature is

$$\mu_0 H_{c1||}(0) = \frac{\Phi_0}{4\pi\lambda_{ab}\lambda_c}\ln(\kappa) = \frac{\Phi_0}{4\pi\gamma\lambda_{ab}^2}\ln(\kappa)$$

Here the anisotropy ratio $\gamma_\lambda = \frac{\lambda_c}{\lambda_{ab}}$.

With $\gamma_\lambda = 5.1(4)$[34] the in-plane lower critical field at zero temperature is $\mu_0 H_{c1||}(0) = 2.9 \pm 1.2$ mT.

Penetration (first vortex entry) field is lower than the first critical field due to the demagnetization factor *N*:

$$B_p = \mu_0 H_{c1} \cdot (1 - N)$$

Demagnetization factor can be calculated using the rectangular prism formula:[35]

$$N = \frac{4ab}{4ab + 3c(a+b)}$$

For Sample IIIb with dimensions 2a = 3.0(1) mm, 2b = 2.3(1) mm and 2c = 1.0(1) mm the demagnetization factor is *N* = 0.6345. Corresponding out-of-plane penetration field is $B_{p\perp}$= 5.3±2.3 mT. For the in-plane geometry $B_{p||}$= 2.1±1 mT.

The algorithm for estimating the low-$T_c$ and high-$T_c$ fractions for sample IIIb is described in Section 2.5 in the main text and below.

### S.4.8.2 High-$T_c$ fraction in samples intercalated with Sr–Ca–Li ternary alloys

In Table S4 summarized the data and results of estimation of high-$T_c$ fractions for Samples IIIb, IIIc and IIId intercalated with Sr–Ca–Li ternary alloys. It shows volume magnetic susceptibilities $\chi_{V\perp}$ (magnetic field *H* perpendicular to faces of flakes responsible for high-$T_c$ transition) and $\chi_{V||}$ (magnetic field *H* parallel to faces of flakes) calculated from behaviour of magnetization $|\Delta m|$ during high-$T_c$ transition (Section 2.5 of the manuscript). Magnetic measurements of a lead reference sample with purity 99.9999 having the same shape and dimensions as sample IIIb yield a susceptibility value of $\chi_V$ = -1.53(1).

Based on the magnetization measurement results for samples IIIc and IIId presented above (Figs. S.4.5.2 and S.4.5.3) and the corresponding change in magnetization $|\Delta m(H)|$ at the high-temperature transition, the volume susceptibilities $\chi_{V\parallel}$ were calculated. The volume susceptibility $\chi_{V\perp}$ (with the magnetic field $H$ perpendicular to faces of flakes) for samples IIIc and IIId was calculated using the ratio $\gamma_\chi = \frac{\chi_{V\perp}}{\chi_{V\parallel}}$ = 4.3(6) of sample IIIb. Based on this assumption, the high-$T_c$ fractions of Samples IIIc and IIId were evaluated.

**Table S4.** Estimation of the volume of low- $T_c$ and high-$T_c$ fractions in samples intercalated with Sr–Ca–Li ternary alloys.

| Sample name | Metal/Alloy (atomic ratio)/ Intercalation temperature (K)/ Intercalation time (min) | Sample volume ($cm^3$) | Volume susceptibilities $\|\chi_{V\perp}\|$ and $\|\chi_{V\parallel}\|$ (SI units) | Volume fraction $f$ (%) |
|---|---|---|---|---|
| Sample IIIb. High-$T_c$ fraction | Sr-Ca-Li (1:2:20)/ 623 K/ 1 min | $6.9(1)\cdot10^{-3}$ | $\|\chi_{V\parallel}\| = 2.3(2)\cdot10^{-5}$ $\|\chi_{V\perp}\| = 1,0(1)\cdot10^{-4}$ | $6.5(7)\cdot10^{-3}$ |
| Sample IIIb. Low-$T_c$ fraction | | | $\|\chi_{V\perp}\| = 1.7(1)\cdot10^{-2}$ $\|\chi_{V\parallel}\| = 2.9(1)\cdot10^{-3}$ | 1.1(1) |
| Sample IIIc. High-$T_c$ fraction | Sr-Ca-Li (1:2:20)/ 473 K/ 10 min | $6.3(1)\cdot10^{-3}$ | $\|\chi_{V\parallel}\| = 2.50(1)\cdot10^{-4}$ $\|\chi_{V\perp}\| = 1.1(2)\cdot10^{-3}$ | $7.2(13)\cdot10^{-2}$ |
| Sample IIId. High-$T_c$ fraction | Sr-Ca-Li (1:2:20)/ 423 K/ 10 min | $5.2(1)\cdot10^{-3}$ | $\|\chi_{V\parallel}\| = 1.13(5)\cdot10^{-4}$ $\|\chi_{V\perp}\| = 4.9(7)\cdot10^{-4}$ | $3.2(5)\cdot10^{-2}$ |

### S.4.9 HOPG samples intercalated by ternary Sr–Ca–Li alloy

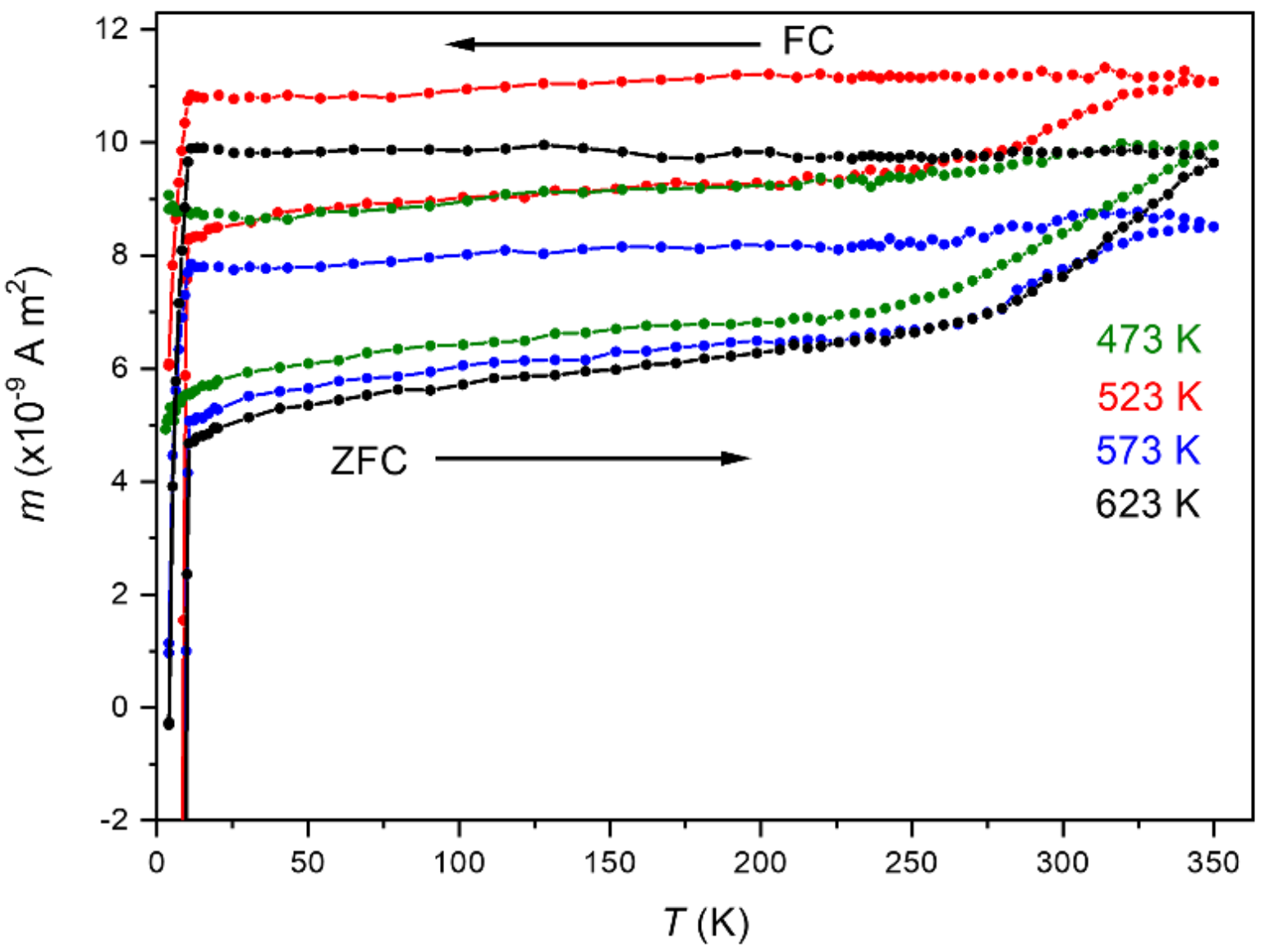

**Figure S27.** Temperature-dependent magnetization, *m(T)*, of HOPG samples. HOPG intercalated by ternary Sr-Ca-Li alloy (atomic ratio 1:2:20) for 1 minute at 473 K (green points), 523 K (red points), 573 K (blue points) and 623 K (black points). ZFC-FC measurements are done in magnetic field of 6 mT. The low-$T_c$ transition near 11.5 K attributed to known superconducting phases of $CaC_6$ or $Ca_2Li_3C_6$. The high-$T_c$ transition around room temperature is significantly smeared.

### S.4.10 Samples intercalated with other alloys: K–Li, Sr–Li, Mg–Li, Zn–Li, and Mg–Zn–Li

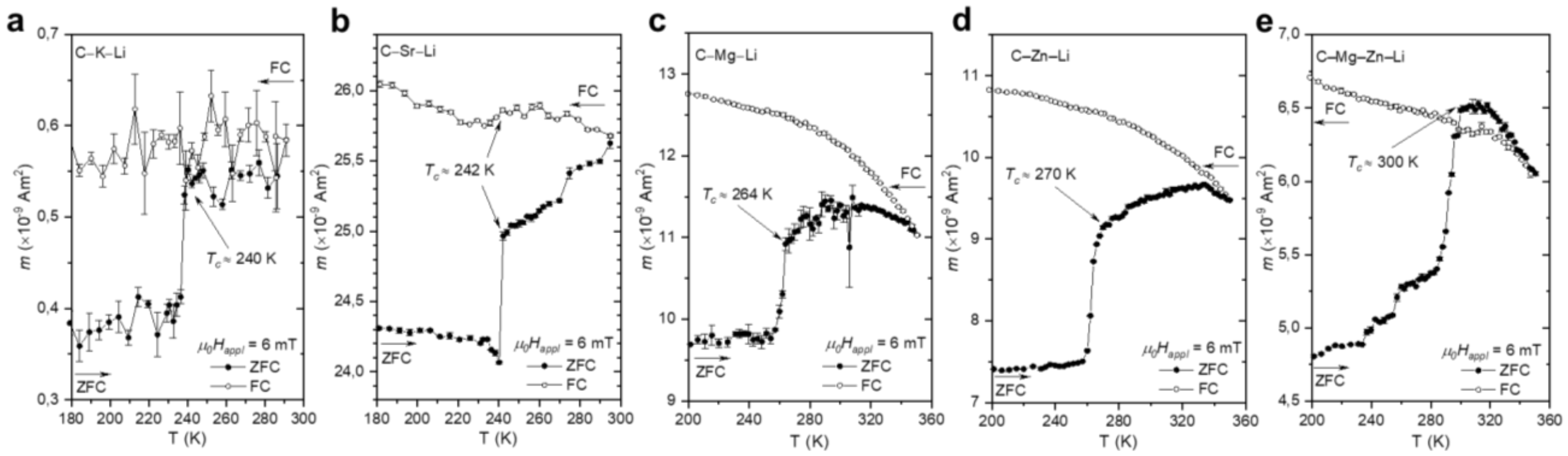

**Figure S28.** Temperature-dependent magnetization, *m(T)*, of graphite samples intercalated with K–Li, Sr–Li, Mg–Li, Zn–Li, and Mg–Zn–Li alloys. a) sample synthesized with K-Li (1:3, 523 K, 30 seconds); high-$T_c$ transition at ≈240 K. b) sample synthesized with Sr-Li (1:3, 623 K, 60 seconds); high-$T_c$ transition at ≈242 K. c) sample synthesized with Mg-Li (1:4, 623 K, 5 minutes); high-$T_c$ transition at ≈264 K d) sample synthesized with Zn-Li (1:4, 573 K, 5 minutes); high-$T_c$ transition at ≈270 K. e) sample synthesized with Mg Zn-Li (1:1:8, 523 K, 4 minutes); high-$T_c$ transition at ≈300 K. Applied magnetic fields were perpendicular to graphene layers in the samples.

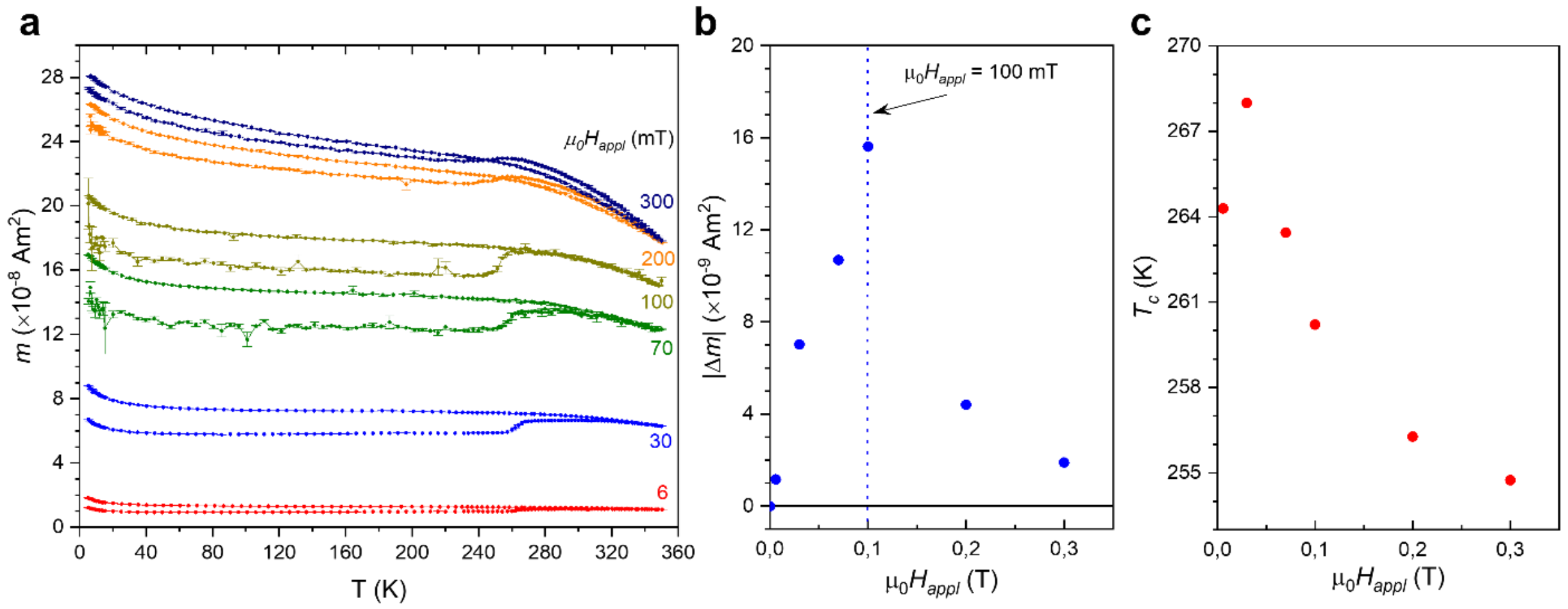

**Figure S29.** Magnetic field dependence of superconducting transitions in graphite samples intercalated with Mg–Li alloy. The sample was synthesized with Mg-Li (1:4) alloy at 523 K for 5 minutes. a) Temperature dependence of magnetization *m(T)* at different magnetic fields $H_{appl}$ applied perpendicular to graphene layers. b) Magnetic field dependence of the magnetization change $|\Delta m|$ during the high-$T_c$ transition. c) Shift of the transition temperature $T_c$ in applied magnetic fields.

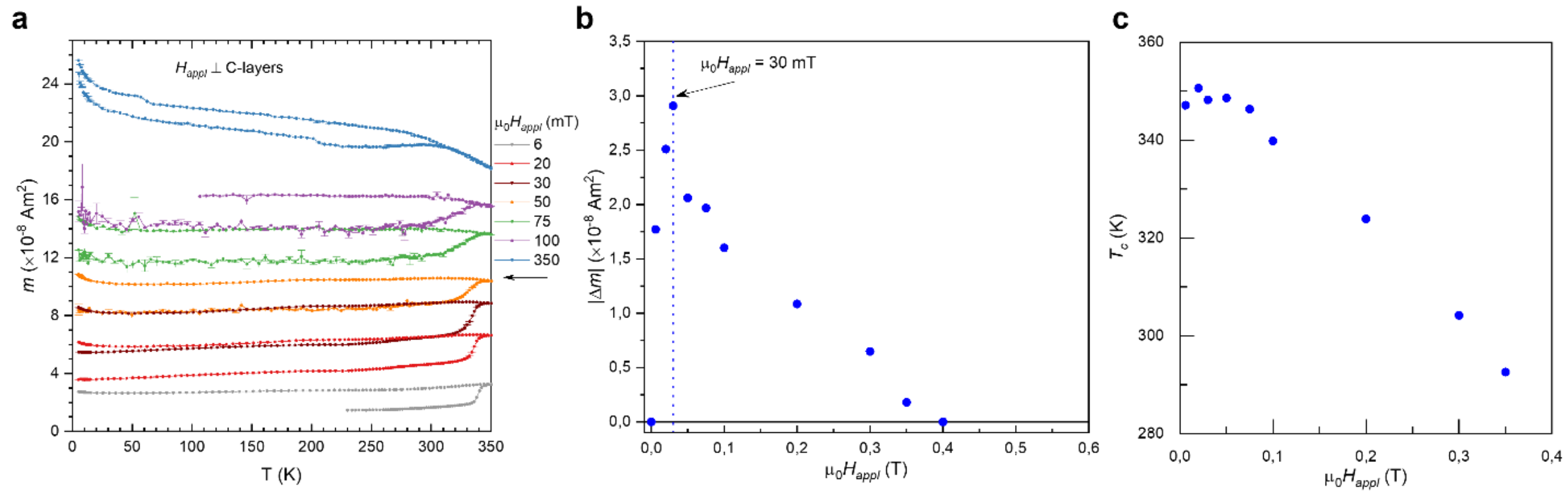

**Figure S30.** Magnetic field dependence of superconducting transitions in graphite samples intercalated with Zn–Li alloy. The sample was synthesized with Zn-Li (1:4) alloy at 573 K for 5 minutes. a) Temperature dependence of magnetization *m(T)* at different magnetic fields $H_{appl}$ applied perpendicular to graphene layers. b) Magnetic field dependence of the magnetization change $|\Delta m|$ during the high-$T_c$ transition. c) Shift of the transition temperature $T_c$ in applied magnetic fields.